\documentclass[final,5p,times,twocolumn,authoryear,usenatbib]{elsarticle}

\usepackage{epsfig}
\usepackage{float}
\usepackage{pdflscape}
\usepackage[dvipsnames]{xcolor}
\usepackage{hyperref}\hypersetup{
    colorlinks=true,
    linkcolor=blue,
    filecolor=magenta,      
    urlcolor=cyan,
    pdftitle={Overleaf Example},
    pdfpagemode=FullScreen,
    }
\usepackage{amssymb}
\usepackage[]{natbib}
\usepackage[varvw]{newtxmath}  

\journal{New Astronomy Reviews}

\begin{document}

\begin{frontmatter}



\title{Observations of pre- and proto-brown dwarfs in nearby clouds:\\
paving the way to further constraining theories of brown dwarf formation}


\author{Aina Palau$^\mathrm{a}$}
\affiliation{organization={Instituto de Radioastronomia y Astrofisica, Universidad Nacional Autonoma de Mexico},
            addressline={Antigua Carretera a Patzcuaro 8701, Ex-Hda. San Jose de la Huerta}, 
            city={Morelia},
            postcode={58089}, 
            state={Michoacan},
            country={Mexico}}

\author{Nuria Hu\'elamo$^\mathrm{b}$}

\author{David Barrado$^\mathrm{b}$}
\affiliation{organization={Centro de Astrobiologia (CAB), CSIC-INTA},
            addressline={Camino Bajo del Castillo s/n}, 
            city={Villanueva de la Canada},
            postcode={28692}, 
            state={Madrid},
            country={Spain}}

\author{Michael M. Dunham$^\mathrm{c,d}$}
\affiliation{organization={Department of Physics},
            addressline={Middlebury College}, 
            city={Middlebury},
            postcode={05753}, 
            state={VT},
            country={USA}}
\affiliation{organization={Department of Physics, State University of New York at Fredonia},
            addressline={280 Central Avenue}, 
            city={Fredonia},
            postcode={14063}, 
            state={NY},
            country={USA}}

\author{Chang Won Lee$^\mathrm{e,f}$}
\affiliation{organization={Korea Astronomy and Space Science Institute (KASI)},
            addressline={776 Daedeokdae-ro}, 
            city={Yuseong-gu},
            postcode={34055}, 
            state={Daejeon},
            country={Republic of Korea}}
\affiliation{organization={University of Science and Technology, Korea (UST)},
            addressline={217 Gajeong-ro}, 
            city={Yuseong-gu},
            postcode={34113}, 
            state={Daejeon},
            country={Republic of Korea}}

\begin{abstract}
Brown Dwarfs (BDs) are crucial objects in our understanding of both star and planet formation, as well as in our understanding of what mechanisms shape the lower end of the Initial Mass Function (IMF). However, since the discovery of the first BD in 1995, there is still an unconcluded debate about which is the dominant formation mechanism of these objects. For this, it is mandatory to study BDs in their earliest evolutionary stages (what we call pre- and proto-BDs), comparable to the `pre-stellar' and `Class 0/I' stages well characterized for the formation of low-mass stars. In this review, the recent efforts aimed at searching, identifying and characterising pre- and proto-BD candidates in nearby star-forming regions are presented, 
and revised requirements for an object to be a promising proto-BD or pre-BD candidate are provided, based on a new, unexplored so far, relation between the internal luminosity and the accreted mass.
%
By applying these requirements, a list of 68 promising proto-BD candidates is presented, along with a compilation of possible pre-BDs from the literature.
In addition, updated correlations of protostellar properties such as mass infall rate or outflow momentum rate with bolometric luminosity are provided down to the low-mass BD regime, where no significant deviations are apparent.
%
Furthermore, the number proto-BD candidates in different clouds of the Solar Neighborhood seem to follow the known relations of number of protostars with cloud properties. 
In addition, proto(star-to-BD) ratios for the different clouds are also explored,
%
unveiling a particular underproduction of low-mass proto-BD candidates 
in Ophiuchus compared to Lupus and Taurus. Possible explanations for this behavior are discussed, including heating of the Ophiuchus cloud by the nearby OB stars. 
The overall results of this work, along with the possibility that the planetary-mass regime of the IMF is subtly shaped by stellar feedback, tend to favor a Jeans-fragmentation process and therefore a star-like formation scenario down to the planetary boundary, of $\sim0.01$~$M_\odot$, below which other mechanisms might be at work.
Future observational constraints, such as those provided by upcoming facilities like the next-generation Very Large Array, or the use of isotopologues ($^{13}$CO, $^{15}$NH$_3$, etc.) based on James Webb Space Telescope data, will provide definite clues to disentangle the origin of BDs in the planetary-mass regime.

\end{abstract}




\end{frontmatter}


\newcommand{\nhu}[1]{\textcolor{Plum}{ #1}} 
\newcommand{\db}[1]{\textcolor{red}{ #1}}   
\newcommand{\md}[1]{\textcolor{OliveGreen}{ #1}} 
\newcommand{\cwl}[1]{\textcolor{NavyBlue}{ #1}} 
\newcommand{\src} {SMM2}
\newcommand{\Td}    {T_\mathrm{d}}
\newcommand{\Tex}   {T_\mathrm{ex}}
\newcommand{\Trot}  {T_\mathrm{rot}}
\newcommand{\Tbol}  {T_\mathrm{bol}}
\newcommand{\Menv}  {M_\mathrm{env}}
\newcommand{\Macc}  {M_\mathrm{acc}}
\newcommand{\Mdyn}  {M_\mathrm{dyn}}
\newcommand{\Minfrate}  {\dot{M}_\mathrm{inf}}
\newcommand{\Mvir}  {M_\mathrm{vir}}
\newcommand{\Mgrav}  {M_\mathrm{grav}}
\newcommand{\MBE}  {M_\mathrm{BE}}
\newcommand{\Mjeans}  {M_\mathrm{Jeans}}
\newcommand{\Lbol}  {L_\mathrm{bol}}
\newcommand{\Lsubmm}  {L_\mathrm{submm}}
\newcommand{\Lint}  {L_\mathrm{int}}
\newcommand{\Lacc}  {L_\mathrm{acc}}
\newcommand{\Lcm}  {L_\mathrm{cm}}
\newcommand{\Fout}  {F_\mathrm{out}}
\newcommand{\Nbd}   {N_\mathrm{proBD}}
\newcommand{\Nvello} {N_\mathrm{VeLLO}}
\newcommand{\Nyso}  {N_\mathrm{YSO}}
\newcommand{\Nproto}  {N_\mathrm{protostar}}
\newcommand{\Nexp}  {N_\mathrm{expected}}
\newcommand{\reff}  {r_\mathrm{eff}}
\newcommand{\Eg}   {E_\mathrm{g}}
\newcommand{\Ek}   {E_\mathrm{k}}
\newcommand{\mum}   {$\mu$m}
\newcommand{\kms}   {km~s$^{-1}$}
\newcommand{\cmg}   {cm$^{2}$~g$^{-1}$}
\newcommand{\cmt}   {cm$^{-3}$}
\newcommand{\jpb}   {$\rm Jy~beam^{-1}$}    
\newcommand{\lo}    {$L_{\odot}$}
\newcommand{\mo}    {$M_{\odot}$}
\newcommand{\ro}    {$R_{\odot}$}
\newcommand{\mj}    {$M_\mathrm{Jup}$}
\newcommand{\co}    {$^{12}$CO}
\newcommand{\tco}    {$^{13}$CO}
\newcommand{\ceo}    {C$^{18}$O}
\newcommand{\nh}    {NH$_3$}
\newcommand{\nth}   {N$_2$H$^+$}
\newcommand{\methanol}  {CH$_3$OH}
\newcommand{\water}  {H$_2$O}
\newcommand{\otcs} {O$^{13}$CS}
\newcommand{\et}    {et al.}
\newcommand{\eg}    {e.\,g.,}
\newcommand{\ie}    {i.\,e.,}
\newcommand{\hii}   {H{\small II}}
\newcommand{\uchii} {UC~H{\small II}}
\newcommand{\hchii} {HC~H{\small II}}
\newcommand{\raun}  {$^\mathrm{h~m~s}$}
\newcommand{\deun}  {$\mathrm{\degr~\arcmin~\arcsec}$}
\newcommand{\supa}  {$^\mathrm{a}$}
\newcommand{\supb}  {$^\mathrm{b}$}
\newcommand{\supc}  {$^\mathrm{c}$}
\newcommand{\supd}  {$^\mathrm{d}$}
\newcommand{\supe}  {$^\mathrm{e}$}
\newcommand{\supf}  {$^\mathrm{f}$}
\newcommand{\supg}  {$^\mathrm{g}$}
\newcommand{\suph}  {$^\mathrm{h}$}
\newcommand{\supi}  {$^\mathrm{i}$}
\newcommand{\phnn}  {\phantom{0}\phantom{0}}
\newcommand{\phe}   {\phantom{$^\mathrm{c}$}}
\newcommand{\phb}   {\phantom{$>$}}
\newcommand{\phnm}  {\phantom{0}\phantom{$.$}}
\newcommand{\phbn}  {\phantom{$>0$}}
\newcommand{\phbnn} {\phantom{$>00$}}
\newcommand{\phnb}  {\phantom{0}\phantom{$>$}}
\newcommand{\phmm}  {\phantom{\pm0.0}}
\newcommand{\phmn}  {\phantom{0\pm0.00}}
\newcommand{\phda}  {\phantom{\dag}}
\newcommand{\apjl}  {Astrophys. Journal Letters}
\newcommand{\apj}   {Astrophys. Journal}
\newcommand{\aap}   {Astronomy \& Astrophysics}
\newcommand{\mnras} {Monthly Notice of the Royal Astronomical Society}
\newcommand{\aj}    {Astron. Journal}
\newcommand{\apjs}  {Astrophysical Journal Supplement Series}
\newcommand{\aapr}  {Astronomy and Astrophysics Reviews}
\newcommand{\zap}  {Zeitschrift fuer Astrophysik}
\newcommand{\araa}  {Annual Review of Astron and Astrophys}
\newcommand{\nar} {New Astronomy Review}
\newcommand{\nat} {Nature}
\newcommand{\pasj} {Publications of the ASJ}
\newcommand{\planss} {Planet.~Space~Sci}









\section{Why to worry about how brown dwarfs form: their relation to planet formation}\label{sec:intro}


Nowadays it is well established that brown dwarfs (BDs) are fundamentally different from low-mass stars because their mass, $<0.075$~\mo\ for solar metallicity \citep[e.g.,][]{Chabrier2023_Hburninglimit},  is too low to sustain prolonged hydrogen fusion reactions. 
These objects were first predicted by \cite{Kumar1963} and \cite{Hayashi1963} and subsequent works such as \cite{Staller1981_BDLuminFunctions} predicted some of their observational properties. They were finally discovered by \citet[]{Nakajima1995_BD}, \cite{Oppenheimer1995} and \cite{Rebolo1995_BD}.
The lack of an internal source of energy in these substellar objects (a part from gravitational contraction) completely determines their fate and makes them, in some sense, objects much closer to planets than to stars. 
In fact, high-quality spectroscopy of BDs can be collected and used as reference for atmospheric studies in exoplanets (see, for instance \citealt[]{Barrado2023}). 
It is commonly assumed that the border line between planets and BDs is at $0.012$~\mo\ or 13~\mj\ \citep[e.g.,][]{LecavelierdesEtangs2022_exoplanetsreview}, because this is the mass limit for deuterium burning.
However, there is probably a more essential process that could set a fundamental difference between BDs and giant planets: their formation mechanism, and several authors have proposed to use a formation-based definition \citep[e.g.,][]{Chabrier2014_PPVI, Kirkpatrick2024_20pcCensus}, which is given below and is adopted in this work. 

Initially it was thought that BDs could not be formed as a scaled-down version of low-mass stars because typical Jeans masses of molecular clouds were thought to be well above the substellar limit.
However, since the beginning of the XXI century, an important number of theoretical and numerical works in the literature proposed that BDs do form in a similar manner as low-mass stars: from fragmentation of (turbulent) molecular clouds \citep[e.g.,][]{Padoan2002, Padoan2004, Hennebelle2008, Chabrier2010_PrestellarMF_IMF, Chabrier2014_PPVI, Haugbolle2018, VazquezSemadeni2019, Dhandha2023_BDformation}. In most of these works, the Jeans mass decreases significantly because turbulence allows the formation of extremely dense density fluctuations, but high densities can also be achieved as a natural outcome of gravitational collapse\footnote{There are a number of problems with the turbulent fragmentation scenario, as summarized in \cite{Lomax2016_turbfrag} and \cite{Whitworth2018}. Alternatively, in scenarios where turbulence does not have such a dominant role \citep[e.g.,][]{Bate2019, VazquezSemadeni2019}, fragmentation proceeds isothermally as the cloud undergoes hierarchical collapse, increasing its density, and the fragmentation is only stopped by the change in the fragmentation regime, from isothermal to adiabatic. This transition takes place when the gas is so dense that it becomes optically thick and does not allow the radiation to escape, and can occur for masses as low as $\sim0.005$~\mo\ \citep[e.g.,][]{Larson1969, Young2023_FHCs}.}.
If BDs form like stars, BDs and giant planets should have a different internal structure. 
Furthermore, several authors proposed that BDs are indeed a scaled-down version of low-mass stars but forming in a very low mass and compact core with the {\it particular property} of having an unusually poor accretion history. Such a particularly poor accretion history could be related to competition with other protostars that prevent them from gaining further mass \citep[e.g.,][]{Bonnell2006_competacc}, to tidal shear and high velocity dispersion within a nascent stellar cluster \citep[e.g.,][]{Bonnell2008_BDs},  to ejection from their natal environments \citep[e.g.,][]{Reipurth2001_ejection, Bate2002, Bate2019, Basu2012_Formation_ProtoBD, Reipurth2015_BinaryBD}, to extreme outflows that can remove the material from the cloud that is available for star formation \citep[e.g.,][]{Kratter2016_InstabilityDisks, Machida2009_BDformation}, or to the strong effects of radiative feedback in the surroundings of OB stars \citep[e.g.,][]{Whitworth2004_photoerosion, Whitworth2007_BDformation, Whitworth2018}.

From an observational point of view, works studying the complete (typically down to $\sim0.01$~\mo) Initial Mass Function (IMF) in the central bulge of the Galaxy or in other complexes such as the Pleiades, 25\,Ori, W3 or NGC\,1333 seem to find results consistent with a lognormal IMF \citep[e.g.,][]{Moraux2003_Pleiades_IMF, Scholz2012, Kirkpatrick2019, Suarez2019, Huston2021, Chabrier2023_BD-IMF}\footnote{Among the works reporting no deviations of the IMF with respect to a lognormal distribution, some of them, such as \cite{Suarez2019} or \cite{Huston2021} do not reach the planetary regime.}.
The lognormal IMF is the expected outcome of turbulence-driven star formation \citep[e.g.,][]{Zinnecker1984_IMF, Chabrier2005_IMF}, and the finding of a lognormal IMF down to $\sim0.01$~\mo\ suggests that BD formation is a continuation of the star formation process.
Other authors fit the low-mass regime of the IMF, down to substellar masses, using one single power-law, which also suggests that the formation of BDs is a continuation of the star-formation process \citep[e.g.,][]{Barrado2002_APer_IMF, Barrado2004_C69_IMF, Moraux2007_Blanco1_IMF, Lodieu2007_UperSco-IMF, Bayo2011_C69_IMF, Damian2023}. This has been recently found using {\it Euclid} down to 4~\mj\ in the $\sigma$ Ori cluster \citep{Martin2024_Euclid}, and is supported by other recent works.
\citet{Parker2023} find that the spatial distribution of planetary-mass objects down to $\sim0.005$~\mo\ in NGC\,1333 is indistinguishable from that of low-mass stars, and that this distribution should be different if it stemed from ejected objects formed in circumstellar discs. 
In the same line, \cite{Scholz2023_NGC1333} find that the disc fraction in the same region does not decline towards smaller masses, suggesting that planetary-mass objects seem to form like stars at least down to $\sim0.01$~\mo. This limit was pushed down to $\sim0.004$~\mo\ by \citet{Langeveld2024_JWST-FFPs}, using the James Webb Space Telescope (JWST) also in NGC\,1333.
The formation of BDs as a scaled-down version of low-mass stars is additionally favored by the recent study of the BD desert with Gaia DR3 \citep[]{Stevenson2023_BDdesert_GaiaDR3} (with some hints of two different parent distributions) and by JWST measurements of the $^{14}N/^{15}N$ ratio in the atmosphere of a BD 
\citep{Barrado2023}.

\begin{figure*}
\begin{center}
\begin{tabular}[b]{c}    
\epsfig{file=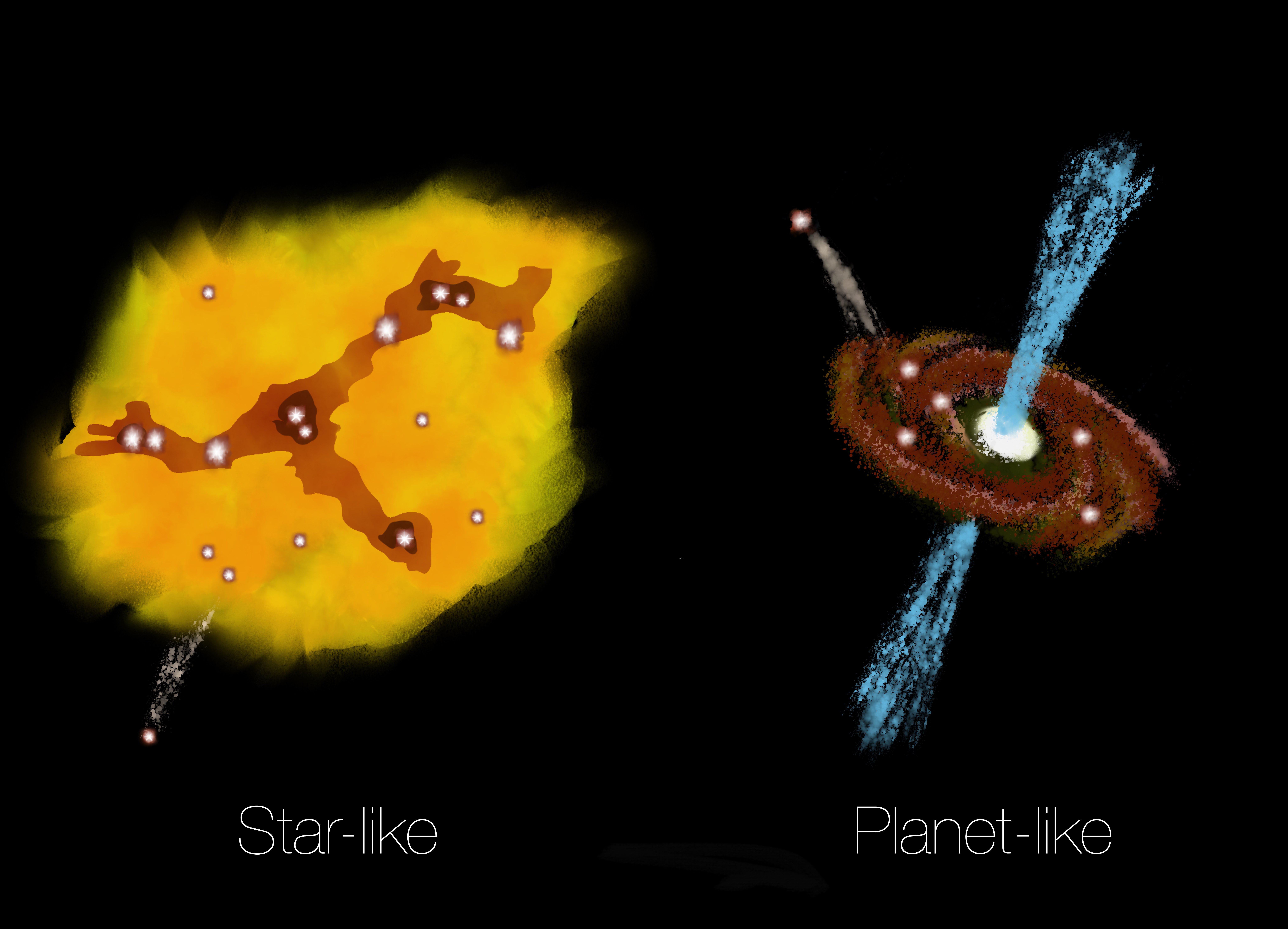, width=14cm,angle=0}\\
\end{tabular}
\caption{
Sketch of the two main scenarios proposed for the formation of isolated BDs. Left: Star-like scenario, which might involve, in some cases, ejection from the cloud. Right: Planet-like scenario, which always requires ejection from a protostellar disc.}
\label{fig:sketch}
\end{center}
\end{figure*}


However, other works report that the very low-mass end of the IMF (from 0.01 to 0.3~\mo) is bimodal, in regions such as NGC\,1333, M35, the Pleiades or the Orion Nebula Cloud \citep{Barrado2001_M35_IMF, Oasa2008_NGC1333, Bouy2015_Pleiades_IMF, Drass2016}, and also
a clear excess of planetary-mass objects, also known as Free-Floating Planets (FFPs, with masses $\lesssim0.01$~\mo) has been recently found in Ophiuchus and Upper Sco \citep[e.g.,][]{MiretRoig2022,MiretRoig2023}, suggesting that additional mechanisms other than the standard star-formation pathways must be at work, at least in certain environments of the Galaxy or for a certain range of masses.
Additional searches of even lower mass objects 
(\citealt{DeRocco2024_terrestrialFFPs} about sub-terrestrial FFPs; 
\citealt{Pearson2023} for alleged binary FFPs in the Trapezium) support this view.

To explain this excess, `planet-like' formation mechanisms such as disc fragmentation and subsequent ejection are invoked, along with stellar encounters or dynamical interactions between the members of multiple systems \citep[e.g.,][]{Stamatellos2007, Stamatellos2009, Basu2012_Formation_ProtoBD, Reipurth2015_BinaryBD, Thies2015, Vorobyov2017_FormationIFFPs, Wang2023FFPs, Chen2023FFPs, Yu2024_FFPs-flybys, Coleman2024_prediction-FFPs-population}\footnote{Alternatively, \citet{Dhandha2023_BDformation} and \citet{Kuruwita2024_lowendIMF-solenoidal} suggest that solenoidal turbulence would naturally produce a BD excess. This would allow to explain the BD excess found in some regions again from fragmentation of molecular clouds.}.
Moreover, the theoretical predictions of binary separation distributions of BDs and stars \citep{Thies2015} do not agree with the numerous observations in the field and in stellar associations \citep[]{Bouy2003_MultiplicityIFFP, Burgasser2003_BinarityBD, Martin2003_BinaryBD_Pleiades, Close2003_BinaryBD_Field}, which are more consistent with the predictions of a dynamical origin of BD binaries being ejected from multiple systems \citep[e.g.,][]{Reipurth2015_BinaryBD}.
Additionally, there are evidences of very-low mass stars ejected by dynamical decay of multiple systems, such as the case of HH24 \citep[]{Reipurth2023_MultiplicityJet_ProtoStar_HH24}. In this case, a border-limit star moves at 25~\kms\ and was ejected about 5800 yr ago. 
All these evidences point to a possible non-negligible role of alternate scenarios to the star-like scenario for BD formation.\\


As a summary, there are two main scenarios for BD formation: the star-like scenario (fragmentation of a molecular cloud + possible ejection) or the planet-like scenario (disc fragmentation + ejection), both sketched in Fig.~\ref{fig:sketch}. It is important to note that in both scenarios the objects might be ejected at some point from their natal places, either a molecular core, a filament, a multiple system or a protostellar disc. But the decisive question to understand is how the original seed of the young BD was formed. The current view of BD formation is that BDs form as low-mass stars in most/some star-forming regions of the Galaxy and probably down to a certain mass, below which they would form like planets. A possible lower-limit for the formation of objects like stars is suggested to be around 10--15~\mj\ \citep[e.g.,][]{Schlaufman2018_upperplanetmass, MiretRoig2022, Kirkpatrick2024_20pcCensus, Scholz2023_NGC1333}, and we will adopt here a tentative upper limit for the planetary regime of $\sim10$~\mj. In turn, there are already evidences (based on the trend of companions around metal-rich hosts, on the ratio of C/O values, and on radial velocity distributions) consistently indicating that objects with mass $<4$~\mj\ do always form like planets \citep[e.g.,][]{Schlaufman2018_upperplanetmass, Kirkpatrick2024_20pcCensus}. In any case, there must be some overlapping in the mass range, where both formation mechanisms could co-exist. In addition, specific regions of the Galaxy might be more prone to the star-like or the planet-like mode of formation due to several factors unexplored so far. Thus, the study of BDs (and FFPs) in their {\it earliest} stages of formation
is essential to constrain the different formation theories of these mysterious objects and to better understand their siblings: both very low-mass stars and planets. \\


A direct method to further constrain theories of BD formation is to study these objects in their earliest evolutionary stages. For example, if BDs form indeed as a scaled-down version of low-mass stars, they should have properties that are a scaled-down version of the properties of the most embedded protostars, the Class 0 and Class I protostars \citep[e.g., following the classification by][]{Adams1987_SEDs, Lada1987_Classes, Andre1993_Class0}. Class 0 and I protostars are known to follow several trends, such as the mass infall rate vs the internal luminosity, the centimeter luminosity vs bolometric luminosity, or the outflow momentum rate vs the bolometric luminosity or envelope mass \citep[e.g.,][]{Bontemps1996, Kim2021_infallVeLLO, Anglada2018}. Thus, proto-BDs should fall in the corresponding low-end of these relations if their formation path was star-like. What is more, in the star-like scenario, the equivalent to pre-stellar cores should be found in the substellar regime, the so-called `pre-BDs'. Pre-stellar cores are gravitationally-bound compact objects at the verge of collapse to form a star, but with no signs of stellar activity yet, such as infrared emission or outflow signatures \citep[e.g.,][]{Pagani2010_prestellar, Simpson2011_prestellar, Maret2013_prestellar}. 
Throughout this work, we will use the term `proto-BD' to refer to deeply embedded BDs whose properties are comparable to the properties of Class 0/I protostars.
In this work we aim at compiling the observational evidence reported so far in the literature about proto-BDs and pre-BDs, so that constraints on the different BD formation scenarios can be explored. Therefore, this review will not include Class II BDs, or young FFPs.

Although it is clear that observations of pre- and proto-BDs are required to constrain theories of BD formation, there is nowadays a rather confusing knowledge on which kind of objects can actually qualify as robust pre- and proto-BD candidates, and there is a lack of consensus in the literature about the precise meaning of these terms.
The advent of `last-generation' telescopes, such as ALMA upgrades or the JWST, and the next-generation VLA and SKA in the future, is progressively unveiling fainter objects in star-forming regions which might well be excellent pre- and proto-BD candidates \citep[e.g.,][]{Kawabe2018,LeeLi2018}. Thus, there is the need and it is the proper time of clearly defining what these terms refer to, where we are, and where we should address our future efforts. \\

This review is organized as follows. In Sec.~\ref{sec:defs} we provide several definitions essential to this work. In Sec.~\ref{sec:searches} we present the main efforts of searching (pre-) and proto- BDs in the literature. In Sec.~\ref{sec:challenges} we mention the main challenges to properly identify these objects. In Sec.~\ref{sec:mainprops} we describe the necessary conditions that a candidate must fulfill to be a robust proto-BD candidate, and apply these criteria to the SUCANES database \citep{PerezGarcia2024_SUCANES} to extract the first statistically significant sample of robust proto-BD candidates. In Sec.~\ref{sec:mainpropspreBD} we present a similar procedure for pre-BDs. In Sec.~\ref{sec:NbdMCproperties} we compare the proto-BD candidate content in different clouds of the solar neighborhood. In Sec.~\ref{sec:constraints2theory} we discuss possible constraints from the compiled information. Finally, in Sec.~\ref{sec:lessons} we outline future research lines in view of the recent and upcoming exciting last-generation telescopes.


\section{Definitions}\label{sec:defs}

Before summarizing the main strategies to search for proto-BDs reported so far in the literature, we define here the most important parameters that are used to characterize these objects.

\paragraph{Effective radius, $\reff$} Given the area $A$ of the millimetre/sub-millimetre source, taken at the 3$\sigma$ contour or as the full-width at half maximum, $\reff$ is the radius corresponding to $A$ assuming that the emission is circular \citep[e.g.,][]{Kauffmann2008_MAMBO_SpitzerCores}:

\begin{equation}
\reff = \sqrt{A/\pi}.    
\end{equation}

\paragraph{Bolometric luminosity, $\Lbol$} The bolometric luminosity results from integrating the Spectral Energy Distribution (SED) for all frequencies:

\begin{equation}
\Lbol = 4\pi\,D^{2}\,\int_{0}^{\infty}F_\nu\,d\nu,    
\end{equation}

\noindent where $F_\nu$ is the flux density of the source at frequency $\nu$. It is important that the targets have a relatively complete and well-sampled SED covering from the infrared to the sub-millimetre regime to have an accurate estimate of $\Lbol$.

\paragraph{Luminosity in the sub-millimetre range, $\Lsubmm$} The sub-millimetre luminosity is the result from integrating the SED from the lowest frequency available up to the frequency corresponding to 350~\mum:

\begin{equation}
\Lsubmm = 4\pi\,D^{2}\,\int_{0}^{c/350\mu m}F_\nu\,d\nu,   
\end{equation}

\noindent where $c$ is the speed of light. The $\Lsubmm/\Lbol$ parameter is considered an evolutionary indicator so that higher values correspond to more embedded (younger) objects \citep[e.g.,][]{Andre1993_Class0, Kauffmann2008_MAMBO_SpitzerCores}.

\paragraph{Bolometric temperature, $\Tbol$} The bolometric temperature is the temperature of a blackbody having the same mean frequency as the observed SED \citep{Myers1993_Tbol_YSO}. 
It can be calculated as \citep[e.g.,][]{Chen1995_Tbol, Dunham2008_c2dVeLLOs}:

\begin{equation}
\Tbol = 1.25\times10^{-11}\,\frac{\int_{0}^{\infty}\nu\,F_\nu\,d\nu}{\int_{0}^{\infty}F_\nu\,d\nu}\, \mathrm{K},
\end{equation}

\noindent and is considered an evolutionary indicator, with $\Tbol<70$~K corresponding to Class 0 objects,  $70<\Tbol<650$~K corresponding to Class I objects, $650<\Tbol<2800$~K corresponding to Class II, and  $\Tbol>2800$~K corresponding to Class III \citep{Chen1995_Tbol, Evans2009_Lifetimes}. The
$\Tbol$ evolutionary indicator anticorrelates with other indicators such as the slope of the SED from 2 to 20~\mum\ \citep[e.g.,][]{Evans2009_Lifetimes}, and correlates with the $\Lsubmm/\Lbol$ indicator \citep[Fig. 10 of][]{Kauffmann2008_MAMBO_SpitzerCores}.

\paragraph{Internal luminosity, $\Lint$} The internal luminosity is the luminosity coming from both the central protostar and a circumstellar disc/envelope, excluding the external luminosity $L_\mathrm{ext}$, which is the luminosity arising from heating of the disc/envelope by the interstellar radiation field (ISRF). Thus, $\Lbol = \Lint + L_\mathrm{ext}$. To estimate $\Lint$, radiative transfer modeling of the SED is required. Therefore, $\Lint=L_\mathrm{star}+L_\mathrm{disc}$ (Keplerian disc) and usually $\Lint$ is around 70\% of $\Lbol$. Consequently, $\Lint$ includes accretion and luminosity from the photosphere but does not include luminosity from external heating by the interstellar radiation field \citep{Vorobyov2017_EffectAccretion_BD}. \citet{Dunham2008_c2dVeLLOs} explore the relation of $\Lint$ vs the flux at different wavelengths and find that the flux at 70~\mum\ correlates very well with $\Lint$ calculated using radiative transfer models. $\Lint$ can be calculated using the following equation \citep{PerezGarcia2024_SUCANES}:
\begin{equation}
\Lint = 3.3\times10^8 \, [F_\mathrm{70\,cgs}\,(D/140)^2]^{0.94},    
\end{equation}
where $\Lint$ is given in \lo, $D$ is the distance (in pc), and $F_\mathrm{70\,cgs}$ corresponds to $\nu\,F_\nu$ at 70~$\mu$m given in cgs units, i.e., erg\,cm$^{-2}$\,s$^{-1}$. In practical units:

\begin{equation}
\Lint = 8.96\times10^{-5} \, [F_\mathrm{70\,mJy} (D/140)^2]^{0.94},    
\label{eq:LintFlux70mic}
\end{equation}

\noindent where $F_\mathrm{70\,mJy}$ corresponds to the flux density at 70~$\mu$m in mJy.


\paragraph{Envelope mass, $\Menv$} The envelope mass is the mass of the core surrounding the protostar (or hydrostatic core). $\Menv$ is usually inferred from observations in the millimetre/sub-millimetre range, which allow to measure the flux at the corresponding wavelength, $F_\nu$. Adopting a dust temperature $T_\mathrm{d}$ and a dust opacity $\kappa_\nu$, $\Menv$ can be estimated as:

\begin{equation}
\Menv = \frac{F_\nu\,D^2}{B_\nu(T_\mathrm{d})\,\kappa_\nu}. \end{equation}

\paragraph{Dynamical or accreted mass, $\Macc$ or $\Mdyn$} The dynamical or accreted mass is the mass of the hydrostatic core or protostar, and, for the most embedded objects, is typically estimated from the velocity pattern tracing Keplerian motions of the gas in the disc (see Sec.~\ref{sec:challenges-Macc}). If the Keplerian motion is not fully resolved and only a velocity gradient is detected, the accreted mass can be estimated from: 

\begin{equation}\label{eq:Macc}
\Macc=\Mdyn = \frac{R\,v_\mathrm{rot}^2}{G},    
\end{equation}

\noindent where $R$ is the distance from the central object, of mass $\Mdyn$, to the position where the observed velocity $v_\mathrm{rot}$ is measured. If this method is used to estimate the accreted mass, the term `dynamical mass' is typically used, to emphasize that it has been obtained from the observed gas kinematics. If other methods are used (as outlined in Sec.~\ref{sec:challenges-Macc}), the term `accreted mass' ($\Macc$) is used.

\paragraph{Accretion luminosity, $\Lacc$} The accretion luminosity corresponds to the energy per unit time radiated as heat by the matter infalling towards the hydrostatic core.
$\Lacc$ is directly proportional to the mass accretion rate:

\begin{equation}\label{eq:Lacc}
L_\mathrm{acc} = \eta_\mathrm{L}\,\frac{G\,\Macc\,\dot{M}_\mathrm{acc}}{R_*},
\end{equation}

\noindent where $\eta_\mathrm{L}$ is the accretion luminosity efficiency with respect to steady spherical infall (for steady accretion through an optically thick disc,  $\eta_\mathrm{L}\sim1/2$, \citealt[]{Hartmann1998_Book}), and $R_*$ is the radius of the central hydrostatic object.

\paragraph{Virial mass, $\Mvir$} The virial mass is the mass that an object needs to have to be 
gravitationally bound, assuming the virial theorem applies, i.e., that $2\Ek = |\Eg|$.  For the most simple case of the virial theorem, and assuming a uniform-density sphere of radius $R$, $\Eg=-3GM^2/5R$ and $\Ek=3M\sigma_\mathrm{tot}^2/2$ \citep[e.g.,][]{Andre2012_PreBD, Camacho2023}:

\begin{equation}\label{eq:Mvir}
\Mvir = \frac{3\,R\,\sigma_\mathrm{tot}^2}{G},
\end{equation}

\noindent where $\sigma_\mathrm{tot}$ is the total 1D velocity dispersion including thermal and non-thermal contributions\footnote{It is important to note that the total 1D velocity dispersion includes both thermal, $\sigma_\mathrm{th}$, and non-thermal, $\sigma_\mathrm{nth}$, contributions, i.e., $\sigma_\mathrm{tot}=\sqrt{\sigma_\mathrm{th}^2 + \sigma_\mathrm{nth}^2}$. However, for pre-BDs the non-thermal contribution is small, and care must be taken to properly measure the thermal contribution, by calculating it using the mean molecular weight per particle of molecular gas, $\mu$, typically taken as 2.3, instead of using the molecular weight of the observed molecular tracer: $\sigma_\mathrm{th} = \sqrt{k T/(\mu\,m_\mathrm{H})}$, with $k$ being the Boltzmann constant, $m_\mathrm{H}$ the mass of the hydrogen atom and $T$ the temperature.}. The virial parameter is then $\alpha_\mathrm{vir}\equiv\Mvir/M_\mathrm{obs} = 2\Ek/\Eg$ and it has the following meaning: if $\alpha_\mathrm{vir}\lesssim1$, the object is strongly self-gravitating; if $\alpha_\mathrm{vir}\sim1$, the object is gravitationally bound; if $\alpha_\mathrm{vir}\gg\,2$, self-gravity is not dominant.

\paragraph{Gravitational mass, $\Mgrav$} The gravitational mass is the mass required to have equipartition between gravitational energy and kinetic energy, i.e., $\Ek = |\Eg|$, or, equivalently, $\alpha_\mathrm{vir}\sim2$.
This is a criterion less severe than the constraint arising from the virial parameter, and has also been used as a measure of the gravitational boundness of a core, for example, by \cite{Pound1993_ProtoBD, Pound1995_protoBD} and \cite{Palau2012_PreBD_Cores_Taurus}:

\begin{equation}\label{eq:Mgrav}
\Mgrav = \frac{3\,R\,\sigma_\mathrm{tot}^2}{2\,G}.
\end{equation}

\paragraph{Bonnor-Ebert mass, $\MBE$} The Bonnor-Ebert mass is the maximum possible mass of an isothermal sphere in hydrostatic equilibrium embedded in a medium exerting an external pressure $p_0$ on it. $\MBE$ can be estimated as \citep[e.g.,][]{Bonnor1956, Ebert1957, Draine2011_Book}:

\begin{equation}\label{eq:MBE-p0}
\MBE = 0.26\,\bigg(\frac{T}{10\,\mathrm{K}}\bigg)^{2}\,\bigg(\frac{10^6\,\mathrm{cm}^{-3}\,\mathrm{K}}{p_0/k}\bigg)^{1/2}
\end{equation}

\noindent in \mo, and with $k$ being the Boltzmann's constant and $T$ the temperature of the sphere. A pressure-bounded isothermal sphere with mass equal to $\MBE$ has a density at the center of $\sim14$ times the density at the surface, $\rho_0$. Taking into account that $\rho_0 = p_0/c_s^2$ (with $c_s$ being the sound speed) and using the number density at the surface of the sphere, $n_0$, equation~\ref{eq:MBE-p0} can be re-written as


\begin{equation}\label{eq:MBE-n0}
\MBE = 2.6\,\bigg(\frac{T}{10\,\mathrm{K}}\bigg)^{3/2}\,\bigg(\frac{n_0}{10^3\,\mathrm{cm}^{-3}}\bigg)^{-1/2},
\end{equation}

\noindent comparable to equation (1) of \cite{Padoan2004}.

\paragraph{Jeans mass, $\Mjeans$} The Jeans mass is the minimum mass that a perturbation, occurring within a uniform, stationary, nonrotating and unmagnetized gas, needs to have to become gravitationally unstable. $\Mjeans$ can be estimated as \citep[e.g.,][]{Jeans1928, Draine2011_Book, Palau2015}:

\begin{equation}\label{eq:Mjeans}
\Mjeans = 0.6285\,\bigg(\frac{T}{10\,\mathrm{K}}\bigg)^{3/2}\,\bigg(\frac{n_\mathrm{H_2}}{10^5\,\mathrm{cm}^{-3}}\bigg)^{-1/2},
\end{equation}

\noindent where $n_\mathrm{H_2}$ is the density of H$_2$ molecules, typically used when considering the dense parts of molecular clouds. $\Mjeans$ is of the order of $\MBE$, with $\MBE \sim 1.2\,\Mjeans$ \citep[e.g.,][]{Draine2011_Book}. Thus, for higher temperatures of the gas, $\Mjeans$ is larger, making more difficult to fragment a cloud down to very small masses.


\section{Systematic searches and serendipitous discoveries}\label{sec:searches}

Up to now, a number of systematic searches have been carried out to search for proto-BDs and pre-BDs.
But also an important number of these discoveries have been serendipitous. Here we briefly summarize each one of these main discoveries, in order to learn from the crucial criteria and characteristics of the proto-BD and pre-BD candidates discovered so far.

\subsection{Systematic searches}\label{sec:searches-system}

Systematic searches of pre- and proto-BDs have been based on two main strategies. On one hand, an important number of works started from a selection criteria based on the identification of deeply embedded infrared sources, and then carried out a follow-up in the millimetre/sub-millimetre bands, where a number of criteria were applied to identify the reddest and least massive objects, typically with additional indicators to reject background contaminants. On the other hand, searches based on millimetre/sub-millimetre imaging have also been carried out. We present below the main works based on each strategy.


\subsubsection{Searches of infrared sources and follow-ups}\label{sec:searches-system-IR}

In this section, a list of the searches of very low-luminosity infrared sources is presented. 
It has been usual in the literature to refer to embedded objects with $\Lint \lesssim 0.1$--0.2~\lo\ as Very Low Luminosity Objects (VeLLOs) \citep[e.g.,][]{Kauffmann2005_Spitzer_VeLLO, DiFrancesco2007_PPV, Dunham2008_c2dVeLLOs, Kim2016_VeLLOs}, and they certainly constitute an excellent base to look for proto-BDs.
We exclude from this review the works searching for millimetre/sub-millimetre emission towards samples of Class II/III BDs \citep[with ages $\gtrsim 1$~Myr; e.g.,][]{ Klein2003, Scholz2006, Spezzi2013, Harvey2012_Herschel-first, Harvey2012_Herschel-second}, and the works searching for embedded protostars in entire clouds, regardless of luminosity \citep[e.g.,][]{Kirk2007_SCUBA, Joergensen2007_PROSAC, Joergensen2008_OphPer, Enoch2007_Bolocam, Enoch2008_PerOphSer, Hsieh2013_YSOs, Harvey2013_AurCal}.
The most relevant studies based on searches of very low-luminosity infrared sources are presented below, listing first the work and then the main telescopes/surveys used for the searches:

\begin{description}

\item[\citet{White2004}, Keck I:] A spectroscopic optical study with the Keck I telescope is presented for 15 Class I and Class II sources driving HH objects, selected from the IRAS catalog in the Taurus-Auriga cloud. All stars in the cloud with Class I–like SED (based on either their mid-infrared spectral index, \citealt[]{Myers1987_IRAS_Cores}) or their $\Tbol$ (\citealt[]{Myers1993_Tbol_YSO}) were selected and masses were inferred based on the spectral type classification and the \cite{Siess2000_evoltracks} isochrones at 1~Myr. They found 3 objects that qualify as Class I proto-BD candidates (IRAS 04158+2805, IRAS 04248+2612, IRAS 04489+3042), given their $\Tbol<650$~K, spectral types later than M5.5 and substellar masses \citep[see Sec.~3.2 and Fig.~11 of][]{White2004}. The sub-millimetre emission was studied in \cite{Motte2001_mmsurvey} with the IRAM\,30m radiotelescope. Follow-up sub-millimetre Array (SMA) observations of IRAS\,04158+2805 were presented in \cite{Andrews2008_I04158}. However, the model fitting the CO\,(3--2) velocity gradient of this source yields $\Macc\,\sim0.15$--0.45~\mo, challenging the substellar nature (see also \citealt[]{Luhman2010_Taurusdisks}).

\item[\citet{Dunham2008_c2dVeLLOs}, Spitzer IRAC+MIPS (c2d):] After the works of \cite{Young2004_VeLLOs} and \cite{DiFrancesco2008_SCUBA}, \cite{Dunham2008_c2dVeLLOs}  base their selection criteria on detections in MIPS bands, IRAC colors characteristic of rising SEDs, infrared luminosity smaller than 0.5~\lo, and not classifying as a galaxy candidate according to \cite{Harvey2007}. A value of $\Lint\lesssim0.1$~\lo\ and association with a dense core were established as the criteria to consider an object a Very Low Luminosity Object, and \cite{Kauffmann2005_Spitzer_VeLLO} and \cite{Dunham2008_c2dVeLLOs} identify 15 VeLLOs, being L1014, IRAM\,04191, L673-7, L1148-IRS, IRAS\,16253$-$2429, L328-IRS, IRAS\,16253$-$2429 among them. 

\item[\citet{SiciliaAguilar2008_VeLLO_Coronet}, VLT/FLAMES + Spitzer/IRS:] A sample of 62 very low-mass stars and BDs, previously detected in X-rays, was observed in the optical and infrared with the VLT/FLAMES and {\it Spitzer}/IRS in the Coronet cluster. The optical spectra reveal spectral types between M1 and M7.5, confirm the Class I nature for 11 of the sources (2 classified as `flared'), and show the presence of accretion and shocks. The IRS spectra, together with infrared photometry from the IRAC/MIPS instruments on {\it Spitzer} and 2MASS, confirm the presence of infrared excesses characteristic of discs around $\sim$70\% of the objects. The Class I sources are: G-6, G-17, G-28 (flare source), G-36, G-43, G-45, G-64 (late M spectral type, shocks), G-101, G-108 (flare source, late M spectral type), G-112, and G-122. This study was followed-up in \cite{SiciliaAguilar2011_Coronet-APEX} where APEX observations were presented. This later study reveals sub-millimetre emission associated with source G-17, whose integrated flux corresponds to a mass of 0.18~\mo\ (assuming a dust temperature of 20~K and a dust+gas opacity of 0.0175~cm$^2$~g$^{-1}$, \citealt{Ossenkopf1994_dustopacity}).


\item[\citet{Barrado2009}, Spitzer IRAC:] The Taurus cloud was surveyed to identify all sources with detections in the four IRAC bands, whose location in color-color diagrams corresponds to Class 0/I sources and whose IRAC1 magnitude was below 11 mag, corresponding to the expected value for a substellar photosphere at the Taurus distance, according to the models of \citet[]{Baraffe2003}. Further criteria to reject extragalactic contamination were applied. This resulted in 12 proto-BD candidates. Follow-ups were carried out by \citet[]{Palau2012_PreBD_Cores_Taurus} and \cite{Morata2015_Jets_ProtoBD}, where millimetre/sub-millimetre emission was studied along with a search for thermal radiojets.

\item[\citet{Bulger2014_BD_Taurus}, Herschel for M4--L0 members:] A sample of Taurus members with spectral type  M4--L0 was built by requiring that they have been observed with {\it Herschel}, yielding a total of 150 objects. Among the 150 members, 7 are Class I or earlier (Sec. 6.7 of \citealt{Bulger2014_BD_Taurus}), and 3 are VeLLOs: [GKH94]41,
IRAS 04191+1523 B, and L1521F-IRS. IRAS~04191+1523B is the secondary component of a binary system in which the primary is of stellar nature. A follow-up work by \cite{DangDuc2016_VeLLO_Taurus} with CARMA at 2.9 mm reveal that [GKH94]41 and IRAS~04191+1523B have $\Macc \lesssim 0.057$~\mo, but subsequent ALMA observations yield $\Macc \lesssim0.120$~\mo, twice the estimate from CARMA \citep{LeeLeeDunham2017_I04191}. This gives an idea of the uncertainty in the estimate of $\Macc$.

\item[\citet{Riaz2015_VeLLos}, UKIDSS+WISE:] A systematic search for proto-BDs in the UKIDSS+WISE data of the $\sigma$-Orionis region was carried out, finding that the sources Mayrit1701117 and Mayrit1082188 are good Class I/Flat proto-BD candidates. Further optical spectroscopic and sub-millimetre observations were used to complement the characterization of these two sources, for which $\Lbol$ in the range 0.16--0.18~\lo, and $\Menv\sim 0.022$--0.036~\mo\ were inferred. Mayrit1701117 was further studied in \citet{Riaz2017_Jet_ProtoBD, Riaz2019_Mayrit1701117} and \cite{Riaz2021_AccretionOutflow_ProtoBD}, who report that the source is driving a Herbig-Haro object and that has an associated pseudo-disc seen with ALMA. This strategy was later applied to the Ophiuchus, Perseus, Taurus and Serpens-Aquila clouds \citep{Riaz2022_D_H_ratio_ProtoBD, Riaz2022_HDCO_D2CO_Class0_I_ProtoBD, Riaz2022_CH3D_Class0_I_ProtBD, Riaz2023_Methanol}, and several studies of the chemistry in these objects were also carried out \citep{Riaz2018_chem-OphSer, Riaz2019b_chem-OphSer, Riaz2021_AccretionOutflow_ProtoBD,Riaz2022_D_H_ratio_ProtoBD, Riaz2022_HDCO_D2CO_Class0_I_ProtoBD, Riaz2022_CH3D_Class0_I_ProtBD, Riaz2023_Methanol}.

\item[\citet{Kim2016_VeLLOs}: Spitzer c2d IRAC + MIPS + Herschel:] The work of \cite{Kim2016_VeLLOs} includes five c2d clouds at distances $<350$~pc, and eleven nearby molecular clouds from the Gould Belt survey\footnote{The five clouds from c2d are Perseus, Serpens, Chamaeleon II, Lupus, and Ophiuchus; and the 11 clouds from the Gould Belt survey are California, Chamaeleon I, Chamaeleon III, Musca, Lupus V, Lupus VI, Ophiuchus North, Aquila, Corona Australis, Cepheus, and IC 5146.}, along with the Orion and Taurus molecular clouds. For these 16 clouds, about 100000 point-sources were detected by {\it Spitzer} and {\it Herschel}, and the criteria of \cite{Dunham2008_c2dVeLLOs} were applied along with the additional criteria: 
$\Lint < 0.2$~\lo\footnote{Although the original definition of VeLLOs requires that $\Lint\leq0.1$~\lo, \cite{Kim2016_VeLLOs} adopted $\Lint\leq0.2$~\lo\ to take into account the dispersion in the $F_\mathrm{70\mu m}$ vs $\Lint$ relation found by \cite{Dunham2008_c2dVeLLOs}.},
detection at 70~\mum, $F_\mathrm{1.65\mu m}/F_\mathrm{70\mu m}<2.8$, and $[8]-[24]>2.2$, to assure that the sources are embedded. 
Furthermore, the criteria of \cite{Harvey2007} to reject extragalactic contaminants were also applied. 
This resulted in 320 sources, of which 95 present evidences of having a dusty envelope associated from {\it Herschel}/SPIRE or James Clerk Maxwell Telescope (JCMT/SCUBA) data. In addition, firm evidence of dense (\nth) gas was found in 44 of the 95 objects, being these objects classified as confirmed VeLLOs. The remaining 51 objects with no dense gas associated were classified as VeLLO candidates (detected only in the sub-millimetre continuum). It is important to note that the \cite{Kim2016_VeLLOs} strategy recovers the bona-fide IC348-SMM2E proto-BD in the confirmed VeLLOs group. 
This study was followed-up in \cite{Kim2019_CO_Outflow_VeLLO} and \cite{Kim2021_infallVeLLO} to look for outflow and inflow motions, making it the most complete and uniform sample of VeLLOs up to date from 16 nearby molecular clouds.

\item[\citet{Riaz2016_ProtoBD_Serpens}: Spitzer c2d IRAC:] The Serpens Main and Serp/G3–G6 clusters were studied using the c2d {\it Spitzer} catalog. The selection criteria were a Class 0/I/Flat SED and $\Lbol \leq 0.3$~\lo. These criteria resulted in 28 very low-mass stars or proto-BD candidates, 13 of which were classified as Class 0/I and 15 as Flat sources, according to the {\it Spitzer} c2d catalog. Most targets listed in this analysis are associated with bright 250~$\mu$m\ {\it Herschel} emission and JCMT 450 and 850~\mum\ emission. These data were complemented with spectroscopy in the near- and mid-infrared, as well as with HCO$^+$ observations with the Caltech Submillimeter Observatory (CSO). From this study, one proto-BD candidate with a significant (non-tenuous) envelope was discovered, J182902. This object is a Class I/Flat source with $\Lint \sim0.03$~\lo, for which $\Macc\lesssim 0.04$~\mo. The detection with the JCMT at 450 and 850~\mum\ corresponds to $\Menv$ in the range 0.01--0.03~\mo.
However, after the work of \citet{Riaz2016_ProtoBD_Serpens}, where a distance of 260 pc was adopted, \citet[]{OrtizLeon2023_Serpens} report an updated distance to the Serpens cluster from H$_2$O maser parallax of $440.7\pm3.5$~pc. In the following, the parameters of the Serpens sources for which a distance of 260 pc had been adopted in previous works, have been recalculated to the new distance of $440$~pc, implying that some of the candidates of \citet{Riaz2016_ProtoBD_Serpens} and \cite{Kim2016_VeLLOs}, initially classified as VeLLOs, are re-classified now as likely protostars. 
Follow-up studies combined this strategy with the strategy carried out by \cite{Riaz2015_VeLLos} and performed a chemical study of the resulting combined sample \citep{Riaz2018_chem-OphSer, Riaz2019b_chem-OphSer, Riaz2021_AccretionOutflow_ProtoBD}.

\end{description}


\subsubsection{Sub-millimetre/millimetre imaging and follow-ups}\label{sec:searches-system-submm}

A second approach to search for pre- and proto-BDs consisted on performing large-scale imaging surveys in the sub-millimetre/millimetre bands of specific regions of molecular clouds. This has the advantage of including pre-BDs and objects more deeply embedded because the detection of an infrared source is not required. We list here a number of this kind of searches in the literature:

\begin{description}

\item[\citet{Pound1993_ProtoBD}, AT\&T Bell Labs and NRAO\,12m:] A sample of 400 clumps in 7 nearby High-Latitude clouds \citep[from][]{Magnani1985} was observed with AT\&T Bell Labs 7m Telescope. The 7 clouds were imaged in CO\,(1--0) and $^{13}$CO\,(1--0) with a  beam of $\sim100''$. Seventy clumps with moderate bright integrated intensity and size $< 3'$ were selected and imaged at higher angular resolution with the NRAO 12m Kitt Peak radiotelescope in CO, $^{13}$CO, and CS, allowing to estimate masses from $^{13}$CO. The estimated masses were compared to the gravitational mass, but all objects had estimated masses smaller than the gravitational mass $\Mgrav$, with only two objects being nearly gravitationally bound, although they were not substellar. This was followed-up by \cite{Pound1995_protoBD} by applying the same strategy to the Ophiuchus (Oph B) and Taurus (B18) molecular clouds. Imaging in C$^{34}$S\,(2--1), DCO$^+$\,(1--0) and DCO$^+$\,(3--2) was performed using the AT\&T and NRAO antennas, and the CSO telescope was used to image the continuum emission at 800~\mum. From this work, 4 objects were found close to the substellar limit and to be gravitationally bound, being Oph-B11 one of the most promising proto-BD candidates. Follow-ups of this work are presented in \cite{Greaves2003_isolatedplanet}, where the JCMT was used to image part of the $\rho$-Oph\,B region, finding 11 compact cores with $\Menv<15$~\mj. Among these 11 compact cores, CO\,(3--2) wings were detected in the Oph-B11 source. This object was later found to be a pre-BD core \citep[][]{Andre2012_PreBD}.

\item[\cite{Tachihara1996_I15398}, Nagoya University telescope :] The proto-BD candidate IRAS 15398$-$3359 was discovered as the driving source of a $^{13}$CO outflow in a survey carried out using the Nagoya University telescope towards the Lupus clouds. Later \cite{Oya2014_I15398} report ALMA observations of the outflow and the driving source, finding that it is a Class 0 source with molecular gas detections in H$_2$CO and C$_2$H. The position-velocity diagrams of some transitions of these molecules suggest a central mass 0.02--0.09~\mo. However, \citet{Joergensen2013_I15398} study the chemistry in this object and report $\Menv\sim1.2$~\mo, and $\Lbol\sim1.8$~\lo, possibly due to an accretion burst. Later \cite{Yen2017_I15398} and \cite{Okoda2018_I15398} present ALMA observations of C$^{18}$O, C$_2$H and SO, from where $\Macc$ was estimated to be in the range 0.007--0.01~\mo. \cite{Tabatabaei2023_I15398} and \cite{Thieme2023_eDisk-I15398} further study the kinematics and magnetic field of the region.

\item[\cite{Motte1998_rhoOph}, IRAM\,30m:] The central region of $\rho$-Ophiuchi was imaged at 1.3~mm with the IRAM\,30m telescope, revealing about 68 clumps, out of which 26 have $\Menv < 0.15$~\mo, and 18 of these 26 have $\Menv>\MBE$, suggesting that they are gravitationally unstable. A follow-up of these 18 least massive gravitationally bound objects of this sample would be very useful.

\item[\cite{Kauffmann2008_MAMBO_SpitzerCores}, IRAM30m:] A sample of 38 cores selected from the initial target list of the c2d {\it Spitzer} survey and having sizes $\lesssim0.5$~pc ($\lesssim5'$ at distances $\lesssim400$~pc) was observed at 1.3 mm with the MAMBO bolometer array at the IRAM\,30m telescope. The combination of these deep millimetre maps with archival infrared data enabled the derivation of $\Lbol$ and $\Tbol$, as well as $\Menv$, and to identify a number of VeLLOs, which were found to have density profiles steeper and with larger central densities than starless cores. This suggests that VeLLO cores are structurally different from starless cores.

\item[\cite{Nakamura2012_SubSTellarCondensations}, JCMT+SMA:] Two prestellar cores (SM1 and B2-N5) in $\rho$-Ophiuchi were studied using sub-millimetre single-dish emission and the SMA, revealing that each prestellar core splits up into 3--4 subcondensations of 0.01--0.1~\mo, and sizes of a few hundred AU. Since the mean densities are higher than 10$^8$~\cmt, their masses are larger than $\MBE$ (eq.~\ref{eq:MBE-p0}), suggesting that the subcondensations are gravitationally unstable and are, consequently, good pre-BD candidates.

\item[\citet{Liu2016_Planck}, Planck:] Dedicated observations towards Planck Galactic Cold Clumps in the $\lambda$-Orionis complex reveal an extremely young Class 0 protostellar object and a proto-BD candidate in the bright-rimmed clump G192.32-11.88, which seems to be externally heated and compressed, reducing the star formation efficiency and core formation efficiency.

\item[\citet{Gahm2013_Rosette}, Onsala+APEX:] The Onsala and APEX telescopes were used to study the molecular gas in the Rosette Nebula, detecting CO\,(3--2), CO\,(2--1), $^{13}$CO\,(3--2), and $^{13}$CO\,(2--1) in narrow lines of about 1~\kms. About 7 globulettes were found with $\Menv < 0.075$~\mo. The study was complemented with broad JHK filters, narrow Paschen$\beta$, and H$_2$ imaging. The dynamical state of these 7 globulettes should be assessed by studying the gas kinematics.

\item[\citet{deGregorio2016_PreProtoBD_ChaII}, APEX:] The Chamaleon II cloud was studied using LABOCA at 850~\mum\ on APEX, and 5 pre-BD candidates and 1 proto-BD candidate, Cha-APEX-L, were found. Regarding the pre-BDs, their masses range from 0.016 to 0.021~\mo, and have no counterpart in MPG, VLT, 2MASS, WISE, {\it Spitzer}, AKARI, {\it Herschel}/PACS. Since they were unresolved, their radius must be $<2500$~au at the distance of the Chamaleon clouds. It was tentatively estimated that these objects could be gravitationally bound by comparing their masses to $\MBE$. Kinematic information on molecular gas would help define their stability. Regarding the proto-BD candidate, Cha-APEX-L, it has $\Lbol\sim0.08$~\lo, and $\Menv\sim0.050$~\mo, and its infrared and sub-millimetre fluxes are comparable to those of other Class 0/I proto-BD candidates, with a clear counterpart at 3.6 and 4.5~\mum. 

\item[\citet{Barrado2018_B30}, APEX:] The star-forming region B30, at the border of the bubble in the $\lambda$\,Ori cluster, was studied using LABOCA on APEX at 850~\mum, resulting in 8 proto-BD candidates, with $\Lbol<0.16$~\lo, and $\Menv\lesssim 0.012$~\mo. However, a follow-up study by \cite{Huelamo2017_B30} with ALMA at 880~\mum\ (with a rms noise of 0.2~\jpb, corresponding to a 5$\sigma$ mass sensitivity of $\sim0.001$~\mo\ for a dust temperature of 15 K) reveal no detections towards these objects, making its nature unclear.
In the \cite{Barrado2018_B30} study, six sub-millimetre cores, with $\Menv$ in the range 0.082--0.106~\mo, have no optical/infrared counterparts. \cite{Huelamo2017_B30} find that 3 out of the 6 starless cores (cores LB8, LB10, LB31) have an ALMA compact millimetre source associated, being these three objects potential pre-BDs. Further observations of dense gas would help constrain the nature of these objects.

\item[\cite{Tokuda2019_preBD}, IRAM\,30m + ACA:] A survey at 1.2 mm was carried out using the IRAM\,30m telescope and the ACA array in the L1495 region of the Taurus cloud. A condensation was identified to be dense and connected to a narrow filament of the L1495 region. This condensation, called MC5-N, was proposed to be a pre-BD candidate. The mass of the core is 0.2--0.4~\mo, and observations of N$_2$H$^+$ and N$_2$D$^+$ reveal a high deuterium fractionation.

\item[\citet{SantamariaMiranda2021_ALMA_Lupus_ProtoBD}, ASTE+ALMA:] The AzTEC 1.1 mm array camera at the Atacama sub-millimetre Telescope Experiment (ASTE) was used to image the Lupus I and Lupus III clouds, followed by ALMA observations. The measured $\Menv$ are below 0.124~\mo, and among the 15 detections, 2 are new Class I/0 proto-BD candidates, and 12 are new possible pre-BDs. In particular, the Class 0/I source ALMA J154229$-$334241 could potentially be a promising proto-BD candidate. 

\end{description}

\subsection{Serendipitous discoveries}\label{sec:searches-serendip}

A number of (isolated) proto-BD candidates were discovered serendipitously, that is, without following a well-defined strategy specifically designed to find these kind of objects. We list here some relevant cases (excluding proto-BD candidates which are clearly part of a (close) multiple system where the primary is a young stellar object, e.g., \citealt{Apai2005_ProtoBD_NICMOS}):

\begin{description}

\item[\cite{Palau2014}, IC348-SMM2E:] About 10 arcsec to the (north)east of the Class 0 source driving the HH\,797 object (also known as IC348-MMS, with $\Lbol\sim1.9$~\lo, and $\Menv\sim10$~\mo, \citealt[]{Hatchell2007_Perseus}), a faint compact core with $\Menv\sim0.03$--0.05~\mo\ also driving a compact outflow and a thermal radiojet was discovered \citep{Palau2014, Forbrich2015_IC348SMM2E}. The SED of the compact object is typical of Class 0 sources, and has $\Lint\sim0.1$~\lo. A velocity gradient perpendicular to the outflow was identified in C$^{18}$O\,(2--1) with the SMA, which suggests $\Macc$ in the range 0.036--0.110~\mo\ for a disc inclination with respect to the plane of the sky of 30--60$^\circ$ and at the revised distance to IC348 of 320~pc \citep{OrtizLeon2018_DistancePerseus}. Since the distance from the proto-BD candidate to the stellar-mass Class 0 source is $\sim3000$~au, it is not likely that this object was formed in the disc of the nearby protostar and rather suggests that it was formed by gravitational fragmentation of the parental structure.

\item[\cite{Whelan2018_Outflow_ISO_Oph200_ProtoBD}, ISO Oph 200:] ISO Oph 200 is a Class I
source with $\Lbol\sim0.09$~\lo, and its 850~\mum\ emission detected with the JCMT indicates $\Menv\sim0.060$~\mo. \cite{Whelan2018_Outflow_ISO_Oph200_ProtoBD} report H$_2$ and [FeII] emission with VLT/SINFONI indicative of outflow activity. This object was also studied by \citet{Riaz2021_Structure_ProtoBD} with ALMA in CO and their isotopologues, who conducted a physical and chemical modelling of the internal structure. The model includes an extended outflow, envelope/pseudo-disc, and an inner pseudo-disc, and indicates that the object could have been misclassified due to its line-of-sight outflow orientation. 

\item[\cite{Kawabe2018}, SM1-A and Source-X in Oph-A:] Two proto-BD candidates were identified about 15 arcsec (or 2000 au) to the north-east of the VLA1623 source, named SM1-A and Source-X \citep{DiFrancesco2004_OphA, Kawabe2018}. Both sources drive compact molecular outflows, have continuum millimetre emission associated an also X-ray emission detected with Chandra. Their $\Lbol$ are very low, of 0.035 and 0.01~\lo, respectively, but their $\Menv$ cannot be easily estimated because both objects are embedded within the brightest sub-millimetre clump of Oph-A.

\item[\cite{PhanBao2022_protoBD}, SMA1627$-$2441:] A proto-BD candidate was serendipitously discovered with the SMA about $\sim20$ arcsec ($\sim3000$ au) to the west of the Class II BD, ISO-Oph102 \citep[e.g.,][]{Greene1992_rhoOph, Natta2002_BDs, PhanBao2008_Oph102}. The source seems to be driving a small-scale bipolar outflow detected in CO\,(2--1).

\end{description}

It is important to note that a significant number of these serendipitous discoveries (at least  IC348-SMM2E, SM1-A, Source-X and SMA1627$-$2441) were found in the surroundings ($\lesssim5000$~au) of well-known Class 0 protostars or BDs (IC348-MMS, VLA1623, and ISO-Oph102, respectively). At least another candidate has been discovered this way \citep{Palau2022_J041757}. 
If proto-BDs were typically found in the surroundings of protostars, this would imply that systematic offsets between the proto-BD positions and column density peaks should be found, as discussed also in \cite{Kauffmann2008_MAMBO_SpitzerCores}. This suggests that a promising strategy to search for proto-BDs is to carry out deep interferometric observations in the immediate surroundings of Class 0--II protostars or BDs.

As a final note, a possibility worth exploring coming from the aforementioned studies is whether there are pre- and/or proto-BDs (preferentially) associated with shells or bubbles. \citet[]{Kim2016_VeLLOs} report about 10 VeLLOs which are not associated with active star-forming regions and are classified as `small cores', isolated and located in different clouds (Table 1 of \citealt[]{Kim2016_VeLLOs}), and one of them, L328, is not associated with a large molecular cloud complex and lies within a shell-like structure, similar to OphB-11 (pre-BD of \citealt[]{Andre2012_PreBD}) and the pre-BDs of B30 \citep{Huelamo2017_B30, Barrado2018_B30}. Thus, a non-negligible fraction of pre- or proto-BDs could potentially reside at the border of bubbles.


\section{Challenges and difficulties to identify promising pre- and proto-BD candidates}\label{sec:challenges}

Given the searches and discoveries presented in the previous section, a number of challenges and difficulties can be identified in the pre- and proto-BD search. We enumerate these challenges and briefly describe them in the following subsections. 

\subsection{Not all VeLLOs or FHCs are good proto-BD candidates}\label{sec:challenges-vellosfhcs}

Proto-BDs can be confused with VeLLOs and First Hydrostatic Cores (FHCs). While VeLLOs are objects whose $\Lint\lesssim0.1$~\lo\ (Sec.~\ref{sec:searches-system-IR}), FHCs are compact objects in quasi-hydrostatic equilibrium, 
resulting from a change in the gravitational collapse regime, from isothermal to adiabatic, 
which occurs when the gas becomes optically thick, preventing the gravitational energy to be released and thus increasing the thermal pressure \citep[e.g.,][]{Larson1969}.
With a lifetime of around 10\,000 yr \citep[e.g.,][]{Bate2014_collapse, Maureira2020_FHCs}, FHCs continuously accrete, increasing their temperature up to 2000 K, when collisional dissociation of H$_2$ consumes energy triggering a second collapse. Observationally, FHCs are objects whose internal luminosity is still very low (emission at 70~\mum\ is marginal or undetected), have a clear millimetre source associated, drive small outflows, and have a very low $\Tbol$, typically below 20 K \citep[e.g.,][]{Commercon2012_FHCs-SEDs, Palau2014, Maureira2020_FHCs, Young2023_FHCs}. A summary of the FHC properties is presented in the wrap-up last figure of this work, along with the pre-stellar, pre-BD, protostellar and proto-BD cases. The FHC stage of the low-mass star-formation process, and the proto-BD phase of the substellar regime share similar properties and can be easily confused, but there are clues that allow us to distinguish between them. In the criteria used to select both VeLLOs or FHCs there is no restriction about the mass of the parental core, which can reach values up to $\sim10$~\mo\ in some cases \citep[e.g.,][]{Kauffmann2011_VeLLO_L1148-IRS, Palau2014, Maureira2020_FHCs}. With such a large $\Menv$, the object could accrete enough mass in the future and become stellar.
Thus, a low value of $\Menv$ is a crucial property of proto-BDs that VeLLOs or FHCs might not necessarily fulfill.



\subsection{Difficulty to find infrared/optical counterparts of submm cores and to constrain their evolutionary stage}

If the strategy to search for proto-BDs is based on sub-millimetre continuum surveys of clouds (Sec.~\ref{sec:searches-system-submm}), then the identification of the infrared/optical counterpart might be very complicated because proto-BDs are intrinsically very faint, and this problem worsens if the object is deeply embedded and highly extincted \citep[e.g.,][]{deGregorio2016_PreProtoBD_ChaII, Barrado2018_B30, SantamariaMiranda2021_ALMA_Lupus_ProtoBD}. This in turn implies that no estimate of $\Lbol$ and $\Tbol$ can be made because of poorly sampled SEDs, leaving the nature of the candidates uncertain \citep[e.g.,][]{Huelamo2017_B30, PhanBao2022_protoBD}.

\subsection{Uncertain accreted mass $\Macc$}\label{sec:challenges-Macc}

In order to assess whether a proto-BD candidate is robust, it is vital to estimate its current accreted mass $\Macc$. There are several methods in the literature used to estimate $\Macc$, that we outline below:

\begin{description}
    
\item[Evolutionary models:] Isochrones and mass tracks from theoretical evolutionary models (such as \citealt{Chabrier2000, Baraffe2017}) can be used to estimate accreted masses \citep{DangDuc2016_VeLLO_Taurus, Riaz2019_Mayrit1701117}. However, these models are typically calculated for too evolved stages (ages $\sim1$ Myr) with little accretion, and this is not consistent with the proto-BD nature. One exception is the work of \citet{Vorobyov2017_Nature_VeLLOs}, who provide  $\Lint$–$T_\mathrm{eff}$ tracks for Class 0 objects, but this method should be more explored. In the cases where both this method and also a rough estimate from the method presented below (disc kinematics) have been applied, the results are not fully consistent. For example, \cite{Riaz2021_Structure_ProtoBD} apply this method to ISO\,Oph200 and find that $\Macc$ is in the range 0.01--0.02~\mo\ while, if the velocity gradient reported in their Fig. 5 is assumed to trace a Keplerian motion, then $\Macc\sim0.23$~\mo, an order of magnitude larger, probably due to the contribution from infall.  

\item[Mass-accretion rate:] If both the mass-accretion rate onto the proto-BD and the time during which the object has been accreting can be determined, then an estimate of $\Macc$ can be made \citep[e.g.,][]{Dunham2010_VeLLO_L673_7, Lee2013_ProtoBD_L328-IRS, Kim2019_CO_Outflow_VeLLO}. However, the determination of the mass-accretion rate, usually inferred from the mass-loss rate of the detected outflow, is a quantity highly uncertain. Mass-loss rates from molecular outflows are affected by the sensitivity of the telescope (the extension of the outflow can vary depending on the sensitivity), assumption of CO abundance, beam dilution (particularly important for proto-BDs with very small and compact outflows), inclination and opacity effects (see discussion by, e.g., \citealt{Kim2019_CO_Outflow_VeLLO}). Similarly, the time of accretion is also uncertain. Thus, this method, although useful to make first and preliminary estimates of $\Macc$, needs to be complemented with another method, as indicated in \cite{Schwarz2012_Outflow_VeLLO}. We compared the result of this method and the result of the disc-kinematics method (described in the next paragraph) for the particular case of IC348-SMM2E, and find that this method yields $\Macc\sim5\times10^{-5}$~\mo\ \citep{Kim2019_CO_Outflow_VeLLO}, while $\Macc$ is 0.038--0.115~\mo\ from the disc-kinematics method \citep[][and this work]{Palau2014}, thus differing by more than two orders of magnitude. 

\item[Disc kinematics:] If a velocity gradient perpendicular to the outflow is identified, it is reasonable to assume that this is tracing a Keplerian motion associated with a disc structure. In this case, $\Macc$, usually referred to as `dynamical mass' or $\Mdyn$, can be estimated following eq.~\ref{eq:Macc}, or fitting the Keplerian curve for well resolved cases. This is routinely done for protostars in all the range of masses \citep[e.g.,][]{Tobin2012_L1527, Guilloteau2014_Mdyn, Maret2020_CALLYPSO, Thieme2023_eDisk-I15398}. However, the typical uncertainties of this method can be large, at least of 50\% \citep[e.g.,][]{Andrews2008_I04158, Maret2020_CALLYPSO}, and in Class 0 objects infall needs to be taken into account. For more evolved objects, a good match between dynamical masses inferred from disc kinematics and masses from evolutionary models has been reported \citep{Guilloteau2014_Mdyn}. Thus, to properly obtain accurate $\Mdyn$ using this method, it is imperative the use of sensitive interferometric observations, to detect the Keplerian motion with detail and constrain the model. If the Keplerian motion is properly resolved, this method can be considered the most reliable to constrain the accreted mass in proto-BDs.

\end{description}

\subsection{Need to determine the mass that will be accreted in the future}\label{sec:challenges-massfuture}

Assuming that $\Macc$ could be determined reasonably well, this is not enough yet to definitely confirm that the object is a bona-fide proto-BD. Since proto-BDs are still undergoing an active accretion phase, it is required to somehow estimate the total amount of mass that the object will accrete in the future. A first approach is based on $\Menv$, which in principle constitutes the mass reservoir of the object. Thus, a good estimate of $\Menv$ is vital to assess whether the object will remain substellar or not after the main accretion phase. For this, it is essential to measure $\Menv$ using a single-dish, because masses inferred using interferometric measurements suffer from filtered flux. In addition, it is required to adopt a certain Star Formation Efficiency (SFE). However, which value to adopt is not well defined yet and probably depends on the large-scale environment. For example, \citet{Alves2007_IMF_MolecularClouds} and \cite{Andre2010_CMF} suggest SFE$\sim30$\%, but in early stages it has been shown that the core formation efficiency depends on the density of the environment \citep{Motte1998_rhoOph, Bontemps2010_CygnusX, Palau2013_FragmentationMassiveCores, Pandian2024_fragmentation-CFE}. It is also important to keep in mind that the material accreted could eventually come from scales larger than the dense core associated with the proto-BD \cite[e.g.,][]{ValdiviaMena2023_accretionstreams}.
Another approach to estimate the mass that will be accreted in the future is to determine the mass-accretion rate and to estimate the remaining time of accretion.
Once the mass that will be accreted in the future is determined, it must be added to the current mass, $\Macc$, and if the total mass remains $<0.075$~\mo, the object can be considered a robust proto-BD.

\subsection{Episodic accretion and variability complicate the estimate of luminosity}\label{sec:challenges-episodicacc}

Episodic mass accretion complicates the classification of proto-BDs because proto-BDs could also be protostars that are in a quiescent state \citep[e.g.,][]{Lee2007_ChemicalEvolution_VeLLO, Schwarz2012_Outflow_VeLLO}. \cite{Dunham2012_LuminProblem} show that accretion bursts are required to solve the so-called `luminosity problem', which refers to the fact that the luminosities measured for embedded protostars are about factors 10--100 smaller than expected given the predicted mass accretion rates in the `standard model' of star formation (assuming that, in the embedded phase, the luminosity mainly comes from accretion as in eq.~\ref{eq:Lacc}). Thus, when detecting a very-low luminosity object, it could be either a very low mass object or a more massive object in a quiescent stage, and it has been proposed that chemistry could be a useful tool to distinguish between both possibilities \citep[e.g.,][]{Lee2007_ChemicalEvolution_VeLLO}.

One of the first cases where chemistry was shown to trace a recent accretion burst in a low-mass protostar is IRAS\,15398$-$3359 \citep[][see Sec.~\ref{sec:mainprops-interferometric} for further details on the chemistry]{Joergensen2013_I15398}.
The burst was supposed to have occurred about 100--1000 years ago, increasing the luminosity by two orders of magnitude. This would naturally explain the high luminosity measured for this source ($\Lbol\sim1.8$~\lo) and the low $\Macc$, of around 0.01--0.09~\mo\ \citep[e.g.,][]{Oya2014_I15398, Yen2017_I15398, Okoda2018_I15398}. 

Another case where chemistry reveals a possible accretion burst is IRAM\,04191+1522 (hereafter IRAM\,04191). \cite{Anderl2020_Chemistry_VeLLO_IRAM04191} find, using the Plateau de Bure Interferometer (PdBI), that the morphology of the N$_2$H$^+$ and C$^{18}$O lines is not consistent with a constant luminosity model using the present-day internal luminosity, and that the N$_2$H$^+$ peaks are rather consistent with an accretion burst corresponding to a luminosity 150 times higher than the current luminosity, that should have taken place about 150 years ago.
Taking into account the accretion rate in the bursts and an accretion rate in the quiescent phase and a number of additional assumptions, a final masss in the Class 0 stage of 0.2--0.25~\mo\ is estimated for this object. However the authors assume an infinite mass reservoir, while IRAM\,04191 is located in an isolated environment (see Fig.~\ref{fig:spadistribTaurusPerseus}-top) and the largest $\Menv$ reported is of $\sim0.5$~\mo\ \citep{Andre1999_Portostar_Taurus}\footnote{$\Menv$ for IRAM\,04191 has been estimated in different ways, being the value reported by \cite{Kim2016_VeLLOs} among the lowest, $\Menv\sim0.05$~\mo.}. This would require a SFE of 50\%, which seems too high. Additionally, the mass loss rate should be taken into account also in the calculation of the final mass. 
Thus, although the precise final mass of IRAM\,04191 is still open, chemistry has already been shown to be able to trace accretion bursts in several cases, including larger samples of VeLLOs \citep[e.g.,][]{Hsieh2018_VeLLOs}. In a very recent work, the NO molecule has been proposed as a potential outburst tracer in the VeLLO DC3272+18 by \cite{Kulterer2024_outburst}.

%
Episodic accretion can also be characterized by studying the molecular outflows or jets. 
For example, \cite{Okoda2021_I15398} and \cite{GuzmanCcolque2024_I15398} find signs of episodic ejection in IRAS\,15398$-$3359, consistent with the chemical results of \cite{Joergensen2013_I15398}. 
Similarly, \citet{Dutta2024_EpisodicAccretion} report episodic accretion based on a study of molecular jets in the Orion Nebula Cloud. They report that mass-loss rates, mass-accretion rates, and periods of accretion events are dependent on $\Lbol$ and $\Menv$, indicating that a large envelope favors the accretion-ejection processes. They determine mean periods of ejection events of 20--175 years in their sample (high-accretion rates may trigger more frequent ejection events), consistent with the assumptions done in \citet{Hsieh2018_VeLLOs}, who adopt about 100-200 years for the duration of the bursts, an a time interval between bursts of 10\,000--20\,000 yr. 

All these studies demonstrate the presence of accretion bursts in protostellar and proto-BD objects, illustrate the complexity of dealing with episodic accretion, and show that chemistry is a powerful tool to distinguish between very low-mass objects and quiescent protostars. Further details on the chemistry of accretion bursts are provided in Sec.~\ref{sec:mainprops-interferometric}.

\subsection{Multiplicity could hinder proto-BD selection}\label{sec:challenges-mult}

In addition to the aforementioned difficulties to identify proto-BDs, it must be kept in mind that the proto-BD candidates could actually be multiple systems, implying that $\Lint$ is an upper limit. This would be expected, given the high multiplicity fraction found for deeply embedded protostars \citep[e.g.,][]{ChenArce2012_IRAM04191, ChenArce2013_multiplesyst, Tobin2018_VANDAM}. On the other hand, it is well known that the stellar binary fraction increases with the mass of the primary \citep[e.g.,][]{Offner2023_PPVII}. Thus, it is of decisive importance to study multiplicity among proto-BDs, and to take this into account when defining upper limits to $\Lint$ for proto-BD selection. Studies of large samples of proto-BDs with high angular resolution would be key to shed light on this topic.

\subsection{Extragalactic contamination}

Finally, if no molecular gas is detected towards the proto-BD candidates, it is necessary to consider the possibility that the object is a background object. There are several tests applied in the literature to assess this possibility, mainly focused on color-color diagrams \cite[e.g.,][]{Harvey2007, Barrado2009}. In addition to this, the study of proper motions (Hu\'elamo et al., in prep.) or centimeter emission \citep[e.g.,][]{Morata2015_Jets_ProtoBD, Rodriguez2017_BDs-VLA, Rodriguez2024_JuMBO} are very useful tools to assess the extragalactic nature.\\

Given the aforementioned list of challenges to properly identify pre- and proto-BDs, it can be easily understood that there are no large samples of pre- and proto-BDs in the literature. Thus, it becomes mandatory to focus our efforts on defining criteria that help to efficiently select bona-fide pre- and proto-BD candidates. In addition, very high angular resolution observations as provided by new-generation telescopes will be crucial to refine or re-define the criteria.

\section{Main properties of proto-BD candidates known so far}\label{sec:mainprops}
\subsection{Requirements for bona-fide proto-BD candidates}\label{sec:mainprops-requirements}

The observational efforts to identify proto-BDs presented in the previous sections 
constitute a very good starting point, but they contain very different approaches and criteria. In this section we synthesize what we consider as the backbones to select promising candidates. These backbones are intended to serve as a guide to build reasonable samples of proto-BD candidates from big data surveys and catalogs. By no means these requirements are enough conditions for an object to be a confirmed proto-BD. The confirmation of the candidates will have to be done using interferometers in follow-up deep, dedicated studies.
Taking into account the criteria applied in the aforementioned strategies, along with the lessons learned from these works, we propose the following four requirements for an object to be a promising proto-BD candidate:

\begin{description}

\item[Requirement 1. Substellar central object, $\Lint\,\leq0.13$~\lo:] Ideally, it would be desirable to have good measurements of $\Macc$ and to use this parameter for proto-BD identification and searches. However, as stated in Sec.~\ref{sec:challenges-Macc}, the determination of $\Macc$ is highly uncertain or difficult to obtain because it would require sensitive interferometric observations of the gaseous disc, or very accurate mass-accretion rates and accretion timescales. This prompts to search for alternate parameters that could be related to $\Macc$. In principle, the luminosity should be a proxy to the accreted mass $\Macc$. For Class 0/I objects, where an important amount of the luminosity comes from accretion, $\Macc\sim\Lacc$ (see eq.~\ref{eq:Lacc}). In turn, $\Macc\sim\Lacc\sim\Lint$ because $\Lint$ corresponds to the luminosity contribution due to only star+disc, without taking into account the external contribution of the ISRF. Since $\Lint$ is easily calculated from observations at 70~\mum\ (eq.~\ref{eq:LintFlux70mic}), establishing a relation between $\Macc$ and $\Lint$ would be of great help in the proto-BD search. In the literature, the criterion of $\Lint \leq 0.2$~\lo\ is used to assess the substellar nature for the most embedded cases (common to almost all the strategies mentioned in Sec.~\ref{sec:searches},  \citealt{Young2004_VeLLOs,Dunham2008_c2dVeLLOs,Kim2016_VeLLOs}), which takes into account the uncertainty in the relation between the flux at 70~\mum\ and $\Lint$ of \cite{Dunham2008_c2dVeLLOs}. 
In the case that 70~$\mu$m\ fluxes were not available, it is  required in some cases that $\Lbol<0.3$~\lo\ \citep[e.g.,][]{Riaz2016_ProtoBD_Serpens}, which is used as an upper limit for $\Lint$. However, the criterion of $\Lint \leq 0.2$~\lo\ is probably too relaxed to select promising proto-BDs for the following reason.
We have compiled a list of protostars and VeLLOs for which $\Lint$ has been estimated following eq.~\ref{eq:LintFlux70mic}, and for which $\Macc$ has been derived from the `disc-kinematics' method outlined in Sec.~\ref{sec:challenges-Macc}, i.e., $\Macc=\Mdyn$.
In this compilation, we have restricted $\Lint$ to be $<10$~\lo, because the relation between flux at 70~\mum\ and $\Lint$ was explored only for $<10$~\lo\ \citep{Dunham2008_c2dVeLLOs}.
This compilation is presented in Table~\ref{tab:LintMdyn} and the $\Lint$--$\Mdyn$ relation is presented in Fig.~\ref{fig:LintMdyn}. A few objects from Table~\ref{tab:LintMdyn} have not been plotted in Fig.~\ref{fig:LintMdyn} for several reasons. The cases of known multiple systems for which $\Lint$ is obtained including all components while $\Mdyn$ was obtained for only one of the components, yielding an obvious mismatch, have not been plotted in the figure. Other causes of exclusion are evidences in the literature that the source is outbursting, or $\Tbol<20$~K\footnote{Sources with $\Tbol<20$~K are not included because for these very early evolutionary stages (considered as FHCs in some papers) the emission at  70~\mum\ could still be increasing \citep[e.g., see Fig.~2 of][]{Young2023_FHCs} and very low $\Tbol$ has been associated also with pre-stellar cores \citep[e.g.,][]{Young2005_EvolutionarySignatures}.}.
The objects of Table~\ref{tab:LintMdyn} that have not been included in the figure, along with the reason for this exclusion, are indicated in column 5 of the table.
%
As can be seen from the figure, there is a strong correlation with a Spearman’s rank correlation coefficient $\rho=0.902$, and $p$ value (the probability that the null hypothesis is true, that is, that the relation arises from a random process) $<0.001$\footnote{We note that a certain dispersion in this relation is expected due to possible episodic accretion or variable mass accretion rate. For example,  
if the source is passing an accretion burst, its luminosity might be higher (by about two orders of magnitude, as recognized for the outbursting case of I15398, e.g., \citealt[]{Joergensen2013_I15398, Anderl2020_Chemistry_VeLLO_IRAM04191}) than expected from the typical accretion rate.}. 
A fit was performed using the Orthogonal Distance Regression (ODR) package of Python, taking into account errors in both axes. The resulting fit is:
\begin{equation}
\mathrm{log}(\Lint) = (0.41\pm0.11) + (1.62\pm0.14)\,\mathrm{log}(\Mdyn),    
\label{eq:LintMdyn}
\end{equation}
where $\Lint$ is given in \lo, and $\Mdyn$ is given in \mo.
It is interesting to note that this relation resembles as a first approximation the relation between photospheric luminosity and accreted mass from the models of \cite{DAntona1994_evolutionarytracks} (Tables 1 and 5 of this paper), for a time step of $5\times10^5$~yr (red thin solid line in Fig.~\ref{fig:LintMdyn}). 
This timescale is the typical timescale estimated for the Class 0+I phases \citep[e.g.,][]{Evans2009_Lifetimes, Dunham2015_YSO_GouldBelt, Heiderman2015_GouldBelt}.
The empirical relation is also similar to the $\Lint$ vs $\Macc$ relation for the hybrid models of \cite{Vorobyov2017_EffectAccretion_BD} for the same time step (blue thin solid line in Fig.~\ref{fig:LintMdyn}). These last set of models take into account the accretion luminosity. The fact that the empirical relation is very close to the theoretical relations suggests that this relation is real.
Once we have found a reasonable relation between $\Lint$ and $\Mdyn$, it is possible to set a $\Lint$ criterion for proto-BDs. In order to select promising proto-BD candidates, we are interested in $\Lint$ corresponding to $\Mdyn<0.12$~\mo. We have chosen $\Mdyn<0.12$~\mo\ instead of $\Mdyn<0.075$~\mo\ because the typical uncertainty in  $\Mdyn$ from the `disc-kinematics' method is around 50\% (Sec.~\ref{sec:challenges-Macc}), implying that an object with $\Mdyn=0.075$~\mo\ might have an error associated of $\sim0.04$~\mo, giving an upper limit of $\Mdyn\sim0.12$~\mo. In addition, adopting $\Macc<0.12$~\mo\ allows to take into account possible binary BD-BD systems, something which cannot be ignored (Sec.~\ref{sec:challenges-mult}). 
From the fit to the data of Fig.~\ref{fig:LintMdyn}, $\Mdyn=0.12$~\mo\ corresponds to $\Lint=0.13$~\lo\ (taking into account two times the uncertainty of the fit, whose corresponding shaded area includes $\sim70$~\% of the points used to infer the correlation). In the figure, the dashed vertical green line indicates $\Mdyn=0.12$~\mo, and the dashed blue horizontal line indicates $\Lint=0.2$~\lo. As can be seen in the figure, if we took $\Lint\,\leq\,0.2$~\lo\ as the criterion to select proto-BD candidates (as in previous searches, e.g., \citealt[]{Kim2016_VeLLOs}), we would be including a significant number of objects with $\Mdyn>$0.2~\mo, making our criterion too relaxed.
Furthermore, our proposed criterion of $\Lint\leq0.13$~\lo\ is also consistent with \cite{Young2004_VeLLOs} and \cite{Kauffmann2011_VeLLO_L1148-IRS}, who show that the restriction of $\Lint\lesssim$0.1~\lo\ limits the current mass to $\Macc\lesssim0.1$~\mo. 
It is important to note that Fig.~\ref{fig:LintMdyn} also shows a number of objects with $\Mdyn\sim0.2$--0.3~\mo\ but $\Lint\lesssim0.1$. According to \cite{Tokuda2017_L1521FIRS}, these protostars could be explained if they are undergoing cold accretion (no significant increase of entropy due to their low mass-accretion rate). Thus, if we used the criterion of $\Lint\leq$0.2~\lo, the number of protostars (not proto-BDs) picked up undergoing cold accretion would probably be too high, while using $\Lint\leq$0.13~\lo\ allows to minimize this effect. 
Finally, our proposed criterion is slightly less strict than the criterion adopted by \citet[]{Burrows2001_BD_Planets} and \cite{Riaz2018_chem-OphSer}, who suggest that a proto-BD must have $\Lbol \lesssim 0.09$~\lo. Thus, our criterion of $\Lint\,\leq\,0.13$~\lo\ is also a compromise between the different criteria adopted in the literature\footnote{We do not consider the use of $\Lbol$ to select proto-BD candidates because this measurement could be highly affected by the ISRF (see Sec.~\ref{sec:mainprops-sample}).}.

\begin{figure}
\begin{center}
\begin{tabular}[b]{c}
    \epsfig{file=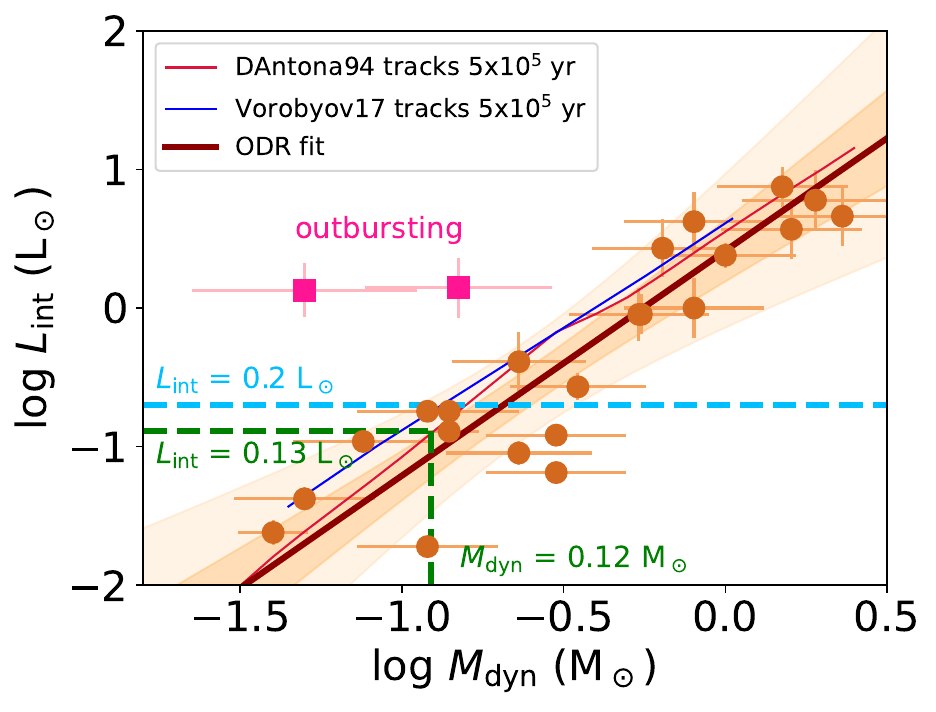, width=9cm,angle=0}\\
\end{tabular}
\caption{$\Lint$ vs $\Mdyn$ for the protostars, VeLLOs and proto-BD candidates for which the accreted mass has been measured from the `disc-kinematics' method using interferometers, and $\Lint$ has been estimated from the 70~\mum\ flux (see Sec.\ref{sec:challenges-Macc}). The correlation is strong, with a Spearman’s rank correlation coefficient $\rho=0.902$, and $p$ value $<0.001$. The dashed vertical green line indicates $\Mdyn=0.12$~\mo, the dashed horizontal green line indicates $\Lint=0.13$~\lo, and the dashed horizontal blue line indicates $\Lint=0.2$~\lo. The linear fit, done with the ODR package of Python and weighting by both x and y errors, is: 
log($\Lint$) = ($0.41\pm0.11$) + ($1.62\pm0.14$)log($\Mdyn$). 
The shaded area corresponds to 2 times (dark) and 5 times (light) the uncertainty of the fit.
%
The red thin solid line corresponds to the relation between photospheric luminosity and accreted mass from the models of \cite{DAntona1994_evolutionarytracks} (Tables 1 and 5 of this paper), for a time of $5\times10^5$~yr. 
The blue thin solid line corresponds to the $\Lint$ vs $\Macc$ relation for the hybrid models of \cite{Vorobyov2017_EffectAccretion_BD} for the same time. 
}
\label{fig:LintMdyn}
\end{center}
\end{figure} 


\item[Requirement 2. Class 0/I SED, $\Tbol < 650$~K:] To consider a proto-BD as such, its SED must necessarily be equivalent to Class 0 or Class I of protostars SEDs. Having a complete well-sampled SED allows to establish the evolutionary stage of the object. A robust proto-BD candidate should show a peak at the far-IR, consistent with a Class-0/I stage. This was quantified by \citet{Chen1995_Tbol}, who proposed to use the $\Tbol$ parameter as an evolutionary indicator (see Sec.\ref{sec:defs}), with Class 0 objects having $\Tbol < 70$~K, and Class I or flat sources having $\Tbol < 650$~K.

\item[Requirement 3. Envelope mass low enough, $\Menv \lesssim 0.3$~\mo:] To warrant that the accreting object will not achieve a stellar mass in the future, the mass reservoir in its surroundings must be low enough (see Sec.~\ref{sec:challenges-massfuture}). This mass reservoir is typically considered as the mass of the envelope around the object, $\Menv$, and should be measured using single-dish observations in the millimetre/sub-millimetre range so that the bulk of the emission is not filtered out. The masses estimated from this kind of observations can be used to estimate the final mass that would be accreted onto the substellar candidates and then decide if they will end up as BDs or very low-mass stars. To do so, one can assume a core formation efficiency or CFE in low-mass dense cores. This quantity has been found to span a wide range of values and to increase with the density of the core \citep[e.g.,][]{Motte1998_rhoOph, Bontemps2010_CygnusX, Palau2013_FragmentationMassiveCores, Pandian2024_fragmentation-CFE}. From these works and taking into account typical densities of proto/pre-BDs cores of $10^4$--$10^5$~\cmt\ \citep[e.g.,][]{deGregorio2016_PreProtoBD_ChaII,Huelamo2017_B30}, a value of CFE$\sim10$\% is obtained. This is close to the SFE value of $30\pm10$\% found by \cite{Alves2007_IMF_MolecularClouds}, the value used in the eDisk program \citep[e.g.,][]{Thieme2023_eDisk-I15398}, or previous works studying proto-BD candidates \citep[e.g.,][adopt 30--50\% for the jet efficiency]{Riaz2021_AccretionOutflow_ProtoBD}. 
The most delicate case is that of Class 0 proto-BDs because these still have to accrete half of their mass. And among the  Class 0 proto-BDs, the most critical cases are those with $\Macc$ already of about 0.030--0.035~\mo. Assuming CFE of 10\%, the maximum mass measured with single-dish should be 0.3~\mo, and the mass that should be added to $\Macc$ would be $\lesssim0.03$~\mo. We note that this is an orientative value and that each individual case should be studied in detail to carefully assess if the proto-BD candidate will remain substellar or not, given the current accreted mass, the mass-accretion rate, the mass reservoir in the envelope, and the extension of accretion, that in some cases can take place from further out of the envelope \citep{ValdiviaMena2023_accretionstreams}.
%

\item[Requirement 4. Tests rejecting a background nature:] Given the intrinsic faint nature of proto-BD candidates, it is mandatory to apply as much tests as possible to rule a background nature out. These tests could consist of i) looking for line emission associated with the known cloud velocity or molecular outflows; ii) proper motion study in case there is infrared data available; iii) searching for extragalactic catalogues, e.g., in the NASA Extragalactic Database, or testing diagnostic color-color diagrams \citep[e.g.,][]{Harvey2007, Barrado2009, Bouy2009_LOri_QKiller}; iv) looking for centimetre thermal emission, very unsual in extragalactic objects \citep[e.g.,][]{Morata2015_Jets_ProtoBD}. 

\end{description}

\begin{figure}[H]
\centering
\begin{tabular}[!h]{c}    
\epsfig{file=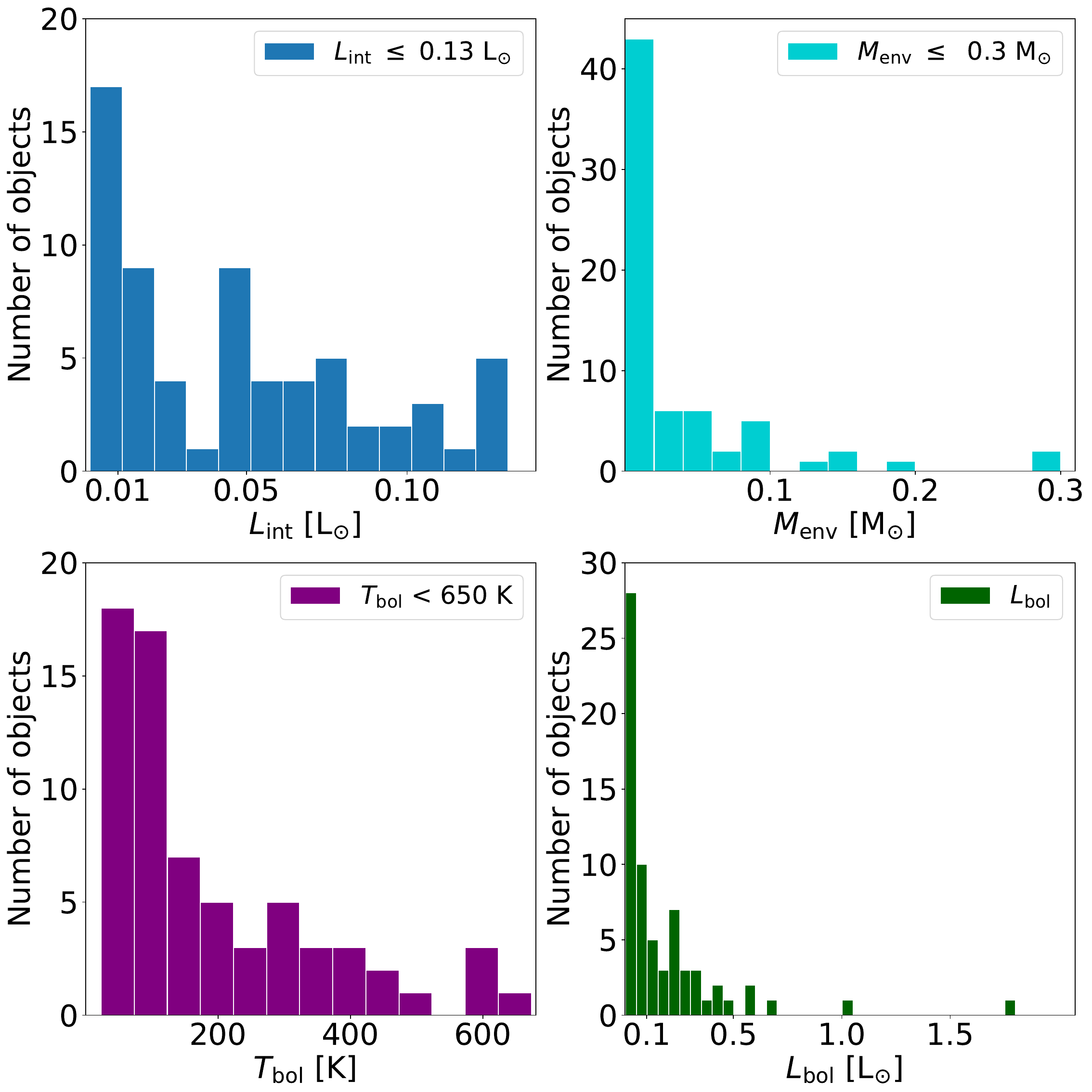, width=8cm,angle=0}\\
\end{tabular}
\caption{Histograms of best proto-BD candidates from SUCANES database following the criteria outlined in Sec.~\ref{sec:mainprops-requirements}.
Top-left: $\Lint$ histogram. 
Top-right: $\Menv$ histogram.
Bottom-left: $\Tbol$ histogram.
Bottom-right: $\Lbol$ histogram.
}
\label{fig:histo}
\end{figure}

We would like to highlight at this point that the four criteria mentioned above are designed to conduct efficient searches of proto-BDs. However, to definitely confirm the most promising candidates as bona-fide proto-BDs, dedicated interferometric studies are required to shed light on the definite nature of the objects. These dedicated studies should be focused on identifying clues of mass ejection and disc-like structures. The interferometric detection of molecular outflows or thermal radiojets clearly associated with the candidate provides a very important indication of an accreting core \citep[e.g., ][]{Bourke2005_Outflow_L1014IRS, Lee2013_ProtoBD_L328-IRS, Morata2015_Jets_ProtoBD}. In addition, the detection of rotating disc-like structures with a substellar dynamical mass is probably one of the most reliable proves of the nature of the object \citep[e.g.,][]{Palau2014, Lee2018_L328_ProtoBD_ALMA, Riaz2021_Structure_ProtoBD}, but this can only be done with interferometers. 

\subsection{A large sample of proto-BD candidates}\label{sec:mainprops-sample}

As explained in Sec.~\ref{sec:searches}, the searches of pre- and proto-BDs have been performed using different observational approaches. As a result, there are hundreds of objects classified as VeLLOs, proto-BD and pre-BD candidates.  All these objects have been gathered in a database that contains all the objects classified as SUbstellar Candidates at the Earliest Stages (SUCANES) and published until the year 2020 (included). All the details about how the database has been built and its contents are described in \cite{PerezGarcia2024_SUCANES}.

In brief, the database\footnote{https://sucanes.cab.inta-csic.es/} contains a total of 174 objects that have been compiled from the literature. The identification and characterization of each object is described in the corresponding paper, and is based on different criteria based on the performed observations. This results in an inhomogeneous sample in terms of properties and observables. 
In any case, the database reflects the advances in the field of low luminosity Class 0/I objects.

All the data gathered in the database come from publications (including all the works mentioned in Sec.~\ref{sec:searches}). The photometry has been complemented with data from  public archives and covers from the optical to centimeter wavelengths. The only physical parameters that have been derived by the database developers are $\Lint$, $\Tbol$ and $\Lbol$. 
They have also updated the distance values for different regions (e.g., Aquila, Perseus, IC348, Ophiuchus, Serpens, $\sigma$ Ori, IC\,5146, Chamaeleon, Cepheus and GF\,9), providing revised values for $\Lint$, $\Lbol$, and $\Menv$ (see distances and references for each cloud in Table 1 of \citealt{PerezGarcia2024_SUCANES}).

As shown in \cite{PerezGarcia2024_SUCANES}, the very young substellar candidates lie in the regions surveyed mainly by {\it Spitzer} and {\it Herschel}, with most of the candidates being identified in the regions of Taurus and Aquila. However, it is worth mentioning that the number of candidates has been substantially reduced when considering the revised distances in the case of Aquila (from 27 to 13), or IC\,5146 (from 8 to 4). 
\\

The criteria outlined in Sec.~\ref{sec:mainprops-requirements} were applied to the SUCANES database with the goal of selecting the best proto-BD candidates. First, we took the subsample of SUCANES with measured $\Lint$ and excluding pre-BD candidates, which results in 135
 objects\footnote{Note that one object fulfilling this criteria, J041847 \citep{Palau2012_PreBD_Cores_Taurus}, was removed from the subsample, since recent {\em Euclid} data has spatially resolved it into a galaxy (Bouy et al., in prep).}. 
 Then, we used the following filters: 

\begin{enumerate}
\item $\Lint < 0.13$~\lo: 
We adopted the criterion that $\Lint$ - err$\Lint$ $< 0.13$ \lo, and err$\Lint$/$\Lint$ $<$ 20\% for objects with $\Lint$ around 0.13~\lo.
This yields 90 objects out of 135 with reported values of $\Lint$ (both detections and upper limits) in SUCANES.


 \item $\Tbol < 650$~K\footnote{While the SUCANES database only includes $\Tbol$ and  $\Lbol$ values if the objects have detections in at least three photometric tables \citep{PerezGarcia2024_SUCANES}, in this review we have relaxed this condition to two photometric tables. The objects for which $\Tbol$ and $\Lbol$ have been obtained using only two photometric tables are marked in Table~\ref{tab:protoBDs}.}. 
This leaves six objects out in total: for five objects (L1448-IRS2E from \citealt{Chen2010_L1448_FHC},  SSTB213 J041913.10+274726.0 and SSTB213 J041726.38+273920.0 from \citealt{Palau2012_PreBD_Cores_Taurus}, ALMAJ153914.996-332907.62 from \citealt{SantamariaMiranda2021_ALMA_Lupus_ProtoBD}, and Source-X from \citealt{Kawabe2018}) the SED was not properly sampled, making the determination of $\Tbol$ uncertain, and were excluded. 
One object with $\Tbol>650$~K (J190418.6-373556 from \citealt{Kim2016_VeLLOs}) was excluded as well, so we ended up with 84 objects after requiring this criterion. 

\item $\Menv \leq 0.3$~\mo. We did not consider here any uncertainty and took a strict 0.3~\mo\ value because this criterion is the most uncertain (see Sec.~\ref{sec:mainprops-requirements} point 3) and the masses were not computed as consistently as $\Lint$, but were just taken from the literature, with each value having its own $T_{\rm dust}$ and dust opacity, etc.
It is worth noting however that 78\% of the sample have $\Menv$ calculated from {\it Herschel}/SPIRE at 250~\mum, and $\Menv$ has been obtained uniformly for all these objects. This 78\% corresponds to the objects from \cite{Kim2016_VeLLOs}, and to the objects of SUCANES that have no $\Menv$ reported in the literature but have a flux at 250~\mum\ (from {\it Herschel}/SPIRE) in the database. 
In these cases, $\Menv$ is calculated in this work adopting $T_{\rm dust}=15$~K and the same dust opacity at 250~\mum\ as \cite{Kim2016_VeLLOs}, to make the estimate of $\Menv$ as uniform as possible\footnote{The uncertainty in $\Menv$ due to dust opacity can be of a factor of 2 or more. For example, \cite{Kim2016_VeLLOs} adopted $\kappa_{250\mu m}=0.19$~cm$^2$\,g$^{-1}$ by interpolating the \cite{Ossenkopf1994_dustopacity} dust opacities for dust with thin ice mantles at 10$^6$~cm$^{-3}$, while a recent estimate of the dust optical depth at 250~\mum\ by \cite{Li2024_dustopacity} yields $\kappa_{250\mu m}=0.09$~cm$^2$\,g$^{-1}$, a factor of 2 smaller, translating into a factor of 2 larger $\Menv$. In other cases, the dust opacity is a factor of 2.5 larger than the thin ice mantle models at 10$^6$~\cmt\ of \cite{Ossenkopf1994_dustopacity} \citep[e.g.,][]{Lin2021_dustopacity}.
It is currently not clear which dust opacities are most appropriate for protostellar, VeLLO or proto-BD envelopes and this constitutes an important source of uncertainty.}. 
After adopting this criterion, we have 68 objects out of 84\footnote{The 16 objects rejected by this criterion are comprised by: 2 objects (J182912.1$-$014845, J182958.3$-$015740) from
\citealt{Kim2016_VeLLOs}, 
for which there are no $\Menv$ in the literature, and 14 objects (see Table~\ref{tab:rejected}) for which $\Menv>0.3$~\mo. Among these 14 rejected objects there is L328-IRS, an object that was considered as a promising proto-BD during many years, but has $\Menv\sim0.9$~\mo\ from {\it Herschel}/SPIRE flux at 250~\mum. Additionally, the mass for the central object of L328-IRS has been estimated to be $\sim0.3$~\mo\ from disc kinematics \citep{Lee2018_L328_ProtoBD_ALMA}, too high for a proto-BD.}. We note that all the 68 objects have $\Menv$ estimated from single dish data.



\end{enumerate}

Table~\ref{tab:rejected} lists the 22 
objects that have been excluded from the 90 initial objects that fulfilled the first criterion of $\Lint < 0.13$~\lo.

Table~\ref{tab:protoBDs} presents the list of the 68 proto-BD candidates with their main properties, 
which are summarized in Fig.~\ref{fig:histo}. The figure shows the histograms for the final sample of 68 proto-BD candidates for $\Lint$, which have peaks around 0.005 and 0.05~\lo;
$\Menv$ has a very well defined peak around 0.01~\mo; $\Tbol$ is $\lesssim100$~K in most of the cases; and $\Lbol$ also has the peak very well defined below 0.1~\lo, with very few exceptions of objects that belong to the Orion Nebula Cluster and lie close to the Trapezium, being thus their $\Lbol$ highly affected by the ISRF of the nearby high-mass stars. The most extreme case of these objects is J053504, for which $\Lbol\sim1.8$~\lo, while  $\Lint=0.10\pm0.01$~\lo, being the one closest to the Trapezium. This illustrates very well why it is recommended to use $\Lint$ to identify proto-BD candidates rather than $\Lbol$.

In order to have a wide view of the spatial distribution of the proto-BD candidates of Table~\ref{tab:protoBDs}, in Figs.~\ref{fig:spadistribOrionSerpens}--\ref{fig:spadistribisolated} of Appendix~\ref{app:spat-distr}, the positions of the SUCANES objects (with the subsample of \citealt{Kim2016_VeLLOs} highlighted), along with the proto-BD candidates of Table~\ref{tab:protoBDs} are indicated on the top of the  {\it Herschel}/SPIRE image at 250~\mum\ for each molecular cloud. Although a deep study of the spatial distribution of proto-BD candidates is beyond the scope of this review, this initial visualization is useful for future studies.

\begin{figure*}
\begin{center}
\begin{tabular}[b]{cc}
    \epsfig{file=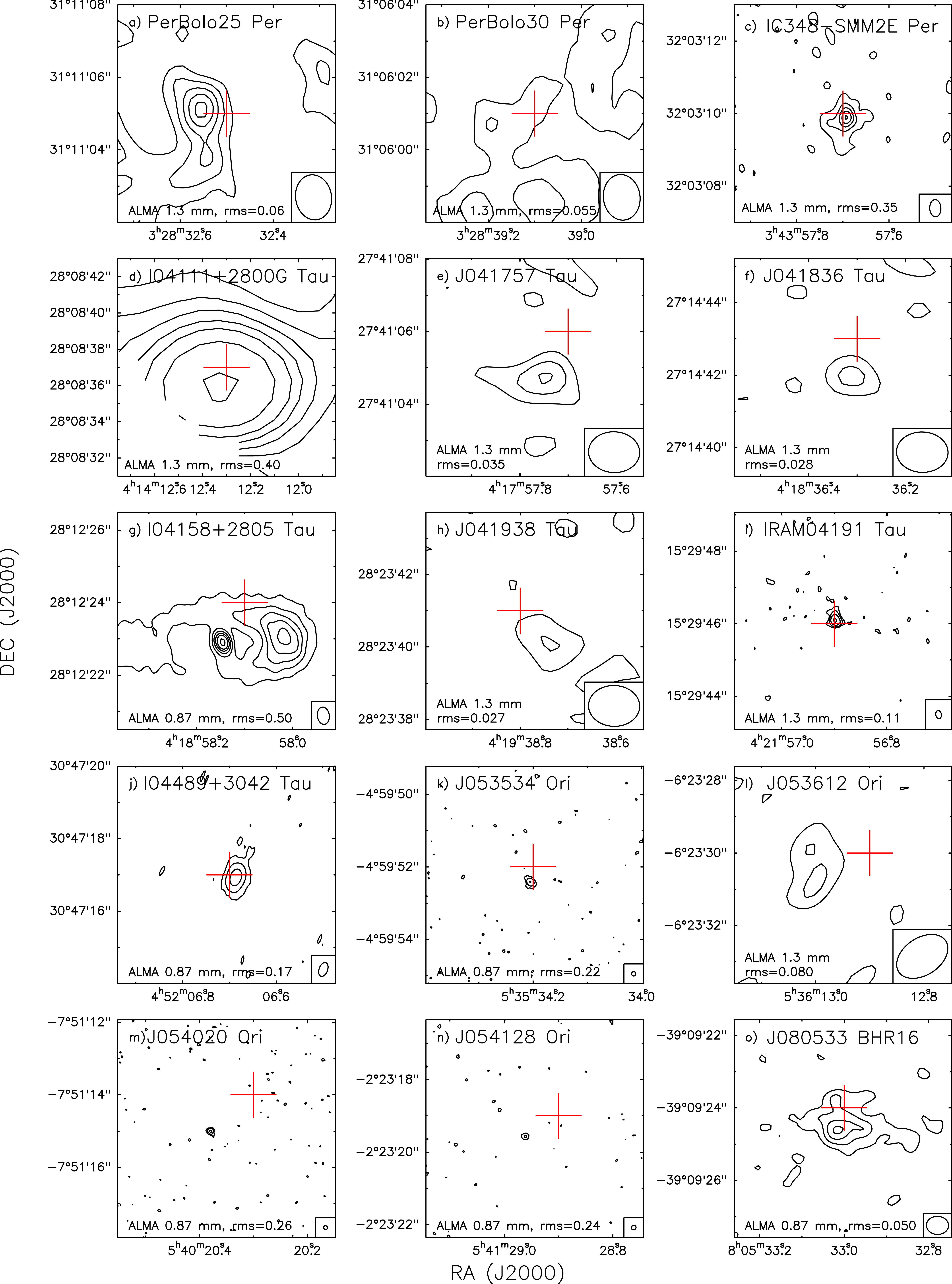, width=16cm,angle=0}\\
\end{tabular}
\caption{Archive ALMA emission of the proto-BD candidates of Table~\ref{tab:protoBDs}. In all the panels, the field of view is $6''\times6''$ (except for panel `d', for which the field of view is $12''\times12''$), the beam is shown in the bottom-right corner, and the contours correspond to the 1.3 mm or 0.87~mm continuum emission (and are $-3$, 3, 6, 9, 12, 15, 20, and 30 times the rms noise, except for IRAM04191 and J054020, which are $-3$, 3, 6, 12, 24 times the rms noise, and for I04489+3042, J053534 and J054128, which are $-3$, 3, 12, 48 times the rms noise). The rms noise of each image is indicated at the bottom of each panel and is given in m\jpb. In all the panels, the red plus sign corresponds to the coordinates of the proto-BD candidate as provided in the SUCANES database \citep{PerezGarcia2024_SUCANES}. The ALMA project codes along with the corresponding references are listed in Table~\ref{tab:protoBDs}.
}
\label{fig:alma}
\end{center}
\end{figure*} 

\setcounter{figure}{3}
\begin{figure*}
\begin{center}
\begin{tabular}[b]{cc}
    \epsfig{file=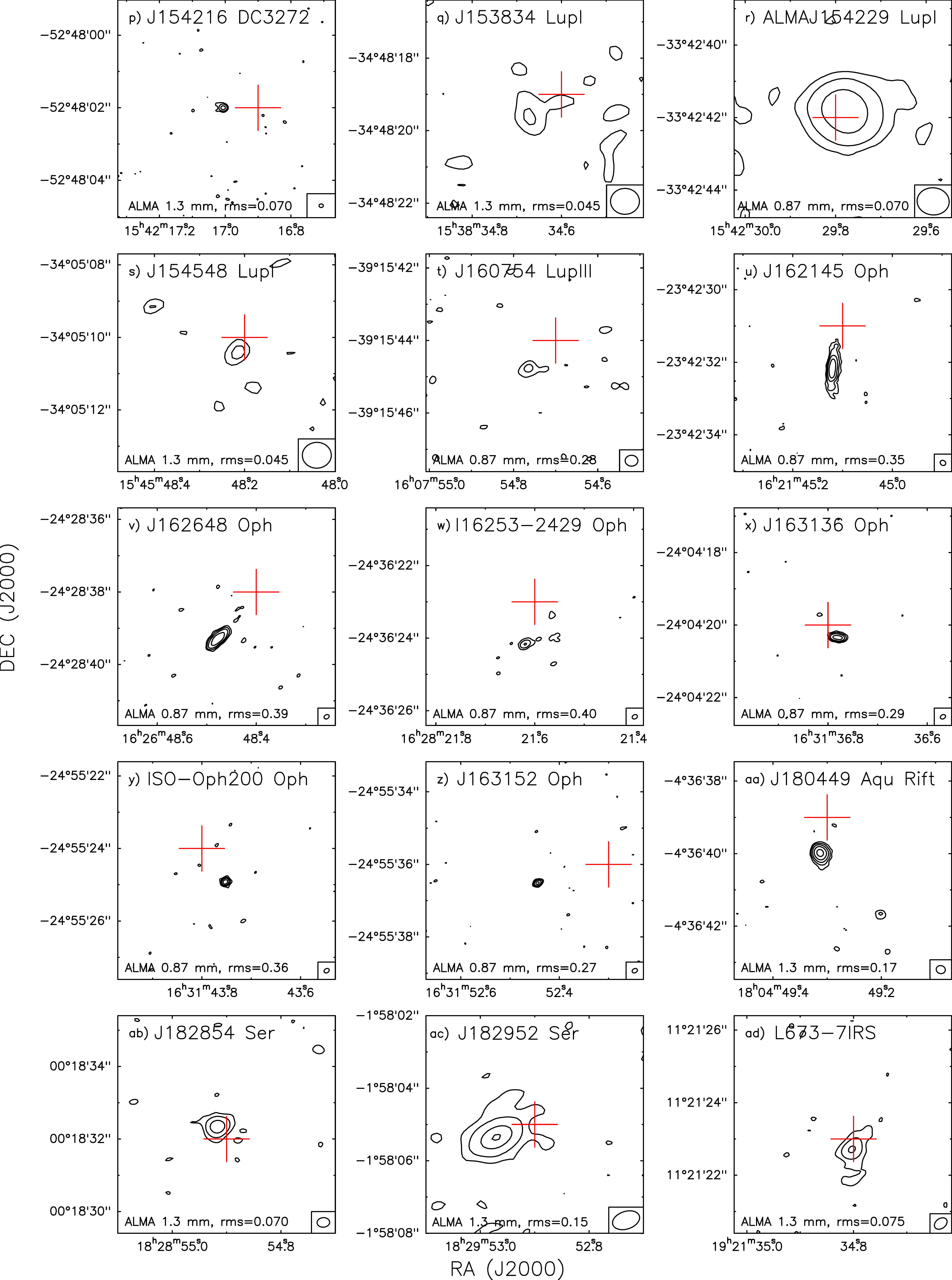, width=16cm,angle=0}\\
\end{tabular}
\caption{(cont.) Archive ALMA emission of the proto-BD candidates of Table~\ref{tab:protoBDs}. In all the panels, the field of view is $6''\times6''$, the beam is shown in the bottom-right corner, and the contours correspond to the 1.3 mm or 0.87~mm continuum emission (and are $-3$, 3, 6, 9, 12, 15, 20, and 30 times the rms noise, except for J154548, which are $-3$, 3, 4 times the rms noise, for J162145, J162648, J163136, ISO-Oph200, J163152 and J180449, which are $-3$, 3, 6, 12, 24 times the rms noise, and for J154216, ALMA-J154229, I16253-2429, J182854, J182952, and L673-7-IRS, which are $-3$, 3, 12, 48, 96 times the rms noise). The rms noise of each image is indicated at the bottom of each panel and is given in m\jpb. In all the panels, the red plus sign corresponds to the coordinates of the proto-BD candidate as provided in the SUCANES database \citep{PerezGarcia2024_SUCANES}.
The ALMA project codes along with the corresponding references are listed in Table~\ref{tab:protoBDs}.}
\label{fig:alma}
\end{center}
\end{figure*}

The ALMA archive was searched for the 68 proto-BD candidates of Table~\ref{tab:protoBDs} and ALMA data were available in 34 cases, with only 4 (out of the 34) being non-detections. The continuum detections are shown in Fig.~\ref{fig:alma}, and the non-detections are shown in Fig.~\ref{fig:alma-nondetections}. Although the images shown in Fig.~\ref{fig:alma} are very inhomogeneous, with a large range of angular resolutions and sensitivities, in an important number of cases the continuum emission is well resolved, and in some Class 0 candidates it is elongated along the outflow direction (e.g., IC348-SMM2E or IRAM\,04191, \citealt{Palau2014}, \citealt{Andre1999_Portostar_Taurus}) while in other Class I/Flat candidates it is tracing a disc structure (e.g., I04158+2805, J162145, J162648, J163136, \citealt{Ragusa2021_I04158}, \citealt{Cox2017_Oph-disks}, \citealt{Encalada2021_Oph-disks}), both properties also seen in protostars.



\subsection{Properties of proto-BD candidates and VeLLOs obtained from deep interferometric observations that can be compared to the properties of protostars}\label{sec:mainprops-interferometric}

In Sec.~\ref{sec:mainprops-requirements}, four criteria were proposed to select proto-BD candidates. These criteria are orientative requirements to build large samples of proto-BD candidates. However, additional deep and sensitive interferometric observations are required to definitely assess the nature of each object. We outline below the main properties that can be studied with dedicated interferometric observations, and the main results obtained so far in the very low-luminosity regime. 

\subsubsection{Molecular outflows} 

In some specific cases, deep interferometric observations have already been carried out in VeLLOs of the literature. For example, \cite{Lee2018_L328_ProtoBD_ALMA} show ALMA data for the VeLLO candidate L328-IRS (excluded here as a proto-BD candidate because of its high reported $\Menv$, see Table~\ref{tab:rejected}) and report the detection of an outflow. Similarly, \cite{Palau2014} report outflow emission from CO\,(3--2) and $^{13}$CO(2--1) with the SMA from the IC348-SMM2E proto-BD candidate. In addition, a PdBI interferometric study of a sample including two VeLLOs (IRAM\,04191+1522 and L1521-F) and Class 0 protostars reveal, through observations of CO, SO and SiO, that the detection rate of jets in SiO and SO was found to increase with $\Lint$ and thus only very few jets of SiO and SO would be expected to be associated with proto-BDs \citep[e.g.,][]{Podio2021_Class0}. Thus, detection of SiO/SO outflows in the deep interferometric studies is challenging.

\subsubsection{Accretion discs and streamers} 

There are proto-BD candidates or VeLLOs for which putative discs have been studied using sensitive interferometric observations, which reveal velocity gradients perpendicular to the outflow \citep[e.g.,][]{Palau2014, Lee2018_L328_ProtoBD_ALMA}. The most reliable cases were gathered and presented in Table~\ref{tab:LintMdyn}, for which $\Macc$ was inferred using the `disc-kinematics' method outlined in Sec.~\ref{sec:challenges-Macc}. Very recently, also accretion streamers and spiral structures (traced by HCO$^+$ and c-HCCCH) near a possible proto-BD were reported \citep{Riaz2024_streamer}, along with hints of molecules (such as HCN) tracing its inner disc \citep{Riaz2024_innerdisk}.

\subsubsection{Chemistry and presence of Complex Organic Molecules} 

The first studies aimed at characterizing the chemistry in VeLLOs were carried out using single-dishes. \cite{Takakuwa2011_OrganicMol_VeLLO_L1511F-IRS} used the Nobeyama\,45m telescope to observe carbon-chain and organic molecular lines towards L1521F-IRS (Table~\ref{tab:rejected}) and IRAM\,04191 (Table~\ref{tab:protoBDs}). These authors report a much stronger detection of CH$_3$CCH, C$_4$H, and C$_3$H$_2$ towards L1521F-IRS compared to IRAM\,04191, and non-detection of CH$_3$OH and CH$_3$CN in any of the two VeLLOs. They attribute the chemical differentiation identified among the two objects to different evolutionary stages.
Another single-dish study focused on VeLLOs/proto-BD candidates is presented in \citet{Riaz2018_chem-OphSer}, who observed N-bearing molecules in a sample of 10 objects with the IRAM\,30m telescope. This work revealed that the chemistry related to CN, HCN and HNC could have a particular behavior in VeLLOs compared to protostars, with the HNC molecule proposed as a tracer of deeply embedded substellar objects.
In a subsequent study, \citet{Riaz2019b_chem-OphSer} focused on the CO isotopologues, H$_2$CO, HCO$^+$, and CS in a sample of 7 proto-BD candidates, also using the IRAM\,30m telescope. These authors found evidences of CO depletion from the gas-phase, and that proto-BD candidates show a factor of 2 higher ortho-to-para H$_2$CO ratio than protostars. In addition, marginal signs of lower abundances of HCO$^+$ and H$_2$CO for lower $\Lbol$ are reported. Follow-up works find evidences of
cold and warm CH$_3$OH \citep{Riaz2023_Methanol}, deuteration \citep{Riaz2022_D_H_ratio_ProtoBD}, H$_2$CO and methane in a sample of proto-BD candidates \citep{Riaz2022_HDCO_D2CO_Class0_I_ProtoBD, Riaz2022_CH3D_Class0_I_ProtBD}.

Regarding interferometric chemical studies, these have been mainly focused on single sources. The most studied objects using interferometers are IRAS\,15398$-$3359 (not included in Tables~\ref{tab:rejected} or \ref{tab:protoBDs} because of its high luminosity, see Sec.~\ref{sec:searches-system}), L1521F (Table~\ref{tab:rejected}), and IRAM\,04191 (Table~\ref{tab:protoBDs}), and in all of them hints of accretion shocks and/or bursts were identified\footnote{The following features are considered signs of accretion bursts: absence of HCO$^+$ where there is CO; extended CH$_3$OH given its current luminosity; warm carbon-chain chemistry on large scales; no massive dust continuum component on small scales \citep[e.g.,][]{Joergensen2013_I15398};  morphology of the N$_2$H$^+$, CO and C$^{18}$O lines inconsistent with a constant luminosity model using the present-day internal luminosity \citep{Hsieh2018_VeLLOs, Anderl2020_Chemistry_VeLLO_IRAM04191}.} \citep[e.g.,][]{Joergensen2013_I15398, Favre2020_Methanol_VeLLO_L1521F, Anderl2020_Chemistry_VeLLO_IRAM04191}, along with complex chemistry in IRAS\,15398$-$3359 unveiled both by ALMA \citep{Okoda2023_I15398} and the JWST \citep{Yang2022_JWST-Class0}.
These hints of burst-induced chemistry, initially found in individual objects, were also found in a sample of 8 VeLLOs studied with ALMA by \cite{Hsieh2018_VeLLOs}. 
The authors also suggest that the interval between accretion episodes in  proto-BD or VeLLO candidates is shorter than that in Class 0/I protostars. 

Chemistry can also provide clues on the formation scenario of BDs by studying isotope ratios. For example, \cite{Barrado2023} measure with the JWST the $^{14}$N/$^{15}$N ratio in a cool equal mass BD-BD binary of 15~\mj\ each component, and find a value consistent with star-like formation by gravitational collapse, being the $^{14}$N/$^{15}$N ratio a very promising tool to distinguish among scenarios. This isotope ratio can be complemented by others, such as $^{13}$C/$^{12}$C \citep[e.g.,][]{Zhang2021_CO_BD, Zhang2021_CO_Planet, GonzalezPicos2024_C-isotopes}.

In summary, chemistry can potentially help identify proto-BDs \citep[e.g.,][]{Riaz2018_chem-OphSer} and trace their formation mechanism \citep[][]{Barrado2023}. In addition, chemistry typical of warm-carbon chain objects, hot corinos and accretion bursts has been identified in VeLLOs and proto-BD candidates \citep[e.g.,][]{Joergensen2013_I15398, Okoda2023_I15398}, indicating that these phenomena must be taken into account when identifying, modeling and interpreting these objects.

\subsubsection{Maser emission} 

Maser emission is a good tracer of energetic processes such as mass-loss and/or accretion and could potentially be a useful tool to study these processes with very high angular resolution in proto-BDs. A first search of water maser emission towards the L1014 VeLLO was conducted by \cite{Shirley2007_L1014-maser}, who report a non-detection.
Later, a survey for water maser emission towards a sample of 44 VeLLOs, including proto-BD candidates, FHCs, and other protostars with $\Lbol\lesssim0.4$~\lo, was carried out by \cite{Gomez2017_Water_BD_VeLLOs} using the Effelsberg telescope. The survey only detected maser emission towards the position of L1448 IRS 2E, which has been classified as a possible FHC or a dense blob interacting with a nearby outflow \citep{Maureira2020_FHCs}. However, follow-up VLA observations showed that the maser emission is associated with the nearby object L1448 IRS2, a Class 0 protostar of $\Lbol\sim4$~\lo. The upper limits for water maser emission determined by \cite{Gomez2017_Water_BD_VeLLOs} are one order of magnitude lower than expected from the correlation between water maser luminosities and $\Lbol$ found for protostars \citep{Furuya2003_watermasers, Shirley2007_L1014-maser}. This suggests that this correlation does not hold at the lower end of the (sub)stellar mass spectrum and calls for further studies to understand this behavior.

\subsubsection{Magnetic fields} 

A few works have studied the magnetic fields in cores harboring proto-BD candidates or VeLLOs. Both \citet{Soam2015_Polarimetry-L328} and \cite{Soam2015_VeLLOs} perform  optical and near-infrared polarization observations toward VeLLOs. \cite{Soam2015_VeLLOs} study five VeLLOs (IRAM 04191, L1521F, L328, L673-7, and L1014), two of which are classified in this work as proto-BD candidates (IRAM 04191 and L673-7, Table~\ref{tab:protoBDs}). The authors find that in 3 out of the 5 sources, the outflows tend to align with the envelope magnetic field, and if the inner magnetic field is taken into account (IRAM 04191) the alignment improves. 

\citet{Soam2015_Polarimetry-L328} focus on L328-IRS, known to be driving a sub-parsec-scale molecular outflow \citep{Lee2013_ProtoBD_L328-IRS, Lee2018_L328_ProtoBD_ALMA}. The magnetic field, with a strength of $\sim$20 $\mu$G, is found to be ordered from 0.02 to 0.2~pc, and parallel to the outflow axis.

In a subsequent work, \citet{Soam2019_Magnetic_VeLLO_L1521F-IRS} undergo the first sub-parsec-scale mapping of magnetic fields in the vicinity of the VeLLO L1521F-IRS, using sub-millimetre polarization measurements at 850~\mum\ with the JCMT. The magnetic fields are found to be ordered on multiple scales. At core scales, the magnetic field strength is 330$\pm$100 $\mu$G, an order of magnitude larger than at clump/envelope scales. This strength implies that the magnetic energies are larger than non-thermal kinetic energies, and corresponds to a mass-to-magnetic flux ratio of 2.3$\pm$0.7, indicative of the core being magnetically supercritical. 

All these results suggest that, even though the magnetic field does not seem to be strong enough to stop the collapse, it must play an important role in launching and collimating the molecular outflows driven by VeLLOs and proto-BDs. Further studies in other proto-BD candidates and VeLLOs need to be carried out to establish this on a more robust base.

Another technique used in the literature to measure the magnetic field strengths in BDs is based on the detection of electron cyclotron emission at radio wavelengths. The first detection of radio emission in a confirmed BD was done by \citet{Berger2001_BDs-cm}. Initially, it was interpreted as due to gyrosynchrotron emission from which the authors inferred a very weak $\sim5$~G magnetic field. Subsequent works showed that some BDs and their higher-mass ultracool dwarf stellar cousins can emit radio emission from the electron cyclotron maser instability \citep{Hallinan2007, Kao2016_BDs-aurora}, a mechanism also responsible for radio aurorae produced by Solar System planets \citep{Zarka2007}. 
Unlike gyrosynchrotron emission, electron cyclotron maser emission occurs at very near the fundamental cyclotron frequency \citep{Treumann2006}, providing a means to directly measure rather than infer magnetic field strengths.  At GHz frequencies, such emission indicates kilogauss magnetic fields
\citep[e.g.,][]{Kao2016_BDs-aurora, Kao2018_BDs-cyclotron-Bfield}, probably related to the electron cyclotron maser instability \citep[e.g.,][]{Pineda2017_BDs-multiwavelength-auroras, Kao2019_BDs-aurora}.  
This technique is very promising, especially for next-generation radio interferometers such as the ngVLA or the SKA (Sec.~\ref{sec:lessons}), which will enable detections of statistically significant samples of BDs associated with electron cyclotron emission and robustly constrain their magnetic field. However, at present, all the objects for which electron cyclotron emission has been detected have ages $>100$~Myr \citep[e.g.,][]{Berger2001_BDs-cm, Kao2016_BDs-aurora, Kao2018_BDs-cyclotron-Bfield, Kao2019_BDs-aurora, Kao2023_BD-radiationbelt}, older than the deeply embedded phases studied in this work ($\lesssim1$~Myr), where the radio emission is supposed to be dominated by thermal radiojets \citep{Forbrich2015_IC348SMM2E, Morata2015_Jets_ProtoBD, Rodriguez2017_BDs-VLA}. It remains an open question whether electron cyclotron emission could be detected at earlier stages.

\subsubsection{Multiplicity} 

It is important to note that multiplicity could be crucial to determine the final mass of an object and therefore its final (sub)stellar nature. Multiplicity could also be of pivotal importance as a tool to distinguish between formation scenarios. For example, if ejections are frequent and constitute an important formation mechanism of isolated BDs, the multiplicity properties of proto-BDs should differ from those of protostars. In particular, in the ejection scenario it would be expected that a smaller number of binary systems survive the ejection, being close (a few au) BD-BD binaries the most successful outcomes \citep{Stamatellos2009, Basu2012_Formation_ProtoBD}. 
On the other hand, in the star-like scenario BD-BD binaries with mass ratios close to unity are expected, with average separations of several tenths of au, and multiplicity fractions of 10--20\% increasing with primary mass \citep{Bate2012, Bate2019}. The star-like scenario is also required to explain the anticorrelation between close binary fraction and metallicity seen in simulations \citep{Bate2019}. What is more, in the star-like scenario a trend similar to what is seen in protostars would be expected for BDs as well, mainly that multiplicity is more frequent for more embedded protostars \citep[e.g.,][]{ChenArce2013_multiplesyst, Tobin2016_VANDAM}. All these features constitute excellent tests to elucidate among the different BD formation scenarios.

Furthermore, crowded star-forming regions could favor dynamical interactions and influence the formation of binary systems in very low-mass objects. A recent study carried out in the Orion Nebula Cluster in the substellar regime reveals an excess of very low mass binaries in this region compared to the Galactic field \citep{DeFurio2022}, consistent with \cite{Pearson2023}, and revealing that the primordial multiplicity properties might be different from those of the field, and that subsequent dynamical interactions are required to obtain the multiplicity properties of the observed field BDs. 


Regarding the most embedded stages, there are a few works that focused on multiplicity including VeLLOs and/or proto-BD candidates. \cite{ChenArce2013_multiplesyst} present SMA observations at 0.85--1.3~mm for a sample of 33 Class 0 protostars in nearby clouds including 1 VeLLO (L1521F) and 2 proto-BD candidates (IC348-MMS and IRAM04191, Table~\ref{tab:protoBDs}). In this work,  64\% of the sample have signatures of multiplicity (multiplicity fraction), with separations in the range 50--5000~au, and circumstellar mass ratios below 0.5. One of the most relevant results of this study is that the multiplicity fraction is two times larger in Class 0 sources than in Class I sources, suggesting that multiplicity properties evolve along with protostars. In the \citet{ChenArce2013_multiplesyst} work, one of the two proto-BD candidates was found to split up into two sources.
In a more recent work, \cite{Maury2019_CALYPSO} undergo a multiplicity study with the PdBI for a sample of 16 Class 0 protostars, including one VeLLO (L1521F) and one proto-BD candidate (IRAM04191, Table~\ref{tab:protoBDs}). This study explores multiplicity at scales similar to the work of \cite{ChenArce2013_multiplesyst}, 100--5000 au, and reports multiplicity fractions $\lesssim57$\%, consistent with \citet{ChenArce2013_multiplesyst}.
%
Important results regarding multiplicity have been reported also from the large-project VLA/ALMA Nascent disc and Multiplicity survey (VANDAM). \cite{Tobin2016_VANDAM, Tobin2018_VANDAM} study multiplicity of all known protostars in Perseus (94 objects in total) using the Very Large Array (VLA) at 0.8, 1, 4 and 6.6 cm and ALMA, including several VeLLOs and FHCs, and find a multiplicity fraction of $(57\pm9)$\% for Class 0 protostars and  $(23\pm8)$\% for Class I, again consistent with previous works. A similar work in 328 protostars of the Orion cloud is presented in \cite{Tobin2022_VANDAM-Orion}, measuring a multiplicity fraction of 30\% for separations of 20--10\,000 au.

Overall, specific studies focusing on the particular multiplicity properties of proto-BD candidates are still lacking, and the aforementioned studies will be an excellent base to compare multiplicity between proto-BDs and higher mass protostars.

\subsection{Comparison of proto-BD candidate properties with the well-known correlations of protostars}\label{sec:mainprops-comparison2protostellar}

A number of works have explored the behavior of the known protostellar correlations in the low-mass end, mainly using single-dish telescopes. We present here four correlations well established for protostellar sources, and overplot the most recent measurements of the corresponding properties for VeLLOs/proto-BDs, in order to explore if they follow the same behavior as their higher mass counterparts.

\subsubsection{The $\Minfrate$ vs $\Lbol$ correlation}

Regarding the infall and accretion process, \cite{Kim2021_infallVeLLO} measured the mass infall rate $\Minfrate$ in the VeLLOs sample of \cite{Kim2016_VeLLOs} and find a relation of $\Minfrate$ with $\Lint$. In order to explore if the relation of \cite{Kim2021_infallVeLLO} is an extension of a possible relation for higher mass protostars, in Fig.~\ref{fig:correls}a we include the \cite{Kim2021_infallVeLLO} measurements along with other measurements reported in the literature \citep[][the entire compilation is listed in Table~\ref{tab:correlMinfrateLbol}]{Belloche2002_IRAM04191, Furuya2009_GF9-2, Wyrowski2016_infall, Keown2016ApJ_L1521F-infall, Yue2021_infall, Pillai2023_infall, Xu2023_infall}, allowing us to explore a range in $\Lbol$ from 0.01 to 10$^6$~\lo\footnote{In panels `a', `b', and `c' of Fig.~\ref{fig:correls}, $\Lbol$ for the VeLLOs and proto-BD candidates is taken as $\Lint$ because the least luminous objects are most affected by the ISRF (as explained in Sec.~\ref{sec:defs}) and the accretion/ejection processes are expected to be driven by the mass of the central object, more closely related to $\Lint$ than to $\Lbol$ \citep[see also][]{AMIConsortium2011_AMIobs-c2d-smallcores}.}.
It is important to stress that Fig.~\ref{fig:correls}a includes only measurements with single-dish telescopes. In the figure, the objects of our compilation that were classified in Sec.~\ref{sec:mainprops-sample} as proto-BD candidates (Table~\ref{tab:protoBDs}) are indicated as red squares. 
Although there is a significant dispersion in the intermediate/high-mass regime, the figure shows a clear trend of smaller values of $\Minfrate$ for smaller $\Lbol$. 
A linear fit to these data was performed, excluding the proto-BD candidates and using the ODR package (weighting by errors in both axes):

\begin{equation}
\mathrm{log}(\Minfrate) = (-4.72\pm0.17) + (0.63\pm0.04)\mathrm{log}(\Lbol),
\end{equation}

\noindent
where $\Minfrate$ is given in \mo\,yr$^{-1}$ and $\Lbol$ is given in \lo.
In all panels of Fig.~\ref{fig:correls}, shaded areas correspond to two times (dark) and five times (light) the uncertainty of the fit.
As can be seen from Fig.~\ref{fig:correls}a, the values measured for the proto-BD candidates (red squares) lie within the shaded area, indicating that the least massive objects of the sample, reaching 0.004~\lo, might share a common process of accretion with more massive protostars. 

Assuming that $\Minfrate$ is related to the mass accretion rate, these results are consistent with previous works reporting
evidence that the relation of mass accretion rate vs stellar
mass extends down to $\sim0.015$~\mo\ \citep[][]{Muzerolle2005_acc-substellar, Alcala2014_Maccrate-Mstar-substellar, AlmendrosAbad2024_Maccrate-Mstar-substellar}, with these works being focused on Class II objects. Using a large compilation of substellar accretion diagnostics, \cite{Betti2023_Maccrate-Mstar} has confirmed these results
but have also suggested that the BD and stellar populations are better described separately, taking into account both mass and evolutionary effects.
This could partially explain the results found by 
\cite{Riaz2021_AccretionOutflow_ProtoBD} in a sample of 6 proto-BD candidates, where accretion luminosity and accretion rates 
are within the range measured for low-mass protostars, and higher than Class II brown dwarfs.
It remains to be explored whether this excess is due to the different methodologies used in each work or it is related to the evolutionary stage.

\subsubsection{The $\Lcm$ vs $\Lbol$ correlation}

Regarding the outflow phenomena, the centimeter luminosity, $\Lcm$, tracing thermal radiojets, is known to be well correlated with $\Lbol$. This was recently reviewed by \cite{Anglada2018} and, in Fig.~\ref{fig:correls}b, the relation of $\Lcm$ vs $\Lbol$ compiled in \cite{Anglada2018} is presented, indicating in red squares the objects that are classified in Sec.~\ref{sec:mainprops-sample} (Table~\ref{tab:protoBDs}) as proto-BD candidates (the entire compilation is listed in Table~\ref{tab:correlLcmLbol}). A linear fit was performed to the data of Fig.~\ref{fig:correls}b, excluding the proto-BD candidates and following the same method used for Fig.~\ref{fig:correls}a: 

\begin{equation}
\mathrm{log}(\Lcm) = (-2.05\pm0.08) + (0.66\pm0.03)\mathrm{log}(\Lbol),
\end{equation}

\noindent
where $\Lcm$ is given in mJy\,kpc$^2$ and $\Lbol$ is given in \lo.
In this case, the least massive proto-BD candidates, of 0.002--0.006~\lo, lie slightly above the shaded area. This could be related to the finding of \cite{Ellerbroek2013_outflows}, who report an excess of mass outflow rate vs  mass accretion rate for the lowest-mass objects of the sample. 
In line with the findings of \citet[][see previous paragraph]{Riaz2021_AccretionOutflow_ProtoBD} and \citet{Betti2023_Maccrate-Mstar}, the deviation from the protostellar correlation could be related with an evolutionary effect and/or variability.
Further studies improving the sampling in the lower end of these diagrams need to be carried out to better establish whether these excesses in the BD regime are significant or not.

\subsubsection{The $\Fout$ vs $\Lbol$ correlation}

Another well-known protostellar correlation concerns the momentum rate or outflow force of the outflowing gas, $\Fout$ and $\Lbol$ and $\Menv$ for Class 0 and I protostars \citep[e.g.,][]{Bontemps1996}. \cite{Schwarz2012_Outflow_VeLLO} study a sample of 39 low-luminosity objects using a single-dish telescope with a beam of $30''$, and obtained a surprisingly low detection rate, suggesting that beam dilution could be important for the low-luminosity objects.
In a more recent study, \cite{Kim2019_CO_Outflow_VeLLO} explore the CO emission in the VeLLOs sample of \cite{Kim2016_VeLLOs} using also single-dish telescopes, and compare the outflow force vs $\Lbol$ and $\Menv$ as in \cite{Bontemps1996}. They find that most of the objects follow the trend reported for protostellar objects (observed also with single-dish), down to $\sim0.05$~\lo, but a number of VeLLOs present much lower outflow forces than expected for this relation. These cases could be related to a group of particularly compact outflows, whose CO emission might have been significantly diluted within the single-dish beam. \cite{vanderMarel2013_Oph-outflows} also performed a similar study in Ophiuchus using the JCMT and again  find a few objects with a deficit of outflow force, more clearly seen in the relation of \cite{Towner2024_ALMAIMF-outflows}. Instead, when outflow emission from VeLLOs is studied with interferometers, their $\Fout$ seems to better follow the trend of protostars, as shown by, e.g.,  \cite{Takahashi2012_OMC3-MMS6-outflows} and \cite{Takahashi2013_Outflow_VeLLO_L1521F-IRS}. However, these works compare the results from interferometric observations for VeLLOs with the results from single-dish observations for higher-mass protostars.

\begin{figure*}
\centering
\begin{tabular}[!h]{c}    
\epsfig{file=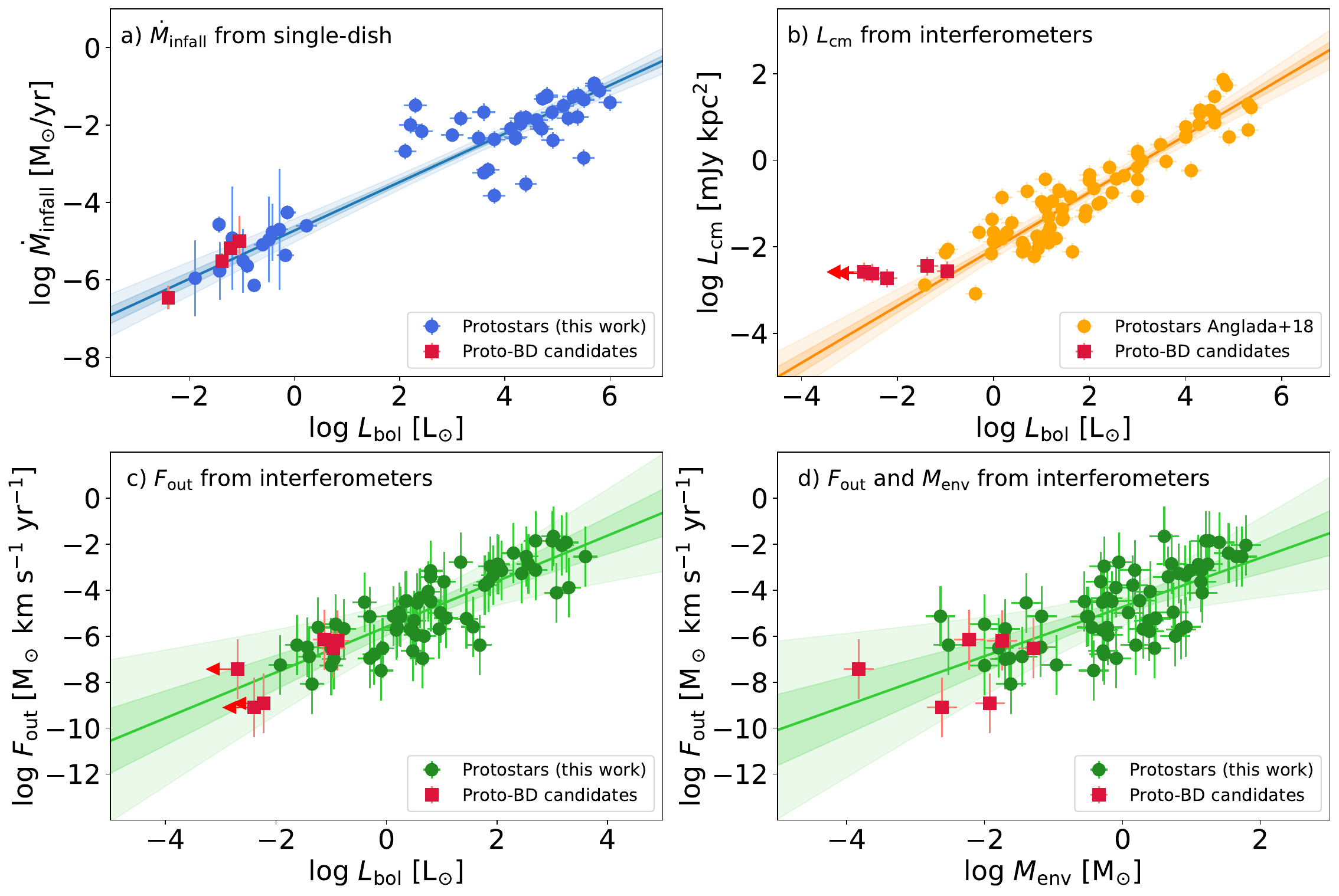, width=17cm,angle=0}\\
\end{tabular}
\caption{
Well-known relations for protostars with the proto-BD candidates of Table~\ref{tab:protoBDs} marked as red squares (also indicated in the last rows of Tables~\ref{tab:correlMinfrateLbol}, \ref{tab:correlLcmLbol} and \ref{tab:correlFout}). The fits have been performed excluding the proto-BD candidates and using the ODR package of Python weighting by both x and y errors.
Shaded areas correspond to two times (dark) and 5 times (light) the uncertainty of the fit.
%
a) Mass infall rate $\Minfrate$ vs $\Lbol$ compiled in this work (Sec.~\ref{sec:mainprops-comparison2protostellar}). All the measurements of $\Minfrate$ have been obtained using single-dish telescopes and are listed in Table~\ref{tab:correlMinfrateLbol}.
The linear fit is: log($\Minfrate$) = ($-4.72\pm0.17$) + ($0.63\pm0.04$)log($\Lbol$). 
b) Centimeter luminosity $\Lcm$ vs $\Lbol$ from \cite{Anglada2018}. All the measurements of $\Lcm$ have been obtained using interferometers and are listed in Table~\ref{tab:correlLcmLbol}.
The linear fit is: log($\Lcm$) = ($-2.05\pm0.08$) + ($0.66\pm0.03$)log($\Lbol$). 
c) Outflow force $\Fout$ vs $\Lbol$ compiled in this work (Sec.~\ref{sec:mainprops-comparison2protostellar}). All the measurements of $\Fout$ have been obtained using interferometers to avoid beam dilution problems for the most compact outflows, and are listed in Table~\ref{tab:correlFout}.
The linear fit is: log($\Fout$) = ($-5.59\pm0.14$) + ($0.99\pm0.08$)log($\Lbol$).
d) Outflow force $\Fout$ vs $\Menv$ compiled in this work (Sec.~\ref{sec:mainprops-comparison2protostellar}). All the measurements of $\Fout$ and $\Menv$ have been obtained using interferometers as in the previous panel, and are listed in Table~\ref{tab:correlFout}.
The linear fit is: log($\Fout$) = ($-4.73\pm0.15$) + ($1.07\pm0.14$)log($\Lbol$).
}
\label{fig:correls}
\end{figure*}

In order to perform a more uniform comparison between VeLLOs and protostars and to avoid beam dilution problems, \citet{Palau2014} compiled a sample of protostellar objects whose outflow parameters were studied using only interferometers, reaching $\Lbol\sim5000$~\lo. In Fig.~\ref{fig:correls}c, the relation of $\Fout$ vs $\Lbol$ of \cite{Palau2014} is updated with subsequent works of the literature \citep{Tobin2016_outflows, Lee2018_L328_ProtoBD_ALMA, Busch2020_ChaMMS1-outflows, Furuya2019_GF9-2, Podio2021_Class0, Vazzano2021_outflows, Dutta2024_EpisodicAccretion}, all based on interferometric observations, spanning a range of $\Lbol$ from 0.002 to $10^4$~\lo. This compilation is provided in Table~\ref{tab:correlFout}. As in Figs.~\ref{fig:correls}a-b, we have marked with red squares the proto-BD candidates of Sec.~\ref{sec:mainprops-sample} (Table~\ref{tab:protoBDs}), and have included two additional, tentative, proto-BD candidates not reported in SUCANES (i.e., in Table~\ref{tab:protoBDs}) that have $\Fout$ measured with interferometers (e.g., \citealt{PhanBao2022_protoBD}, \citealt{Palau2022_J041757}). Following the same method as in previous panels, a linear fit was performed to the data excluding the proto-BD candidates, 
and a shaded area corresponding to two and five times the uncertainty of the fit is indicated in the figure. The linear fit is:

\begin{equation}
\mathrm{log}(\Fout) = (-5.59\pm0.14) + (0.99\pm0.08)\mathrm{log}(\Lbol), 
\end{equation}

\noindent
where $\Fout$ is given in \mo\,\kms\,yr$^{-1}$ and $\Lbol$ is given in \lo.
This panel shows that $\Fout$ of the least massive proto-BD candidates, of $\lesssim0.006$~\lo, falls within the shaded area of the fit uncertainty in the protostellar relation.

\subsubsection{The $\Fout$ vs $\Menv$ correlation}

Finally, Fig.~\ref{fig:correls}d presents the $\Fout$ vs $\Menv$ relation for the same sample of the previous panel. In this case, $\Menv$ has been taken from the same interferometric studies used to estimate $\Fout$, thus the filtering of the interferometer affects equally both parameters. The linear fit and shaded areas are performed as in the previous panel, excluding the proto-BD candidates:

\begin{equation}
\mathrm{log}(\Fout) = (-4.73\pm0.15) + (1.07\pm0.14)\mathrm{log}(\Menv), 
\end{equation}

\noindent
where $\Fout$ is given in \mo\,\kms\,yr$^{-1}$ and $\Menv$ is given in \mo.
As can be seen in Fig.~\ref{fig:correls}d, $\Fout$ of the least massive proto-BD candidates also lies within the shaded area of the $\Fout$ vs $\Menv$ protostellar relation. This is consistent with the pioneering work of \cite{Greaves2003_isolatedplanet}, and provides a larger statistical sample.\\

To sum up, although more proto-BD candidates should be added in the future to properly populate the low-luminosity end of each panel of Fig.~\ref{fig:correls}, the four protostellar relations preliminary studied here seem to extend down to $\Lint\sim0.006$~\lo, corresponding to masses of $\Macc\sim0.024\pm0.007$~\mo\ according to the fit of Fig.~\ref{fig:LintMdyn}.


\section{Main properties of pre-BDs}\label{sec:mainpropspreBD}

The study and the census of pre-BDs could also potentially shed critical light on the formation mechanism of BDs, because if the planet-like scenario had a dominant role, the number of pre-BDs should be very small. However, given the difficulties to identify true pre-BDs, reliable statistics for these kind of objects are still lacking.

As shown in Sec.~\ref{sec:searches-system-submm}, most of the detections of pre-BD candidates are based on millimetre/sub-millimetre observations of objects that do not show counterparts at any other wavelength (e.g. \citealt[]{Andre2012_PreBD, Palau2012_PreBD_Cores_Taurus,Huelamo2017_B30,SantamariaMiranda2020_ParLup34}). In general, they are identified as dusty cores with masses (dust+gas) around the substellar limit ($\lesssim$0.1~\mo) in single-dish data. Then, continuum interferometric observations can help confirm the presence of a compact source inside the core,
while molecular gas observations allow to assess the stability of the core, by comparing $\Menv$ with $\Mvir$ or $\Mgrav$ (Sec.~\ref{sec:defs}). One of the first pre-BD candidates reported in the literature is Oph\,B-11 \citep[]{Andre2012_PreBD}, in which interferometric dust and gas observations reveal a 3.2 mm continuum source with substellar mass. \nth\ was found associated with the compact millimetre source, and a virial analysis indicates that the object is gravitationally unstable \citep{Andre2012_PreBD}. 
From these pioneering works, the main requirements for an object to be a promising pre-BD candidate can be identified as:

\begin{description}

\item[No evidence of (sub)stellar activity:] Similar to the case of pre-stellar cores, which are cores on the verge to collapse but with no hydrostatic core yet \citep[e.g.,][]{Caselli2011_prestellar}, pre-BDs do not have a hydrostatic core either, implying that should not have any infrared/optical counterpart, or signs of ejection activity such as a molecular outflows or jets. 

\item[Gravitationally bound core, $\Menv>\Mgrav$:] In order to assure that the object will form a compact object in the future, the object should be gravitationally bound and preferentially gravitationally unstable. This can be assessed by measuring different quantities. First, the Jeans mass could be estimated and compared to the observed mass of the core, which should be larger than the Jeans mass if gravitationally unstable. However, the Jeans criterion assumes linear density perturbations, while this is not expected if the cores are assembled by turbulent flows, which naturally yield non-linear density perturbations bounded by the shock ram pressure \citep[][]{Padoan2004}. Thus, an estimate more consistent with the \cite{Padoan2004} scenario is to compare the observed mass with the Bonnor-Ebert mass, $\MBE$ (see eq.~\ref{eq:MBE-p0}). If the observed mass is larger than $\MBE$, then the object is gravitationally unstable. The stability study can also be performed with the virial analysis. In this case, the observed mass should be larger than $\Mvir$ as given in eq.~\ref{eq:Mvir} \citep[e.g.,][]{Andre2012_PreBD, Tokuda2019_preBD}. Alternatively, $\Mgrav$ could be used instead of $\Mvir$, following eq.~\ref{eq:Mgrav}, which is a less strict constraint and was used by the first pioneering works and subsequent works \citep[e.g.,][]{Pound1993_ProtoBD, Pound1995_protoBD, Palau2012_PreBD_Cores_Taurus}. It is important to note that the assessment of gravitational instability based on the aforementioned simplified criteria (comparing $\Menv$ vs a critical mass) might not be fully accurate, as it might depend on the scale over which the measurements are performed \citep[e.g.,][]{Gomez2021_gravcollapse}. Thus, the association of the single-dish core with a compact source detected with interferometers can be considered an evidence of on-going gravitational contraction as well \citep{Huelamo2017_B30, SantamariaMiranda2021_ALMA_Lupus_ProtoBD}, as explained below.

\item[Envelope mass low enough, $\Menv<0.15$~\mo:] In principle, if we assumed a 10\% of SFE as in Sec.~\ref{sec:mainprops-requirements}, $\Menv$ for pre-BDs could be as high as 0.7~\mo, given that the hydrostatic core is not formed yet or it still has a negligible mass. However, this criterion has been used in the literature much more strictly to assure that the object will remain substellar, and is typically adopted around 0.1~\mo\ \citep[e.g.,][]{Pound1993_ProtoBD, Pound1995_protoBD, Andre2012_PreBD, Palau2012_PreBD_Cores_Taurus, deGregorio2016_PreProtoBD_ChaII, Barrado2018_B30, Huelamo2017_B30, Tokuda2019_preBD, SantamariaMiranda2021_ALMA_Lupus_ProtoBD}. We propose here the conservative case that $\Menv\leq0.15$~\mo, to consider a CFE of 50\%. The mass limit of $\Menv\leq0.15$~\mo\ is also of the order of the mass of the core that \cite{Machida2009_BDformation} used as initial condition for their similuation of a BD in a star-like scenario, of $\sim0.2$~\mo. By adopting this criterion, and assuming that the object is gravitationally unstable, the probability of ending with a substellar object is very high.

\end{description}

These main properties of pre-BDs relative to other stages of the star/BD formation process are summarized in Fig.~\ref{fig:sketch2}.
In Table~\ref{tab:preBDs} we present all the pre-BD candidates compiled in SUCANES, which amount to 26 candidates. We note that SUCANES just lists the pre-BD candidates reported in the literature, and that in many cases the criteria outlined above could not be applied because of molecular gas detection is missing. Thus, further more dedicated and sensitive interferometric studies need to be carried out to confirm the nature of the objects in Table~\ref{tab:preBDs} as pre-BDs. In particular, the second criteria is one of the most difficult to assess. According to SUCANES, there are $\sim$28 pre-BD candidates mainly identified in surveys in Lupus I, III clouds \citep{SantamariaMiranda2021_ALMA_Lupus_ProtoBD}, Chamaeleon II \citep{deGregorio2016_PreProtoBD_ChaII}, and Barnard~30 region \citep{Huelamo2017_B30, Barrado2018_B30}, Ophiuchus \citep{Pound1995_protoBD}, and Taurus \citep{Palau2012_PreBD_Cores_Taurus, Tokuda2019_preBD}. All these objects have been detected in single-dish observations, and part of them have been further confirmed with interferometers.


In the cases in which both single-dish and interferometric observations are available, this comparison has resulted in `contradictory' results, since many of the cores with compact sources at their centers, indicative of on-going collapse, were classified as gravitationally stable.
\cite{Gomez2021_gravcollapse} propose an alternative scenario to explain this contradiction, and consists on assuming that the cores with compact sources at their centers are already the result of gravitational contraction, constituting the `tip of the iceberg' of a larger-scale collapse. As explained in \cite{Huelamo2017_B30}, in this scenario, the collapse is expected to occur from the outside-in and it  develops self-consistently a near $r^{-2}$ density profile. Hence, a comparison between the dust masses at different core radii can help test the validity of this prediction, along with the search of infall signatures in molecular tracers at large scales.

Overall, since the identifications of the first pre-BDs by several authors \citep[][]{Andre2012_PreBD, Nakamura2012_SubSTellarCondensations, Palau2012_PreBD_Cores_Taurus}, a number of candidates, although not confirmed, have been added to the list, making their search more difficult than previously thought. This suggests that their presence is either very difficult to identify or that they are rarer than expected for the star-like scenario. Further efforts, both from the numerical and observational side, need to be carried out to elucidate how common are pre-BDs in molecular clouds.

\section{Proto-BD candidates in different clouds of the Solar Neighborhood}\label{sec:NbdMCproperties}

A study of the spatial distribution or the number of proto-BD candidates in the nearby clouds could also give important hints on the formation mechanisms of BDs. In Figures~\ref{fig:spadistribOrionSerpens}--\ref{fig:spadistribisolated}, the candidates listed in Table~\ref{tab:protoBDs} are overplotted on the corresponding {\it Herschel} image at 250~\mum, and in the further subsections we analyse the potential relations between the number of proto-BD candidates and the cloud properties.

\begin{figure*}
\centering
\begin{tabular}[p]{c}    
\epsfig{file=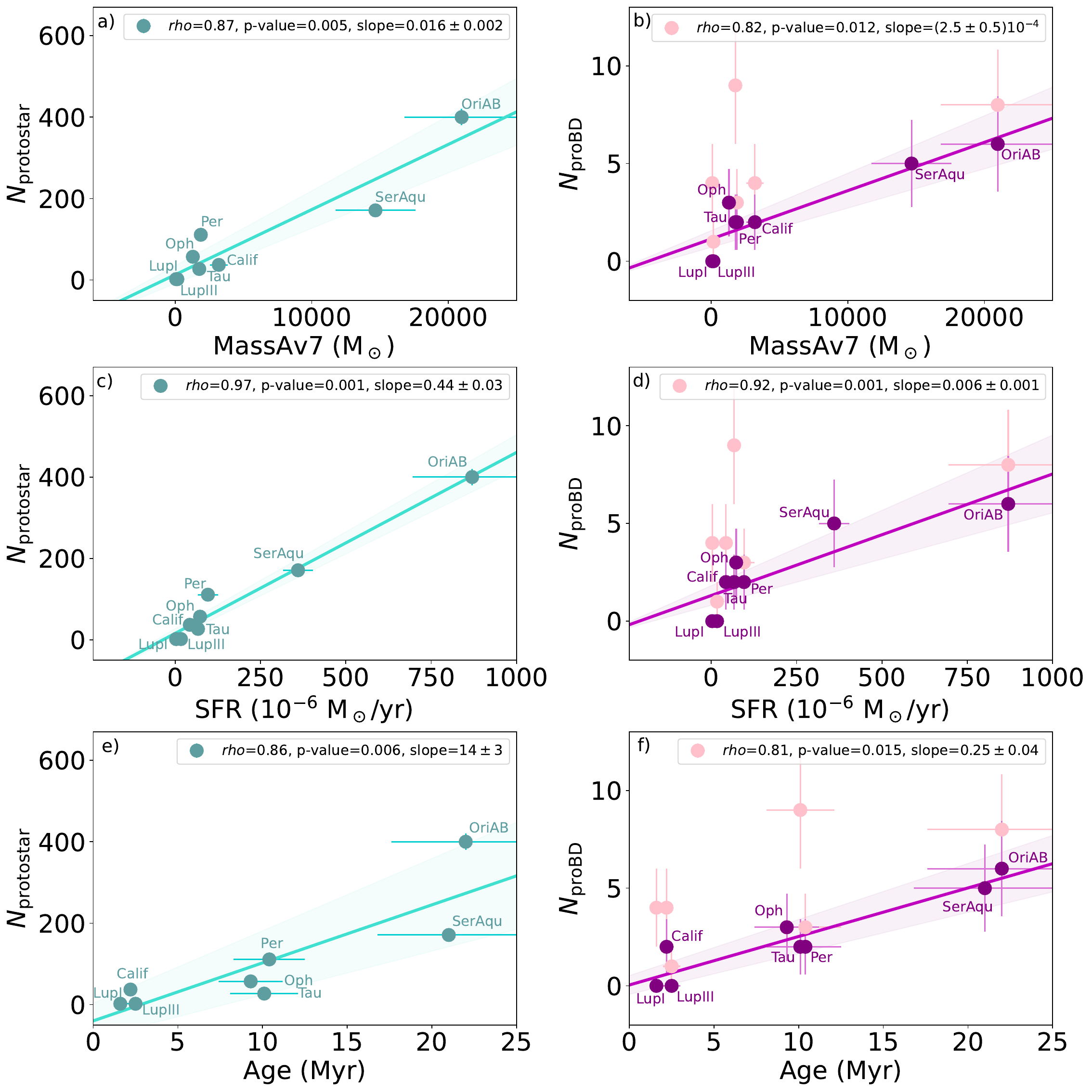, width=16cm,angle=0}\\
\end{tabular}
\caption{Relation between number of protostellar sources $\Nproto$ (left) or $\Nbd$ (right) vs several cloud properties.
In all panels on the right, the y-axis shows $\Nbd$ calculated in two ways. The pink dots correspond to all $\Nbd$ of Table~\ref{tab:protoBDs} for each region that belong to \cite{Kim2016_VeLLOs} catalog. The purple dots correspond to the same $\Nbd$ but computed down to the same completeness threshold for all regions, of $\Lint=0.04$~\lo\ ($N_\mathrm{proBD-c004}$).
(a) $\Nproto$ vs. dense gas mass.
(b) $\Nbd$ (pink) or $N_\mathrm{proBD-c004}$ (purple) vs. dens gas mass.
(c) $\Nproto$ vs. SFR.
(d) $\Nbd$ (pink) or $N_\mathrm{proBD-c004}$ (purple) vs. SFR.
(e) $\Nproto$ vs. cloud dynamical age.
(f) $\Nbd$ (pink) or $N_\mathrm{proBD-c004}$ (purple) vs. cloud dynamical age.
In all panels, the solid line corresponds to the linear fit (done with curvefit in python), and the shaded areas correspond to the 1$\sigma$ uncertainty of the fit.
The uncertainty in the mass, dynamical age, and SFR of the cloud are assumed to be of 20\%. The uncertainty in the number counts of protostars or proto-BD candidates are adopted as the square root of the corresponding number.
}
\label{fig:Nbd}
\end{figure*}

\begin{figure*}
\centering
\begin{tabular}[p]{c}    
\epsfig{file=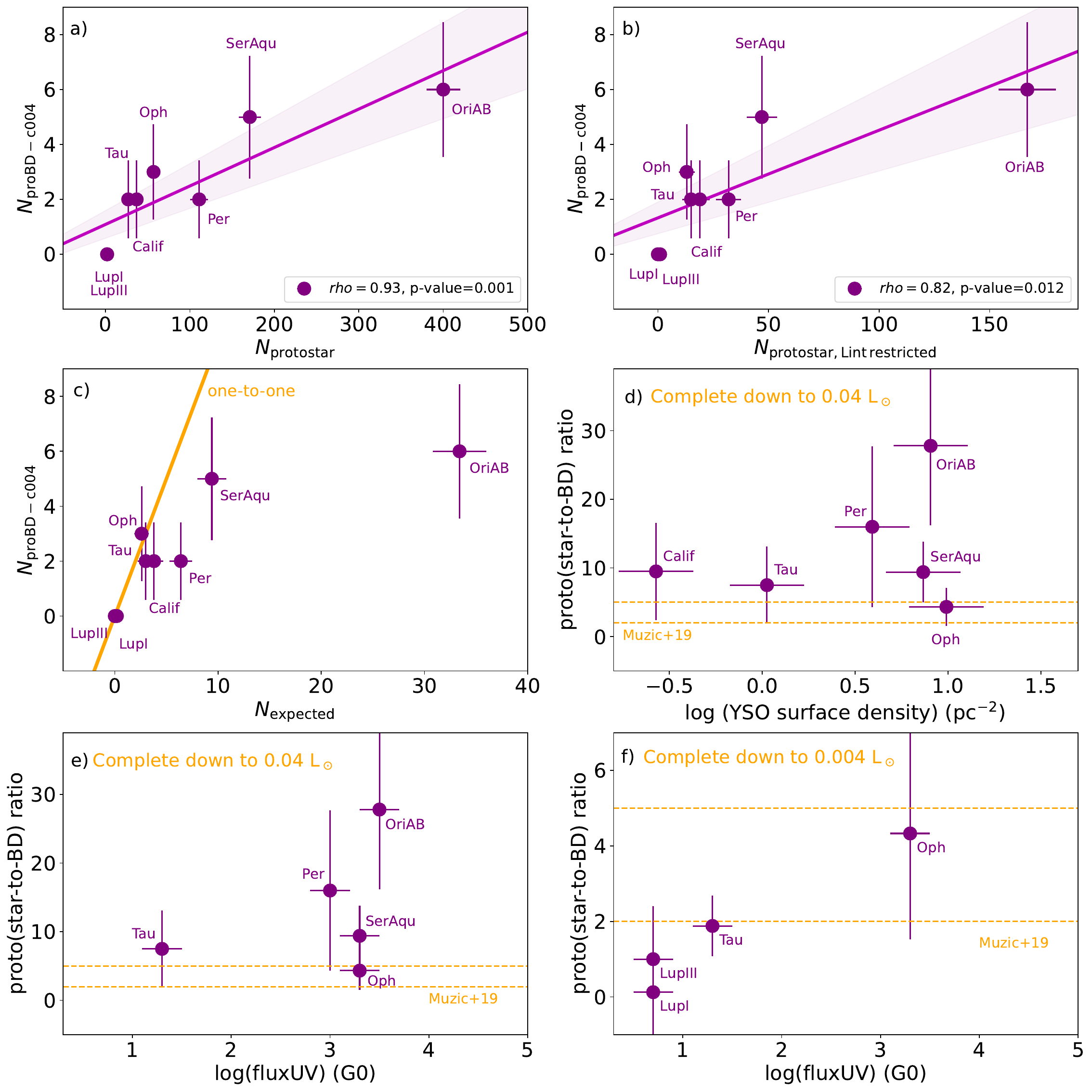, width=15cm,angle=0}\\
\end{tabular}
\caption{
(a) $N_\mathrm{proBD-c004}$ (complete down to 0.04~\lo) vs $\Nproto$.
(b) $N_\mathrm{proBD-c004}$ (complete down to 0.04~\lo) vs $N_\mathrm{proto-Lrestrict}$ ($\Nproto$ within the luminosity range 0.13--1~\lo).
(c) $N_\mathrm{proBD-c004}$ (complete down to 0.04~\lo) vs $\Nexp$ (where $\Nexp=N_\mathrm{proto-Lrestrict}/5$, see Sec.~\ref{sec:NbdMCproperties-star2BD004}). The orange solid line corresponds to the one-to-one relation. 
(d) Proto(star-to-BD) ratio vs log of protostar surface density. This plot is comparable to Fig.~15 of \cite{Muzic2019_NGC2244} or Fig.~11 of \cite{AlmendrosAbad2023_NGC2264}. 
(e) Proto(star-to-BD) ratio vs log of FUV flux.
(f) Proto(star-to-BD) ratio vs log of FUV flux, with the proto(star-to-BD) ratio estimated for the most nearby clouds and therefore down to 0.004~\lo, near the planetary boundary (see Sec.~\ref{sec:NbdMCproperties-star2BD0004}).
}
\label{fig:Nbdexp}
\end{figure*} 

\subsection{Are proto-BD candidates following the well-known relations of star formation in molecular clouds?}\label{sec:NbdMCproperties-Lada10relations}

A well-known property of molecular clouds is that there is a tight relation between the number of protostars and the dense gas mass (defined as the mass above $A_\mathrm{V}$=7~mag, equivalent to $A_\mathrm{K}$=0.8~mag), reported by \cite{Lada2010_SFRs}, and later confirmed by many further investigations \citep[e.g.,][]{Hacar2024_fibernetworks}, some of which suggest that the relation is a proof of molecular cloud collapse \citep[e.g.,][]{BallesterosParedes2024_SFlaws, ZamoraAviles2024_SFlaws}.

In order to study these kind of relations, in Table~\ref{tab:clouds1} we list, for a number of molecular clouds, their distance, the cloud area (taken from \citealt{Heiderman2010_Nyso} and \citealt{Carpenter2000}), the total number of young stellar objects or $\Nyso$ (taken from \citealt{Heiderman2010_Nyso} and \citealt{Kim2016_VeLLOs}, and typically including young stellar objects in all evolutionary stages, from Class 0 to Class III, e.g., \citealt{Rygl2013_Lupus}), the total mass of the cloud, the mass of dense gas, the Star Formation Rate (SFR, estimated using equation (1) of \citealt{Lada2010_SFRs}), the surface density of young stellar objects (calculated using $\Nyso$ and the area given in the table), the logarithm of the FUV flux (taken from \citealt{Gupta2024_IC1396} and \citealt{Xia2022_fluxuv}),
and the assigned evolutionary or dynamical age of each cloud according to the model of \citet{VazquezSemadeni2018_cloudages} (which essentially adopts the ratio of dense gas vs total mass as a proxy for evolutionary stage of the cloud, see details in Table~\ref{tab:clouds1}). 

Table~\ref{tab:clouds2} summarizes the number of protostars and proto-BD candidates for each cloud computed in a number of ways. The table first lists $\Nproto$ (corresponding to the number of young stellar objects for which $\Tbol\leq650$~K, thus including both Class 0 and Class I), and $N_\mathrm{proto-Lrestrict}$, which corresponds to the subset of $\Nproto$ whose $\Lint$ is in the range 0.13--1~\lo\ (this will allow to compare our data to \citealt{Muzic2019_NGC2244}). The table also lists $\Nbd$, the total number of proto-BD candidates listed in Table~\ref{tab:protoBDs} that belong to the catalog of \cite{Kim2016_VeLLOs}. This requirement is necessary because \cite{Kim2016_VeLLOs} searched for proto-BDs uniformly across all these clouds, making the comparison among clouds feasible and sound.

In order to take into account possible incompleteness effects, especially in the regions that are further away, we re-calculated $\Nbd$ including only the objects down to the same $\Lint$ threshold for all the clouds. The $\Lint$ threshold was adopted from Fig. 6 of \cite{Dunham2008_c2dVeLLOs} for which the average 3$\sigma$ sensitivity from the c2d {\it Spitzer} program can be estimated as $\Lint = 4\times10^{-3}\,(D/140\, \mathrm{pc})^2$~\lo. The $\Lint$ threshold in our case will be given by the region with largest distance (Serpens-Aquila at 436 pc), corresponding to a completeness limit of 0.04~\lo\ for $\Lint$. The number of proto-BD candidates in Table~\ref{tab:protoBDs} that belong to \cite{Kim2016_VeLLOs} and whose $\Lint\geq0.04$~\lo\ is designated as $N_\mathrm{proBD-c004}$.
According to the correlation between $\Lint$ and $\Mdyn$ of Fig.~\ref{fig:LintMdyn}, $\Lint=0.04$~\lo\ corresponds to $\Mdyn=77\pm20$~\mj. 
This is about a factor of 2 larger than the threshold used by \cite{Muzic2019_NGC2244} and \cite{AlmendrosAbad2023_NGC2264} to estimate the star-to-BD ratio in more advanced (Class II/III) evolutionary stages (of $\sim30$~\mj). When applying this completeness criterion, the most affected clouds are Taurus and Lupus I (and III), while, as expected, Serpens-Aquila and OrionA-B remain almost with the same number of objects. We will further discuss this in Secs.~\ref{sec:NbdMCproperties-star2BD0004} and \ref{sec:explanationsOphdeficit}.

In the left panels of Fig.~\ref{fig:Nbd}, the relation between $\Nproto$ vs dense gas mass, SFR and dynamical age for all the clouds of Table~\ref{tab:clouds1} are shown. For each plot, the Spearman’s rank correlation coefficient $\rho$ and the $p$ value, are indicated in the top-right corner. From the left panels of the figure, the well-known  correlations of $\Nproto$ with cloud properties are recovered, as expected \citep[e.g.,][]{Lada2010_SFRs, Hacar2024_fibernetworks}. Although the three relations are well-known and very strong, the $p$ values indicate that the strongest relation is $\Nproto$ vs SFR ($p$ value 0.001), followed by $\Nproto$ vs dense gas mass ($p$ value 0.005), and followed by $\Nproto$ vs cloud dynamical age ($p$ value 0.006).

The panels on the right of Fig.~\ref{fig:Nbd} present the behavior of $\Nbd$ vs the same cloud properties of the left panels. In the panels on the right, the pink dots correspond directly to $\Nbd$, the total number of proto-BD candidates reported in Table~\ref{tab:protoBDs} that belong to \cite{Kim2016_VeLLOs}. As can be seen from the figure, the pink dots do not present any correlation with cloud properties. However, once $\Nbd$ is corrected for completeness, i.e., when $N_\mathrm{proBD-c004}$ is plotted (purple dots), strong correlations are found of $N_\mathrm{proBD-c004}$ with cloud properties, as for the case of $\Nproto$, again being the relation of $N_\mathrm{proBD-c004}$ vs SFR the strongest ($p$ value 0.001), followed by the relation with dense gas mass ($p$ value 0.012), and followed by the relation with cloud dynamical age ($p$ value 0.015), being the same behaviour as protostars. Although the values of $N_\mathrm{proBD-c004}$ are very small and more dedicated studies should be carried out to enlarge the samples, this preliminary result suggests that the formation of objects down to $\Lint\sim0.04$~\lo\ seems to take place similarly to the star formation process.

\subsection{Proto(star-to-BD) ratio down to $\Lint\sim0.04$~\lo\ among the different nearby clouds}\label{sec:NbdMCproperties-star2BD004}

In this subsection we consider the possibility that the number of proto-BD candidates depends on the cloud environment. 
Variations in the luminosity function in the sub-stellar regime of different clouds were explored by several authors (Sec.~\ref{sec:intro}), but no clear conclusions can be inferred yet from these works.
To further explore this, Table~\ref{tab:clouds2} additionally provides $\Nexp$, the number of expected proto-BDs, calculated as $\Nexp=N_\mathrm{proto-Lrestrict}/5$, where $N_\mathrm{proto-Lrestrict}$ is the number of protostars in each region with $\Lint$ between 0.13 and 1~\lo. In this expression, using $N_\mathrm{proto-Lrestrict}$ is required to make our analysis comparable to the analysis of \cite{Muzic2019_NGC2244} and \cite{AlmendrosAbad2023_NGC2264}, who calculate the star-to-BD ratio by restricting the stars to the range of 0.075--1~\mo. The factor of 5 applied is taken also from \cite{Muzic2019_NGC2244} and \cite{AlmendrosAbad2023_NGC2264}, and corresponds to the upper-end of the star-to-BD ranges found in these works. The table also provides the proto(star-to-BD) ratio for each cloud, which is the equivalent to the star-to-BD ratio (used in \citealt{Muzic2019_NGC2244, AlmendrosAbad2023_NGC2264}) but in the deeply embedded phase, and is estimated as $N_\mathrm{proto-Lrestrict}/N_\mathrm{proBD-c004}$. 

In Fig.~\ref{fig:Nbdexp}-a we present the relation between $N_\mathrm{proBD-c004}$ and $\Nproto$, which is found to be highly correlated ($\rho=0.93$, $p$ value 0.001). Such a strong correlation is the reason why in Fig.~\ref{fig:Nbd} (previous section) the behavior of $N_\mathrm{proBD-c004}$ with cloud properties is the same as the behavior of $\Nproto$. The fact that $N_\mathrm{proBD-c004}$ correlates so strongly with $\Nproto$ suggests that the formation of objects down to $\Lint\sim0.04$~\lo\ ($\Macc\sim77\pm20$~\mj) is intimately linked to the star formation process.

Fig.~\ref{fig:Nbdexp}-b presents the same relation but with $N_\mathrm{proto-Lrestrict}$ instead of $\Nproto$, which is slightly worse than in the previous case because of the restriction in luminosity.
Fig.~\ref{fig:Nbdexp}-c shows $N_\mathrm{proBD-c004}$ vs $\Nexp$, with the one-to-one relation indicated as an orange solid line. While there are a number of clouds that lie close to the one-to-one relation, there are also a few clouds, such as Perseus, Serpens-Aquila, and most extremely OrionA-B, for which $N_\mathrm{proBD-c004}\ll\Nexp$. 
This can be further seen in Figs.~\ref{fig:Nbdexp}-d-e, where the proto(star-to-BD) ratio is shown vs protostar surface density and the FUV flux. Fig.~\ref{fig:Nbdexp}-d is comparable to Fig.~15 of \cite{Muzic2019_NGC2244} or Fig.~11 of \cite{AlmendrosAbad2023_NGC2264}. In Figs.~\ref{fig:Nbdexp}-d-e, the range of 2--5, measured for the star-to-BD ratio in \cite{Muzic2019_NGC2244} and \cite{AlmendrosAbad2023_NGC2264}, is marked with dashed orange lines. While these two panels show no trend of the proto(star-to-BD) ratio with protostar surface density or FUV flux, they clearly show that in the embedded (Class 0/I) phase the ratio calculated here is generally much larger than in the more evolved phase studied by \cite{Muzic2019_NGC2244} and  \cite{AlmendrosAbad2023_NGC2264}. This is related to the fact that we are sensitive to more massive proto-BD candidates (mass $\gtrsim0.077\pm20$~\mo) than in the aforementioned works (mass $\gtrsim0.03$~\mo), and/or that we are still missing an important fraction of Class 0 proto-BDs due to the current identification strategies.

\subsection{Proto(star-to-BD) ratio down to $\Lint\sim0.004$~\lo}\label{sec:NbdMCproperties-star2BD0004}

As a last attempt to study a possible relation of the number of proto-BD candidates with their environment, we focused on the least massive objects of our compilation. These could only be detected in the most nearby clouds: Taurus, Lupus I, Lupus III and Ophiuchus, for which we are sensitive down to $\Lint=0.004$~\lo\ (following the same argument as in Sec.~\ref{sec:NbdMCproperties-Lada10relations}). Table~\ref{tab:clouds2} lists $N_\mathrm{proBD-c0004}$, the number of proto-BD candidates included in \cite{Kim2016_VeLLOs} which are complete down to $0.004$~\lo. Here a very different behavior for these clouds can be appreciated: while for Ophiuchus $N_\mathrm{proBD-c0004} = N_\mathrm{proBD-c004}$, meaning that all the proto-BD candidates in Ophiuchus have $\Lint\geq0.04$~\lo, for Taurus and Lupus I, $N_\mathrm{proBD-c0004} \gg N_\mathrm{proBD-c004}$, meaning that most of the proto-BD candidates in these two clouds have $\Lint<0.04$~\lo. Actually, the median $\Lint$ of Taurus and Lupus I is very low, of 0.0075 and 0.010~\lo, respectively. This is also shown in Fig.~\ref{fig:Nbdexp}-f, where the proto(star-to-BD) ratio down to 0.004~\lo\ is estimated as $N_\mathrm{proto-Lrestrict}/N_\mathrm{proBD-c0004}$. Again following the fit of Fig.~\ref{fig:LintMdyn}, $\Lint=0.004$~\lo\ corresponds to $\Macc=18\pm5$~\mj. This corresponds to low-mass BDs and is close to the planetary boundary of $\sim10$~\mj.

Consequently, Fig.~\ref{fig:Nbdexp}-f indicates that there is a deficit of low-mass proto-BD candidates in Ophiuchus compared to Taurus and Lupus (or conversely, an excess in Taurus and Lupus). A deficit of BDs in Ophiuchus was also suggested by \cite{Rieke1989_rhoOph}, \cite{Erickson2011_rhoOph-IMF}, and measured by \cite{AlvesdeOliveira2012_rhoOph-IMF}. In particular, \cite{AlvesdeOliveira2012_rhoOph-IMF} report a BD-to-star ratio of $0.09^{+0.04}_{-0.02}$ for Ophiuchus and of 0.18 for Taurus, i.e., Ophiuchus seems to have a number of BDs relative to stars that is a factor of 2 lower than the same number for Taurus. Although \cite{AlvesdeOliveira2012_rhoOph-IMF} warn that this measurements should be regarded with caution due to poor accuracy of spectral classification of substellar objects, the fact that we find here a significant result showing the same trend suggests that the early findings by \cite{Rieke1989_rhoOph}, \cite{Erickson2011_rhoOph-IMF}, and \cite{AlvesdeOliveira2012_rhoOph-IMF} could be real.

\subsection{Possible explanations for the deficit of planetary-mass proto-BD candidates in Ophiuchus relative to Taurus and Lupus}\label{sec:explanationsOphdeficit}

There are several scenarios that could explain the deficit of proto-BD candidates in Ophiuchus relative to Taurus/Lupus. First, it is interesting to note that a very recent episode of star formation, consistent with the short cloud dynamical age estimated from the model of \cite{VazquezSemadeni2018_cloudages} (see Fig.~\ref{fig:Nbd}-f), has been proposed for Lupus I \citep{Rygl2013_Lupus, Benedettini2018_Lupus}, because it has a large number of prestellar objects with respect to more evolved objects, compared to other clouds. 
This would suggest that the two phenomena, a recent star formation episode and an excess of low-mass proto-BD candidates, could be related. If this was the case, the excess of proto-BD candidates in clouds with shortest dynamical ages would be indicative that the proto-BD phase is shorter than the protostellar phase, in line with the simulations of \cite{Bate2019}, where BDs accrete from the cloud for a period of time shorter than protostars do, making the typical `lifetime' of a proto-BD much shorter than the typical `lifetime' of a protostar \citep[e.g.,][]{Vorobyov2010_Lifetime_LowMass}. 
While this idea (proto-BD lifetimes shorter than protostars lifetimes) is worth exploring in future studies, this is not fully consistent with the fact that we also do see an excess of low-mass proto-BD candidates in Taurus, compared to Ophiuchus, and no recent star formation episode has been identified in Taurus (actually, Taurus and Ophiuchus share the same cloud dynamical age, Fig.\ref{fig:Nbd}-f). This might suggest that the important stellar feedback associated with Ophiuchus could be related to the deficit of low-mass proto-BD candidates in this cloud.

Therefore, a second scenario to explain the result of Fig.~\ref{fig:Nbdexp}-f is that the feedback from nearby OB stars shortens the proto-BD phase in Ophiuchus compared to this phase in Taurus and Lupus clouds. 
This would be consistent with a star-like scenario, followed by a very short timescale of the proto-BD phase due to the photo-erosion mechanism, that is expected to quickly deprive a contracting core from their associated envelope of gas and dust \citep[e.g.,][]{Whitworth2004_photoerosion, Whitworth2018}.
In this case, regions with strong stellar feedback should present high proto(star-to-BD) ratios (because of the short embedded phase, as Ophiuchus in Fig.~\ref{fig:Nbdexp}-f) but very low (evolved) star-to-BD ratios, since the lifetime of later stages would be longer and BD formation would be favored by photo-erosion.  
Actually, the recent findings of an excess of BDs (i.e., a very low star-to-BD ratio) in NGC\,2264, a region with a high number of OB stars, and for which BDs are preferentially distributed closer to the OB stars in the cluster than stars \citep[see Sec.~4.3 of][]{AlmendrosAbad2023_NGC2264}, support the photo-erosion scenario in this region. 
However, it is not likely that photo-erosion is at work in Ophiuchus. The photo-erosion mechanism has been found to be very inefficient, as it requires a massive core, of about $\sim75$~\mo\ to form one single BD of 75~\mj\ at a distance of 0.1 pc of the ionizing star \citep[following eq.~60 of][]{Whitworth2018}. In addition, the typical distance of BDs to OB stars in NGC\,2264 is only 0.3--0.5~pc, while the distance of the massive OB stars (earlier than B1) of the Sco OB2 association to the border of the Ophiuchus cloud facing them (comprising OphA, C, E and F) is more than 5 pc (the stars closer to Ophiuchus are all of B-type: $\sigma$-Sco at $\sim4$~pc is a B1, and both HD\,147889 at $\sim0.5$~pc, and $\rho$-Oph at $\sim2$~pc are B2). Thus, NGC\,2264 and Ophiuchus are very different cases and does not seem likely that photo-erosion could strongly affect the BD content in the Ophiuchus cloud.


Finally, a third possibility is that the OB stars in the surroundings of the Ophiuchus cloud efficiently heat the cloud, increasing the average temperature of the gas and dust of the cloud. 
Such an effect has already been reported in previous works \citep[e.g.,][]{SanchezMonge2013_propsdensecores, Rumble2021_radiativeheating, Kahle2024_Lagoon-stfeedback}.
%
If the Ophiuchus cloud was on average warmer than Taurus and Lupus, the Jeans mass in Ophiuchus would be larger, naturally suppressing the formation of the least massive objects, as proposed by \cite{Bate2009_radfeedback} and consistent with \cite{Bate2023_highredshift}.
In order to test this hypothesis, we performed preliminary estimates of the average dust temperatures in these clouds. Dust temperature images for Gould Belt clouds are provided by the {\it Herschel} Gould Belt survey key project \citep{Andre2010_CMF}. To estimate average dust temperatures, we selected regions in each cloud above a column density of $4\times10^{21}$~cm$^{-2}$, which is around the star formation threshold reported by \cite{Heiderman2010_Nyso} and \cite{Whitworth2016_SFthreshold}. For the Ophiuchus cloud, the average and standard deviation dust temperature measured in the map of \cite{Ladjelate2020_Herschel-Oph} above the aforementioned  threshold is $17\pm2$~K, while for Taurus this is $12.9\pm0.7$~K (using the dust temperature map of \citealt{Andre2010_CMF, Konyves2015_Cores_Aquila, Marsh2016_Herschel-Taurus}) and for Lupus I this is $13.6\pm0.7$~K (using the dust temperature map of \citealt{Rygl2013_Lupus, Benedettini2018_Lupus}). The fact that the Ophiuchus cloud seems to be slightly warmer than other clouds and heated by the nearby massive stars has also been recognized by other authors \citep[e.g.,][]{Friesen2009_Oph-NH3, Kirk2007_SCUBA, Rumble2021_radiativeheating}.
Assuming a density fluctuation of $10^9$~\cmt\ and a temperature of $\sim13$~K, the corresponding Jeans mass is of 9~\mj\ (following eq.~\ref{eq:Mjeans}). But if the temperature increases up to 19~K due to feedback, then the Jeans mass increases up to 17~\mj, suppressing the formation of low-mass BDs in the warmest regions, as suggested by other authors \citep[e.g.,][]{Rumble2021_radiativeheating}.
This is consistent with the result of \cite{Benedettini2018_Lupus}, who show that, in the Lupus (and Taurus) clouds, the star formation activity is associated with lower column densities compared to other clouds. This can be explained in terms of Jeans fragmentation: if a cloud is colder on average, then the minimum density required to form stars of a given mass is also smaller because the Jeans mass goes with the inverse square root of the density. Given that Jeans fragmentation is known to regulate the star formation process in the dense parts of molecular clouds \citep[e.g.,][]{Motte1998_rhoOph, Gutermuth2011_fragmentation, Palau2015, Palau2021, Kainulainen2016_fragmentation, Beuther2018_fragmentation, Saha2022_fragmentation, Morii2024_fragmentation}, the results outlined here are consistent with the same formation mechanism for BDs and stars, not only for high-mass BDs but also for the low-mass BDs near the planetary boundary, of $\sim10$~\mj. This possibility is consistent also with the recent findings of \citet{Panwar2024AJ_IMF-B59} in Berkeley\,59.

In addition, a particularly strong magnetic field permeating the Ophiuchus cloud could also contribute to increasing the critical mass required for an object to undergo collapse. Actually, \citet{Le2024_Oph-SOFIA} have recently reported that Ophiuchus seems to be a magnetically dominated cloud.
It would be important to further explore this possibility (low-mass BDs are suppressed in regions with higher dust temperatures and/or higher magnetic field strengths) in other clouds with different average dust temperatures or magnetic field strengths and to explore a potential relation between spatial distribution and these quantities in the cloud.

\section{Putting everything together: constraints on the BD formation theories from the current observations}\label{sec:constraints2theory}

In the paragraphs below, we list the main results presented in this work:

\begin{itemize}

    \item A correlation between $\Lint$ and $\Mdyn$ has been found, which is consistent with the \cite{DAntona1994_evolutionarytracks} and \cite{Vorobyov2017_EffectAccretion_BD} models at $5\times10^5$~yr, a typical lifetime for the Class 0/I phase. In the following, this relation has been used to estimate the accreted mass given $\Lint$.

    \item As a  preliminary result, the well-known correlations associated with infall and outflow phenomena in protostars extend reasonably well in general down to $\Lbol\sim0.006$~\lo, corresponding to low-mass BDs of $\Mdyn\sim0.024\pm0.007$~\mo\ (Sec.~\ref{sec:mainprops-comparison2protostellar}). When using single-dishes, a deficit of outflow force has been reported in some cases, probably related to beam dilution effects.

    \item The systematic searches performed so far to identify pre-BDs have provided candidates that need to be confirmed (Sec.~\ref{sec:mainpropspreBD}) in most of the cases. The scarcity of robust pre-BD candidates is probably due to the challenging nature of these observational projects, requiring very deep interferometric observations to obtain conclusive results.

    \item The number of proto-BD candidates with $\Lint\gtrsim0.04$~\lo\ present the same correlations with molecular cloud properties as the number of protostars (Sec.~\ref{sec:NbdMCproperties-Lada10relations}). This arises from the fact that these two numbers are highly correlated (Sec.~\ref{sec:NbdMCproperties-star2BD004}). Although larger samples of proto-BDs are mandatory to confirm it, this preliminary result suggests that the formation of low-mass objects down to 0.04~\lo\ (corresponding to very low-mass stars and massive BDs, $\Macc\sim0.077\pm0.020$~\mo) seems to take place in a star-like manner. Improving the sensitivity to detect less massive BDs would allow to be complete down to lower masses. However, this does not warrant that the correlation of $\Nproto$ vs $\Nbd$ would improve, because for low-mass BDs the environmental effects could become important, as shown in Sec.~\ref{sec:explanationsOphdeficit} and next paragraphs. This constitutes an excellent test for the future.

    \item The proto(star-to-BD ratio) computed for $\Lint\gtrsim0.04$~\lo\ in general yields values larger than the star-to-BD ratio measured in more evolved objects (Sec.~\ref{sec:NbdMCproperties-star2BD004}), due to the fact that we are sensitive only to the most massive proto-BD candidates.

    \item The proto(star-to-BD ratio) including low-mass BDs ($\Lint\gtrsim0.004$~\lo\ or $\Macc\sim0.018\pm0.005$~\mo), which could be determined only for Taurus, Lupus and Ophiuchus clouds, is significantly lower for Taurus and Lupus than for Ophiuchus. This suggests that there is a deficit of low-mass BDs in Ophiuchus (compared to Taurus and Lupus), which could be related to the measured higher average dust temperature in Ophiuchus, suppressing fragmentation near this mass range ($\sim0.018$~\mo) because of the higher Jeans mass. This indicates a possible star-like scenario down to the planetary boundary regime (Secs.~\ref{sec:NbdMCproperties-star2BD0004} and \ref{sec:explanationsOphdeficit}).
    
\end{itemize}

\begin{figure*}
\begin{center}
\begin{tabular}[b]{c}    
\epsfig{file=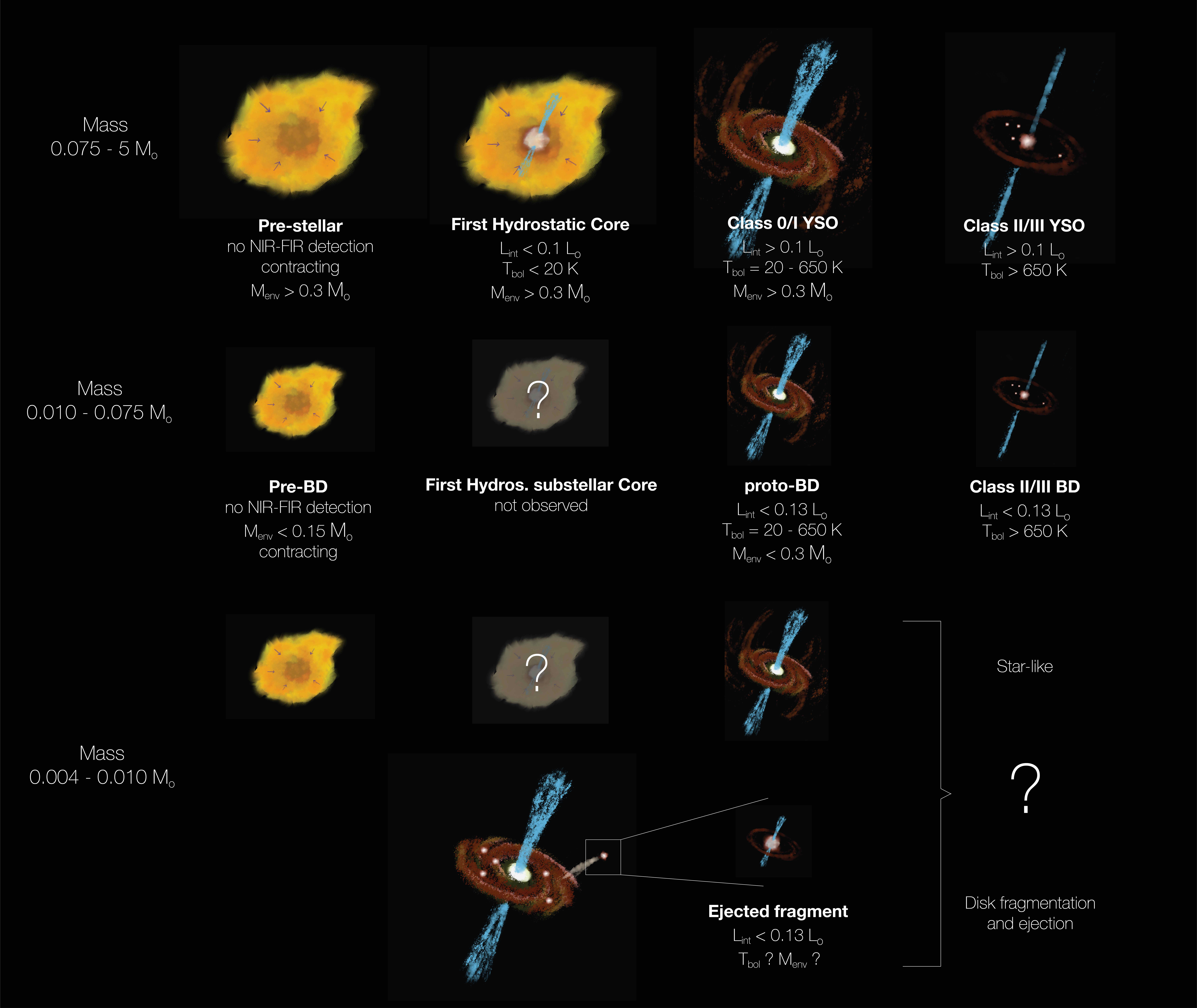, width=17cm,angle=0}\\
\end{tabular}
\caption{
Sketch of the probable/dominant formation process in the different mass regimes. 
The top row corresponds to the process of low-mass star formation, covering masses from 0.075 to $\sim5$~\mo, and undergoing the pre-stellar, FHC, Class 0/I and Class II/III phases. 
The second row corresponds to the process of BD formation, including masses from $\sim0.010$ to 0.075~\mo, and most likely undergoing the pre-BD, proto-BD and Class II/III phases (see Sec.~\ref{sec:constraints2theory}). 
The FHC phase of a substellar object should exist if they form as low-mass stars, but has not been observed yet.
The third and fourth rows correspond to the process of FFP formation, including masses from 0.004 to 0.010~\mo. The third row corresponds to a star-like scenario, while the forth row corresponds to the disc fragmentation + ejection scenario. The question mark indicates the current debate about the role of each scenario in this mass range.}
\label{fig:sketch2}
\end{center}
\end{figure*}

There are at least two caveats in the results presented above. First, each one of these evidences should be further studied in future projects focused at confirming the current proto- and pre-BD candidates and enlarging their statistics, so that the results are settled on a more robust base.
Second, it is important to keep in mind that in Sec.~\ref{sec:mainprops-requirements}, a requirement for a robust proto-BD candidate was that its SED must be similar to Class 0 or Class I SEDs of protostars. However, if proto-BDs are formed from disc fragmentation and subsequently ejected, such a SED shape could change, with the object being equally young. Thus, up to a certain point, it is natural that, given our requirements, we have picked up only the objects that form like stars\footnote{However, the fact that proto-BD candidates with $\Lint>0.04$~\lo\ follow the same behavior as protostars in several molecular clouds (Sec.~\ref{sec:NbdMCproperties-Lada10relations}) suggests that at least for the massive proto-BDs, the contribution from ejection might not be highly significant.}. Further theoretical work about the expected SEDs for ejected objects from discs would be very helpful to deal with those possible biases.

In spite of these caveats, the compiled, in some cases preliminary results presented in this work seem to consistently point towards a star-like scenario for BDs down to the planetary boundary, $\sim0.01$~\mo. 
This is consistent with reported properties of the least massive BDs, like e.g. the finding that BDs with masses down to 10~\mj\ show discs as those detected around stars, reinforcing the idea that they are a scaled-down version of stellar objects \citep[see][and references therein]{Scholz2023_NGC1333}.
Given the increasing evidence that BDs seem to form like stars at least down to the planetary-boundary mass regime, the next question is what happens below this boundary (i.e., $\lesssim0.01$~\mo).

Although the data gathered in the present work are not well suited to answer which is the main formation mechanism below $\sim0.01$~\mo, there are recent works shedding light on this question. In Sec.~\ref{sec:intro}, a number of works focused on the study of the low-mass end of the IMF were mentioned. Most of the 20 IMF works cited in Sec.~\ref{sec:intro} cover masses $>0.01$~\mo, and only three of them include the planetary regime down to $\sim0.005$~\mo: \cite{MiretRoig2022}, \cite{Parker2023}, and \cite{Chabrier2023_BD-IMF}. On one hand, both \cite{Parker2023} and \cite{Chabrier2023_BD-IMF} find that for the galactic bulge and NGC\,1333, the mass distribution, spatial distribution and kinematics of FFPs are consistent with a star-like scenario. On the other hand, \cite{MiretRoig2022} report a rich population of FFPs down to 5~\mj\ in Upper Scorpius and Ophiuchus clouds, which seems to excede the predictions from numerical simulations of the star-like scenario by \cite{Haugbolle2018} and \cite{Bate2019}. This is interpreted as an evidence that other mechanisms have to be at work in this regime, such as disc fragmentation and subsequent ejection, although the numerical simulations of the star-like scenario should be further explored to really assure that cannot reproduce these observations, as indicated by \cite{Chabrier2023_BD-IMF}.

In conclusion, a combined scenario for BD formation, where objects above the planetary regime ($>0.01$~\mo) form mainly like stars, while objects below this regime (0.004--0.010~\mo) form through a combination of both star-like and planet-like scenarios, seems feasible and consistent with the data presented in this work and previous recent findings. Fig.~\ref{fig:sketch2} summarizes this combined scenario. The top row of the figure describes the main observational stages that lead to the formation of a low-mass star, starting with the pre-stellar core, passing through the FHC, Class 0/I and Class II/III stages. The second row describes the formation of BDs that range from 0.010 to 0.075~\mo\ approximately. For BDs, it seems that the formation process is a scaled-down version of low-mass stars (with the only exception that the FHC has not been observed yet). Finally, for objects below the planetary-boundary of 0.01~\mo, the dominant process could be different in different environments, suggesting that both star-like and planet-like scenarios could be at work, being this an open question. 

In this sense, one of the main results found in this work might be worth exploring in the future: the role of the environment in shaping the low-mass end of the IMF. Although this is usually considered to be negligible, in Sec.~\ref{sec:NbdMCproperties-star2BD0004} and \ref{sec:explanationsOphdeficit} it was shown that Ophiuchus seems to under-produce low-mass BDs, and that one possibility to explain this deficit is the heating of the cloud by the nearby OB stars of the Sco OB2 association, naturally increasing the Jeans mass and thus suppressing fragmentation below a certain mass. This opens the possibility that the environment plays a subtle role in shaping the IMF, most apparent in the planetary-boundary and below, precisely the mass range that has been typically uncovered by most of the studies of the IMF, both observational and theoretical. An excellent test to further explore this hypothesis would be to apply the same methodology of \cite{MiretRoig2022} to Taurus and Lupus: we would expect a lower star-to-BD ratio in the planetary boundary in these two clouds compared to Ophiuchus, something which was marginally found by \cite{AlvesdeOliveira2012_rhoOph-IMF} but needs to be confirmed with deeper studies in a large number of clouds. If confirmed, this would favor the fact that the environment where a cloud is embedded can affect the very low-mass end of the IMF by suppressing/allowing the formation of the lowest-mass BDs in a Jeans-like, therefore star-like, manner. 


To sum up, taking into account all these considerations and the observational evidence presented in this work, it would be very useful for observers that the following aspects were taken into account from the theoretical and numerical side: i) predictions of number of pre-BD cores in molecular clouds; ii) differences between expected SEDs for star-like proto-BDs and proto-BDs ejected from discs; iii) simulations that fully sample the planetary-mass regime\footnote{The simulations from \cite{Haugbolle2018} sample masses down to $\sim0.02$~\mo, while the simulations from \cite{Bate2019} sample masses down to $\sim0.01$~\mo, and do not reach statistically the lowest masses observed by \cite{MiretRoig2022}, of 0.005~\mo.}; iv) simulations that take into account heating of molecular clouds by external feedback from OB stars in realistic conditions. These studies would allow to contrast on a more robust base the observations and the theories in the decisive regime where planets and BDs overlap.
\section{What can we learn from the current observations to design future strategies}\label{sec:lessons}

In this section we discuss how we could use {\it Spitzer}, {\it Herschel} (archival), ALMA, JWST, {\it Euclid}, the ngVLA, SKA and future facilities to efficiently search, identify and characterize proto-BDs.

ALMA will be upgraded to a much higher sensitivity, and the ngVLA is planned to start operations in 2032 and its observing bands range from 1.3~GHz to 116~GHz. Thus, it will be feasible to search for faint dusty envelopes associated with proto-BDs with these new generation of interferometers. 

In Sec.~\ref{sec:searches-serendip}, a number of proto-BD candidates were reported to be found serendipitously in the surroundings ($\lesssim5000$~au) of other protostars or proto-BDs, suggesting that a promising strategy to search for proto-BDs would be to carry out deep interferometric observations in the immediate surroundings of Class 0 protostars. In this sense, the ngVLA will be a great instrument, since it will be able to reach sensitivities at 3~mm of $\sim1$~$\mu$\jpb\ in only 1 hour, allowing to observe envelope masses as low as 0.3~\mj, or about 6 times the mass of Neptune, at the distance of Orion. In addition, one single pointing at 3~mm will provide a field of view of $\sim34$~arcsec, corresponding to $\sim5000$~au at the distance of Taurus, making the searches of proto-BDs near protostars very efficient because no mosaicing would be needed. What is more, these deep observations around Class 0 protostars would also allow to explore the possibility that proto-BDs are triggered by the jets from their nearby protostars, which would compress the surrounding dense gas, favoring the fragmentation of very low-mass objects. If this scenario is at work, the deep millimetre and centimetre observations from next-generation interferometers would allow to gather statistically significant numbers of proto-BDs near protostars, and robustly test whether they are preferentially formed along the outflow directions.

Another important test that could be performed with the next-generation millimetre interferometers is studying the role of ambipolar diffusion in the formation process of BDs. Ambipolar diffusion could play an important role if BDs form as a scaled-down version of stars, because it would allow to build the  dense gravitationally bound condensations that will end up as proto-BDs. This could be observationally tested by observing molecular lines from both ionized and neutral species because, if ambipolar diffusion is important, a significant velocity drift between ionized and neutral species is expected \citep[e.g.,][]{Yen2018_B335-ambipolardiff, Tritsis2023_ion-neutral-drift}. This cannot be easily studied nowadays because detecting molecular gas around proto-BDs is challenging, but with the orders-of-magnitude increase in sensitivity of the next-generation interferometers, this will be a feasible test.

Deep interferometric studies will also be essential to elucidate the nature of the current pre-BD candidates (such as those listed in Table~\ref{tab:preBDs}) and to search for new candidates in the solar neighborhood clouds. Most of the candidates of Table~\ref{tab:preBDs} still lack very deep interferometric observations to: look for potential compact sources associated, detect the gas and assess gravitational instability, study the density profile and study possible infall motions associated with the objects. Since pre-BD candidates are so faint, very deep observations are required and next-generation interferometers will be decisive in this sense.

In addition, the ngVLA and SKA will allow to search for extremely faint thermal radiojets such as the one found in IC348-SMM2E or the candidates identified in Taurus \citep{Forbrich2015_IC348SMM2E, Morata2015_Jets_ProtoBD, Rodriguez2017_BDs-VLA}. Taking the flux density measured in IC348-SMM2E at 3.3\,cm, of 23~$\mu$\jpb, which was detected at the 3$\sigma$ level, the ngVLA in the Main array could detect it in one hour with a signal-to-noise ratio of 90, a factor of 30 larger than the current marginal detection (the sensitivity of the Main array at 3.6\,cm or 8\,GHz in 1 hour is of 0.25~$\mu$\jpb). This also means that we could detect these proto-BD radiojets up to 1.5 kpc (at 4$\sigma$), opening a window to search them well outside the solar neighborhood. Furthermore, these new-generation interferometers, along with ALMA in its most extended configurations, will also allow to study multiplicity, including very compact companions, in large samples of proto-BDs, a crucial piece to understand their formation process. For example, the ngVLA at 3~mm will reach an angular resolution of 0.08~mas, corresponding to $\sim0.01$~au at the distance of Taurus.

Moving to shorter wavelengths, the {\it Euclid} mission, whose goal is to establish the nature of dark matter and dark energy, will also play a role in the search and characterizations of BDs. Several deep observations have been already taken in the $\sigma$, $\lambda$ Orionis, and Taurus star-forming regions, among others, during its first operational months \citep[see][]{Martin2024_Euclid}. These data should extend the IMFs down to the mass of Jupiter, and should confirm whether there is a  cut-off and its relation with the proposed formation mechanism. 
In particular, these observations would be very useful to test if the population of BDs in the planetary-boundary presents a deficit in Ophiuchus compared to Taurus and Lupus, as expected from the results of Sec.~\ref{sec:NbdMCproperties-star2BD0004} and ~\ref{sec:explanationsOphdeficit}.
In addition, these {\it Euclid} observations will be important to validate previously identified proto-BD candidates and, as a wide-field survey, will provide the opportunity of many serendipitous discoveries in several star-forming regions.
However, since the longest wavelengths of {\it Euclid} lie in the near-infrared, it will probably still miss the most embedded and least massive proto-BDs, whose SED peak lies near 70--100~\mum. For the extreme most embedded cases, the {\it Herschel} or James Webb Space telescopes, observing at longer wavelengths, would be required.

JWST, launched in December 2022, is providing an extraordinary wealth of information in multiple fields. We have already mentioned the recent works by \citet{Pearson2023} and \cite{McCaughrean2023}, who  have found hundreds of BD and FFP candidates in the Trapezium cluster using deep near- and mid-IR imaging. A detailed follow-up is expected, using its spectroscopic capabilities. Thus, detailed properties will be derived, such as  kinematics, including the outflows (following the steps of \citealt[]{Ray2023_HH211} in the case of the well-known outflow HH\,211) or atmospheric compositions. 
Thus, the JWST is revealing an otherwise hidden population of BDs in the planetary-mass regime, opening the opportunity to test the different formation scenarios. In particular, the star-like formation scenario can be tested through the study of chemical signatures of BD atmospheres \citep[]{Zhang2021_CO_BD,Zhang2021_CO_Planet}. As an example, \cite{Barrado2023} have recently presented JWST observations of a cool BD-BD binary with 15~\mj\ components, and conclude that it has most probably formed in a star-like manner by using the $^{15}$NH$_3$ as a diagnostic tool to distinguish among formation scenarios. However, some caveats should be taken into account, since very high quality data and further improvements in the modelling are needed. The extent of these chemical studies to larger samples will definitely help to confirm the star-like formation scenario for these objects.

The deep observations of the JWST, along with {\it Euclid}, and further exploitation of {\it Spitzer} and {\it Herschel} data are required to build samples of much less luminous objects compared to the current samples. Our requirement that $\Lint<0.13$~\lo\ necessarily includes a fraction of `contaminant' protostars, because $\Lint<0.13$~\lo\ corresponds to $\Mdyn<0.12$~\mo\ according to Fig.~\ref{fig:LintMdyn}. The $\Lint<0.13$~\lo\ requirement was used in Sec.~\ref{sec:mainprops-requirements} as a first best-effort basis to take into account the typical uncertainties in $\Mdyn$ of about 50\%, but strictly speaking and following the relation of Fig.~\ref{fig:LintMdyn}, one should be using the requirement of $\Lint<0.04$~\lo\ to select $\Mdyn<0.075$~\mo\ and, to do this, new techniques, strategies and observatories are required to efficiently identify candidates that are much less luminous than what has been explored so far.

Finally, as mentioned in Sec.~\ref{sec:explanationsOphdeficit}, a promising research line could be the study of molecular cloud properties, such as dust temperature or magnetic fields, with the spatial distribution of low-mass BDs, because if they form in a star-like manner it would be expected some kind of relation based on Jeans fragmentation. To do this, the combined observations of {\it Herschel} and JWST will be pivotal, because the JWST will allow to detect embedded objects with masses of the order of the Jupiter mass in clouds such as Orion, about one order of magnitude lower than the mass sensitivity achieved in this work using {\it Spitzer} data. However, the JWST longest wavelengths reach 28~\mum, and the most embedded objects are expected to lie in extended and colder regions emitting at the Herschel/PACS frequencies. Thus, a selection strategy combining criteria using both JWST and {\it Herschel} bands stands as very promising to have complete counts of the most embedded objects down to a few Jupiter masses.
This in turn might shed important clues on how the low-mass end of the IMF is shaped.


\section*{Acknowledgements}
The authors are grateful to the referee for his valuable and detailed comments that helped complete and clarify the paper.
The authors are grateful to Herv\'e Bouy, N\'uria Miret-Roig, Koraljka Muzic, Matthew Bate, Enrique V\'azquez-Semadeni, Javier Ballesteros-Paredes, Melodie Kao, and \`Oscar Morata for insightful and stimulating discussions.
The authors are grateful to Kazi Rygl and Milena Benedettini for kindly sharing the column density and dust temperature maps of the Lupus I cloud from {\it Herschel} data, and to Eduard Vorobyov for kindly sharing their models published in 2017.
The authors are grateful to Ana Mar\'ia P\'erez-Garc\'ia for her invaluable help on the SUCANES database, to Sundar Srnivasan and Pedro Mas-Buitrago for support on data analysis methods, and to Miguel Agust\'in S\'anchez Vald\'es for support on the graphical design of Figs.~\ref{fig:sketch} and ~\ref{fig:sketch2}.
AP acknowledges financial support from the UNAM-PAPIIT IN111421 and IG100223 grants, the Sistema Nacional de Investigadores of CONAHCyT, M\'exico.
DB and NH have been funded by grants No. MDM-2017-0737 Unidad de Excelencia “María de Maeztu”- Centro de Astrobiología (CSIC/INTA) and by the Spanish grants MCIN/AEI/10.13039/501100011033 PID2019-107061GB-C61 and PID2023-150468NB-I00.
CWL acknowledges support from the Basic Science Research Program through the NRF funded by the Ministry of Education, Science and Technology (NRF- 2019R1A2C1010851) and by the Korea Astronomy and Space Science Institute grant funded by the Korea government (MSIT; project No. 2024-1-841-00).
This work has made use of the SUCANES database, a joint project funded by ESA Faculty under contract No. 4000129603/19/ES/CM, ISDEFE and CSIC,
and has made use of data from the Herschel Gould Belt survey (HGBS) project (http://gouldbelt-herschel.cea.fr). The HGBS is a Herschel Key Programme jointly carried out by SPIRE Specialist Astronomy Group 3 (SAG 3), scientists of several institutes in the PACS Consortium (CEA Saclay, INAF-IFSI Rome and INAF-Arcetri, KU Leuven, MPIA Heidelberg), and scientists of the Herschel Science Center (HSC).
This paper makes use of the following ALMA data: ADS/JAO ALMA\#2013.1.00157.S, 
2013.1.00220.S, 
2013.1.00465.S, 
2013.1.00474.S, 
2015.1.00041.S, 
2015.1.00186.S, 
2015.1.00223.S, 
2015.1.00310.S, 
2015.1.00741.S, 
2015.1.01510.S, 
2016.1.00039.S.
2016.1.00085.S, 
2016.1.00460.S, 
2017.1.01462.S, 
2017.1.01693.S, 
2018.1.00126.S, 
2018.1.00756.S, 
2019.1.00218.S, 
2019.1.00847.S, 
2019.1.01813.S, 
and 2022.1.01270.S.
ALMA is a partnership of ESO (representing its member states), NSF (USA) and NINS (Japan), together with NRC (Canada), NSTC and ASIAA (Taiwan), and KASI (Republic of Korea), in cooperation with the Republic of Chile. The Joint ALMA Observatory is operated by ESO, AUI/NRAO and NAOJ.



\begin{table*}
\caption{Compilation of Fig.~\ref{fig:LintMdyn} of $\Lint$ and $\Mdyn$ for proto-BD candidates, VeLLOs and protostars with $\Lint<10$~\lo, and for which a velocity gradient has been observed}
\label{tab:LintMdyn}
\begin{center}
{\small
\begin{tabular}{lccc cc}
\noalign{\smallskip}
\hline\noalign{\smallskip}
&$\Mdyn$\supa
&$\Lint$\supa
&$\Tbol$\supa
&Excluded
&
\\
Source
&(\mo)
&(\lo)
&(K)
&from fit\supb
&Refs.\supc
\\
\noalign{\smallskip}
\hline\noalign{\smallskip}
L328-IRS        &$0.27\pm0.15$	&$0.121\pm0.020$     &67     &$-$        &L18\\
I04158+2805	    &$0.30\pm0.15$	&$0.065\pm0.010$     &337    &$-$        &A08,Ra21\\
I16253$-$2429   &$0.14\pm0.03$	&$0.129\pm0.030$     &37     &$-$        &H19,A23\\
Mayrit1701117   &$0.35\pm0.17$	&$0.27\pm0.06$       &100    &$-$        &R19\\
ISO-Oph200      &$0.23\pm0.12$	&$0.090\pm0.018$     &329    &$-$        &Ri21\\
I15398$-$3359   &$0.05\pm0.04$	&$1.34$              &122    &outbursting &O14,Y17, O18\\
L1451-mm        &$0.06\pm0.03$	&$0.020\pm0.004$     &19     &$\Tbol<20$~K&Mau20\\
Cha-MMS1        &$0.03\pm0.01$	&$0.024\pm0.004$     &20     &$-$        &Mau20\\
IC348-SMM2E     &$0.07\pm0.04$	&$0.109\pm0.020$     &23     &$-$        &P14\\
IRAM04191       &$0.05\pm0.03$	&$0.042\pm0.008$     &59     &$-$        &H24\\
I04191+1523A    &$0.14\pm0.07$	&$0.18\pm0.03$       &$-$    &$-$       &L17\\
I04191+1523B    &$0.12\pm0.06$	&$0.18\pm0.03$       &$-$    &$-$       &L17\\
L1448-C         &$1.5\pm0.7$	&$7.5\pm2.5$         &$-$    &$-$       &Mar20,H10\\
L1527-IRS       &$0.55\pm0.15$	&$0.9\pm0.1$         &$-$    &$-$       &T12,Mar20\\
L1448-IRS3B     &$1.0\pm0.5$	&$2.4\pm0.5$         &$-$	 &$-$       &T16\\
Per-Bolo58      &$0.12\pm0.06$	&$0.019\pm0.002$     &20	 &$-$       &Mau20\\
Lupus3-MMS      &$0.23\pm0.11$	&$0.41\pm0.20$       &39	 &$-$       &Y17\\
TMC-1A          &$0.64\pm0.32$	&$2.7\pm1.3$         &$-$	 &$-$       &Y17\\
RCrA-IRS7B      &$2.3\pm1.1$	&$4.6\pm2.3$         &$-$	 &$-$       &Y17\\
IRS43           &$1.9\pm1.0$	&$6\pm3$             &$-$	 &$-$       &Y17\\
L1489-IRS       &$1.6\pm0.8$	&$3.7\pm1.8$         &$-$	 &$-$       &Y17\\
L1551NE         &$0.8\pm0.4$	&$4.2\pm2.1$         &$-$	 &$-$       &Y17\\
IRS63           &$0.8\pm0.4$	&$1.0\pm0.5$         &$-$	 &$-$       &Y17\\
TMC1            &$0.54\pm0.27$	&$0.9\pm0.4$         &$-$	 &$-$       &Y17\\
B335            &$0.15\pm0.1$	&$1.4$              &$-$	 &outbursting&Y17,E23\\
VLA1623         &$0.2\pm0.1$	&$1.1\pm0.6$         &$-$	 &multiple  &Y17\\
N1333-IRAS4A2   &$0.08\pm0.04$	&$1.9\pm0.9$         &$-$	 &multiple  &Y17\\
HH211           &$0.05\pm0.02$	&$1.0$              &$-$	 &shocks at 70~\mum &Y17,P21\\
L1521F-IRS      &$0.18\pm0.09$	&$0.037\pm0.003$     &16	 &$\Tbol<20$~K &T17\\

\noalign{\smallskip}
\hline\noalign{\smallskip}
\hline
\end{tabular}
\begin{list}{}{}
\item[$^\mathrm{a}$] $\Lint$ is taken from SUCANES database when available or from the references listed in the table. Its uncertainty is taken as 20\%. For $\Mdyn$, a typical uncertainty of 50\% is adopted. $\Mdyn$ for ISO-Oph200 is re-calculated in this work using Fig. 5 of \cite{Riaz2021_Structure_ProtoBD}. For IC348-SMM2E, both $\Mdyn$ and $\Lint$ are re-calculated to the new distance of 320 pc \citep{OrtizLeon2018_DistancePerseus}. In addition, since IC348-SMM2E presents emission at 4.5~\mum\ associated with the blueshifted lobe \cite{Palau2014}, suggesting that the outflow is inclined with respect to the plane of sky, a range of inclinations 30--60\,$^\circ$ were adopted, corresponding to 36--110~\mj. For I16253$-$2429, the results of \cite{Yen2017_I15398} were also taken into account, but we finally adopted the $\Mdyn$ from \cite{Hsieh2019_VeLLO_IRAS16253_ALMA} because these authors include the \cite{Yen2017_I15398} work in their analysis and in addition obtain results consistent with the more recent work by \cite{Aso2023_eDisks-I16253}.  For Mayrit1701117, ISO-Oph200, and I04191+1523A,B,  $\Lint$ is recalculated from the flux at 70~\mum\ from the SED of the reference paper, and following eq.~\ref{eq:LintFlux70mic}. $\Lint$ for IRAS15398 is 1.34~\lo\ using the measured flux at 70~\mum\ (24 Jy) from \cite{Benedettini2018_Lupus}, consistent with the IRAS fluxes. However, \cite{Joergensen2013_I15398} argue that this object is outbursting and that its true $\Lint$ is two orders of magnitude smaller. Given these extreme characteristics, I15398$-$3359 has been excluded from the fit of Fig.~\ref{fig:LintMdyn}.
$\Tbol$ is taken from SUCANES if the source is included in this database. 
\item[$^\mathrm{b}$] Reason why the object has been excluded from the fit of Fig.~\ref{fig:LintMdyn}. Wherever `multiple' is indicated, it refers to the fact that $\Lint$ includes all components of the multiple system while the velocity gradient has been identified in only one of the components, yielding an evident mismatch between $\Lint$ and $\Mdyn$. Objects with $\Tbol<20$~K have been excluded from the $\Lint$ vs $\Mdyn$ relation because very low $\Tbol$ can be associated also with pre-stellar cores \citep[e.g.,][]{Young2005_EvolutionarySignatures}.
\item[$^\mathrm{c}$] References: 
A08: \cite{Andrews2008_I04158};
A23: \cite{Aso2023_eDisks-I16253};
E23: \cite{Evans2023_B335};
H10: \cite{Hirano2010_L1448C};
H19: \cite{Hsieh2019_VeLLO_IRAS16253_ALMA};
H24: Huelamo et al. (2024);
L17: \cite{LeeLeeDunham2017_I04191};
L18: \cite{Lee2018_L328_ProtoBD_ALMA};
Mar20: \cite{Maret2020_CALLYPSO};
Mau20: \cite{Maureira2020_FHCs};
O14: \cite{Oya2014_I15398};
O18: \cite{Okoda2018_I15398};
P14: \cite{Palau2014};
P21: \cite{Pezzuto2021_Perseus-Herschel};
R19: \cite{Riaz2019_Mayrit1701117};
Ra21: \cite{Ragusa2021_I04158};
Ri21: \cite{Riaz2021_Structure_ProtoBD};
T12: \cite{Tobin2012_L1527};
T16: \cite{Tobin2016_L1448IRS3B};
T17: \cite{Tokuda2017_L1521FIRS};
Y17: \cite{Yen2017_I15398}.
\end{list}
}
\end{center}

\end{table*}

\begin{landscape}
\begin{table}[htbp]
\caption{Objects from SUCANES with $\Lint \le 0.13$~\lo\ that do not fulfill the $\Tbol$ and/or $\Menv$ criteria of Sec.~\ref{sec:mainprops-requirements}}
\label{tab:rejected}
\begin{center}
{\small
\begin{tabular}{llcccccc cccl}
\noalign{\smallskip}
\hline\noalign{\smallskip}
&&Dist.
&&
&$L_\mathrm{int}$
&$T_\mathrm{bol}$
&$M_\mathrm{env}$\supa
&$L_\mathrm{bol}$
&
&
\\
Source
&Region
&(pc)
&R.A.
&Dec.
&(\lo)
&(K)
&(\mo)
&(\lo)
&Class
&Type\supb
&Refs.\supc
\\
\noalign{\smallskip}
\hline\noalign{\smallskip}
L1451-mm        &Perseus    &293    &03:25:09.5 &$+$30:23:51    &$<0.020$       &19     &0.48    &0.099  &0      &VeLLO/FHC  &E06,Mau20,P11\\
L1448-IRS2E     &Perseus    &293    &03:25:22.4 &$+$30:45:13    &$<0.033$       &$-$    &0.047     &$-$    &0      &VeLLO/FHC  &C10,Mau20\\
Per-Bolo58      &Perseus    &293    &03:29:25.5 &$+$31:28:15    &$0.019\pm0.002$&20     &1.64   &0.174  &0      &FHC       &E10\\
B1b-N           &Perseus    &293    &03:33:21.2 &$+$31:07:44    &$<0.014$       &15     &1.20   &0.433  &0      &VeLLO/FHC  &H14,G17,H99\\
B1b-S           &Perseus    &293    &03:33:21.4 &$+$31:07:26    &$0.058\pm0.038$&18     &1.20   &0.757  &0      &VeLLO/FHC  &H14,G17,H99\\
\hline\noalign{\smallskip}
J041726         &Taurus     &137    &04:17:26.4 &$+$27:39:20    &$<0.001$       &$-$    &0.005      &$-$    &I      &protoBD   &P12,M15\\
J041913         &Taurus     &137    &04:19:13.1 &$+$27:47:26    &$<0.001$       &$-$    &0.0007    &$-$    &I      &protoBD   &P12,M15\\
L1521F-IRS      &Taurus     &137    &04:28:38.9 &$+$26:51:36    &$0.037\pm0.001$&16     &0.40    &0.479  &0      &protoBD   &D08,K16\\
\hline\noalign{\smallskip}
G192-S-B30      &$\lambda$-Ori&380  &05:29:54.4 &$+$12:16:30    &$0.080\pm0.005$&113    &0.66    &0.339  &I      &protoBD &L16\\
\hline\noalign{\smallskip}
Cha-MMS1        &Cha I      &192    &11:06:31.7 &$-$77:23:33    &$0.041\pm0.005$ & 20     &2.36   &0.246  &0      &FHC        &Mau20,B20\\
\hline\noalign{\smallskip}
ALMA-J153914    &Lupus I    &153    &15:39:15.0 &$-$33:29:08    &$<0.002$       &$-$    &0.017     &$-$    &I      &protoBD    &S21\\
J160115         &Lupus IV   &150    &16:01:15.5 &$-$41:52:35    &$0.099\pm0.002$&129    &0.40    &0.131  &I      &protoBD   &D08,K16\\
\hline\noalign{\smallskip}
Source-X & Ophiuchus & 137 & 16:26:27.4 & -24:24:18 & $<0.082$ &
 $-$ & 0.04  & $-$ & 0 & protoBD & K17,K18,J00\\
SM1-A & Ophiuchus & 137 & 16:26:27.8 & -24:23:59 & $<0.082$ & 21   & 4.6  & 0.214 & 0 & protoBD & K17,K18,J00\\ 
L328-IRS        &Ophiuchus  &217    &18:16:59.5 &$-$18:02:30    &$0.121\pm0.005$&67     &0.90    &0.126  &0      &protoBD   &L09,L13,K16,L18\\
\hline\noalign{\smallskip}
J182912	        &Aqu-Ser    &436    &18:29:12.1 &$-$01:48:45    &$0.025\pm0.003$&310    &$-$    &0.038  &I      &VeLLO	    &K16\\
J182958         &Aqu-Ser    &436    &18:29:58.3 &$-$01:57:40    &$0.101\pm0.009$&112    &$-$    &0.070  &I      &VeLLO      &K16\\
\hline\noalign{\smallskip}
J190418         &CrAus      &150    &19:04:18.6 &$-$37:35:56    &$0.014\pm0.001$&672    &0.0002    &0.016  &II     &VeLLO     &K16\\
\hline\noalign{\smallskip}
L1148-IRS       &Cepheus    &330    &20:40:56.7 &$+$67:23:05    &$0.112\pm0.004$& 122    &0.58    &0.164  &I      &protoBD    &K05,D08,K16\\
J222933         &Cepheus    &339    &22:29:33.3 &$+$75:13:16    &$0.036\pm0.003$&324    &1.02    &0.104  &I      &VeLLO      &K16\\
\hline\noalign{\smallskip}
L1014-IRS        &Cygnus    &258    &21:24:07.5 &$+$49:59:09    &$0.100\pm0.037$&65     &0.70    &0.451  &0      &protoBD   &K08,D08,K16,H06\\ \hline\noalign{\smallskip}
J214457 & IC\,5146 & 600 & 21:44:57.0 & $+$47:41:52 &$0.085\pm0.031$ & 211     &0.74    &0.632  & I      & VeLLO   & K16\\

\hline\noalign{\smallskip}
\end{tabular}
\begin{list}{}{} 
\item[$^\mathrm{a}$] Envelope masses taken from \citet[]{Kim2016_VeLLOs}. For objects not included in the compilation of \citet[]{Kim2016_VeLLOs} the envelope masses are those measured with single-dish as reported in the literature, except for the \cite{Morata2015_Jets_ProtoBD} sources, for which $\Menv$ was estimated from the emission at 250~\mum, following Sec.~4.2 of \cite{Kim2016_VeLLOs}. For J222933, we calculated $\Menv$ from the flux at 1.2~mm given by \cite{Dunham2015_YSO_GouldBelt}, of $\sim0.35$~Jy, and assuming a dust temperature of 15~K and a dust opacity per mass of gas and dust of 0.01~cm$^2$\,g$^{-1}$ \citep{Ossenkopf1994_dustopacity}.
\item[$^\mathrm{b}$] The type corresponds to the classification assigned in previous works.
\item[$^\mathrm{c}$] References: 
B20: \cite{Busch2020_ChaMMS1-outflows};
C10: \cite{Chen2010_L1448_FHC};
D08: \cite{Dunham2008_c2dVeLLOs};
E06: \cite{Enoch2006};
E10: \cite{Enoch2010_FHC};
G17: \cite{Gerin2017_B1bNS};
H99: \cite{Hirano1999}
H06: \cite{Huard2006_L1014};
H14: \cite{Hirano2014_Barnard1};
J00: \cite{Johnstone2000_Oph-JCMT};
K05: \cite{Kauffmann2005_Spitzer_VeLLO};
K08: \cite{Kauffmann2008_MAMBO_SpitzerCores};
K16: \citet[]{Kim2016_VeLLOs}; 
K17: \citet[]{Kirk2017_Ophiuchus};
K18: \citet[]{Kawabe2018}
L09: \cite{Lee2009_VeLLO_L328};
L13: \cite{Lee2013_ProtoBD_L328-IRS};
L16: \cite{Liu2016_Planck};
L18: \cite{Lee2018_L328_ProtoBD_ALMA};
M15: \citet{Morata2015_Jets_ProtoBD}; 
Mau20: \cite{Maureira2020_FHCs};
P11: \cite{Pineda2011_FHC_VeLLO_L1451};
P12: \cite{Palau2012_PreBD_Cores_Taurus};
S21: \cite{SantamariaMiranda2021_ALMA_Lupus_ProtoBD}.
\end{list}
}
\end{center}

\end{table}
\end{landscape}
\pagestyle{plain}

\begin{landscape}
\begin{table}[htbp]
\caption{68 proto-BD candidates that fulfill criteria of Sec.~\ref{sec:mainprops-requirements}}
\label{tab:protoBDs}
\begin{center}
{\small
\begin{tabular}{llcccccc cccll}
\noalign{\smallskip}
\hline\noalign{\smallskip}
&&Dist.
&&
&$L_\mathrm{int}$
&$T_\mathrm{bol}$
&$M_\mathrm{env}$\supb
&$L_\mathrm{bol}$
\\
Source\supa
&Region
&(pc)
&R.A.
&Dec.
&(\lo)
&(K)
&(\mo)
&(\lo)
&Class
&Type
&Refs.\supc
&ALMA Data\supc
\\
\noalign{\smallskip}
\hline\noalign{\smallskip}
Per-Bolo25      &Perseus    &293    &03:28:32.5 &$+$31:11:05    &$0.061\pm0.009$    &113    &0.122    &0.224  &I      &protoBD    &D08,K16,C16  &2017.1.01693, H19\\
Per-Bolo30      &Perseus    &293    &03:28:39.1 &$+$31:06:01    &$0.022\pm0.004$    &31     &0.154    &0.136  &0      &protoBD    &D08,K16,C16 &2017.1.01693, H19\\
IC348-SMM2E     &Perseus    &320    &03:43:57.7 &$+$32:03:10    &$0.109\pm0.015$    &23     &0.051     &0.287  &0      &protoBD    &P14,K16     &2017.1.01462, Y21\\
\hline\noalign{\smallskip}
I04111+2800G    &Taurus     &137    &04:14:12.3 &$+$28:08:37    &$0.061\pm0.014$    &74     &0.020     &0.095  &0/I    &protoBD    &K16     &2018.1.00756, Tok20\\
J041740         &Taurus     &140    &04:17:40.3 &$+$28:24:15    &$0.0014\pm0.0003$  &397    &0.0006    &0.003  &I      &protoBD    &P12,M15     &$-$\\
J041757         &Taurus     &140    &04:17:57.7 &$+$27:41:06    &$<0.002$           &158    &0.0024     &0.003  &I      &protoBD    &B09,P12,PB14,M15   &2013.1.00465, P22\\
J041828         &Taurus     &140    &04:18:28.1 &$+$27:49:11    &$0.0015\pm0.0004$  &125\supd    &0.0014    &0.003\supd  &I      &protoBD    &P12,M15     &$-$\\
J041836         &Taurus     &140    &04:18:36.3 &$+$27:14:43    &$<0.003$           &281\supd    &0.0012    &0.005\supd  &I      &protoBD    &P12,M15     &2013.1.00465\\
J041840         &Taurus     &137    &04:18:40.2 &$+$28:29:25    &$0.0041\pm0.0007$  &86\supd     &0.002    &0.002\supd  &I      &VeLLO      &K16      &$-$\\
I04158+2805     &Taurus     &140    &04:18:58.1 &$+$28:12:24    &$0.065\pm0.009$    &337    &0.030     &0.210  &I      &protoBD    &W04,A08     &2016.1.00460, Rag21, V20\\
J041938         &Taurus     &140    &04:19:38.8 &$+$28:23:41    &$0.006\pm0.002$    &170    &0.0012    &0.005  &I      &protoBD    &P12,M15     &2013.1.00465\\
{[GKH94]}41       &Taurus     &140    &04:19:46.6 &$+$27:12:55    &$0.017\pm0.002$    &292    &0.0044    &0.081  &I      &protoBD    &D16, this work         &$-$\\
J042019         &Taurus     &140    &04:20:19.2 &$+$28:06:10    &$<0.002$           &322\supd    &0.0009    &0.003\supd  &I      &protoBD    &P12,M15      &$-$\\
J042118         &Taurus     &140    &04:21:18.4 &$+$28:06:41    &$<0.002$           &173    &0.0007    &0.002  &I      &protoBD    &P12,M15      &$-$\\
IRAM04191       &Taurus     &140    &04:21:56.9 &$+$15:29:46    &$0.042\pm0.001$    &59     &0.050     &0.065  &0      &protoBD    &A99,D08,K16 &2015.1.00186\\
J042513         &Taurus     &137    &04:25:13.2 &$+$26:31:45    &$0.008\pm0.002$    &54\supd     &0.001    &0.006\supd  &0      &VeLLO      &K16          &$-$\\
J043202         &Taurus     &137    &04:32:02.0 &$+$25:16:41    &$0.006\pm0.001$    &90\supd     &0.001   &0.004\supd  &I      &VeLLO      &K16           &$-$\\
J043411         &Taurus     &137    &04:34:11.5 &$+$24:03:41    &$0.013\pm0.002$    &96\supd     &0.100    &0.008\supd  &I      &VeLLO      &K16          &$-$\\
J043909         &Taurus     &137    &04:39:09.0 &$+$26:14:49    &$0.011\pm0.002$    &87\supd     &0.004    &0.006\supd  &0/I    &VeLLO      &K16          &$-$\\
J044022         &Taurus     &137    &04:40:22.4 &$+$25:58:32    &$0.002\pm0.001$    &102\supd    &0.001    &0.005\supd  &I      &VeLLO      &K16          &$-$\\
J044209         &Taurus     &137    &04:42:09.4 &$+$25:16:35    &$0.007\pm0.001$    &89\supd     &0.0001    &0.005\supd  &I      &VeLLO      &K16         &$-$\\
I04489+3042     &Taurus     &140    &04:52:06.7 &$+$30:47:17    &$0.123\pm0.016$    &304    &0.003    &0.418  &I      &protoBD    &W04          &2019.1.00847\\
\hline\noalign{\smallskip}
J040129         &Califor    &450    &04:01:29.0 &$+$41:21:28    &$0.032\pm0.003$    &74\supd     &0.010     &0.027\supd  &0/I     &VeLLO      &K16        &$-$\\
J042815         &Califor    &450    &04:28:15.1 &$+$36:30:28    &$0.048\pm0.010$    &399    &0.100    &0.314  &I      &VeLLO      &K16          &$-$\\
J042818         &Califor    &450    &04:28:18.6 &$+$36:54:35    &$0.014\pm0.004$    &122\supd    &0.003    &0.027\supd  &I        &VeLLO      &K16        &$-$\\
J043014         &Califor    &450    &04:30:14.9 &$+$36:00:08    &$0.079\pm0.010$    &79     &0.300    &0.260  &0/I    &VeLLO      &K16          &$-$\\
\hline\noalign{\smallskip}
\end{tabular}
}
\end{center}

\end{table}
\end{landscape}
\pagestyle{plain}

\setcounter{table}{2}
\begin{landscape}
\begin{table}[htbp]
\caption{(cont.) 68 proto-BD candidates that fulfill criteria of Sec.~\ref{sec:mainprops-requirements}}
\label{tab:protoBDs}
\begin{center}
{\small
\begin{tabular}{llcccccc cccll}
\noalign{\smallskip}
\hline\noalign{\smallskip}
&&Dist.
&&
&$L_\mathrm{int}$
&$T_\mathrm{bol}$
&$M_\mathrm{env}$\supb
&$L_\mathrm{bol}$
\\
Source\supa
&Region
&(pc)
&R.A.
&Dec.
&(\lo)
&(K)
&(\mo)
&(\lo)
&Class
&Type
&Refs.\supc
&ALMA Data\supc
\\
\noalign{\smallskip}
\hline\noalign{\smallskip}
J053504         &Orion      &420    &05:35:04.7 &$-$05:37:12    &$0.100\pm0.020$    &30\supd     &0.300    &1.78\supd   &0      &VeLLO      &K16          &2015.1.00041, Tob20\\
J053534         &Orion      &420    &05:35:34.2 &$-$04:59:52    &$0.049\pm0.011$    &50\supd     &0.100    &0.590\supd  &0      &VeLLO      &K16          &2015.1.00041, Tob20\\
J053612         &Orion      &420    &05:36:12.9 &$-$06:23:30    &$0.104\pm0.025$    &241\supd    &0.003    &0.212\supd  &I      &VeLLO      &K16          &2019.1.01813, vT22\\
J053630         & Orion     &420    &05:36:30.3 & -04:32:16     &$0.148\pm0.028$    &306    &0.200     &1.03  & I      &VeLLO      &K16         &$-$\\
J054020         &Orion      &420    &05:40:20.3 &$-$07:51:14    &$0.129\pm0.025$    &69     &0.020     &0.327  &0       &VeLLO      &K16        &2015.1.00041, Tob20\\
J054128         &Orion      &420    &05:41:28.9 &$-$02:23:19    &$0.167\pm0.032$    &74     &0.030     &0.377  &0       &VeLLO      &K16        &2015.1.00041, Tob20\\
J054239         &Orion      &420    &05:42:39.2 &$-$10:01:47    &$0.006\pm0.002$    &128\supd    &0.002    &0.037\supd  &I      &VeLLO      &K16          &$-$\\
J055418         &Orion      &420    &05:54:18.4 &$+$01:49:03    &$0.018\pm0.004$    &127\supd    &0.050     &0.049\supd  &I       &VeLLO      &K16        &2015.1.00041, Tob20\\
\hline
J080533         &BHR16      &440    &08:05:33.0 &$-$39:09:24    &$0.079\pm0.009$  &264  &0.001    &0.104  &I          &VeLLO      &K16          &2022.1.01270\\
\hline\noalign{\smallskip}
J105959         &Cha I      &192    &10:59:59.7 &$-$77:11:18    &$0.019\pm0.002$  &583  &0.0013    &0.016  &Flat       &VeLLO      &K16         &$-$\\
J121406         &Cha III    &193    &12:14:06.5 &$-$80:26:25    &$0.043\pm0.005$  &59\supd   &0.0042    &0.028\supd  &0          &VeLLO      &K16         &$-$\\
Cha-APEX-L      &Cha II     &198    &12:52:55.8 &$-$77:07:35    &$<0.014$         &30\supd   &0.059     &0.071\supd  &0           &protoBD    &dG16       &$-$\\
J125701         &Cha II     &198    &12:57:01.5 &$-$76:48:34    &$0.081\pm0.005$  &119\supd  &0.0002    &0.076\supd  &I          &VeLLO      &K16         &$-$\\
\hline\noalign{\smallskip}
J154216	        &DC3272     &250    &15:42:16.9 &$-$52:48:02    &$0.043\pm0.017$  &42   &0.040     &0.180  &0           &protoBD    &D08,K16    &2016.1.00085\\
\hline\noalign{\smallskip}
J153834         &Lupus I    &150    &15:38:34.6 &$-$34:48:19    &$0.006\pm0.002$  &184\supd  &0.0001    &0.005\supd  &I          &VeLLO      &K16         &2015.1.01510, V21\\
J154051         &Lupus I    &150    &15:40:51.6 &$-$34:21:04    &$0.029\pm0.007$  &52\supd   &0.0001    &0.019\supd  &0/I        &protoBD    &D08,K16     &2015.1.01510, V21\\
L1YSO\,5    & Lupus I & 153 & 15:42:14.6 &$-$34:10:25     &$0.050\pm0.006$  & 161    & 0.0074 & 0.052   &I      &VeLLO    & M17, this work     &2013.1.00474\\
ALMA-J154229    &Lupus I    &153    &15:42:29.8 &$-$33:42:42    &$<0.002$         &52\supd   &0.010     &0.002\supd  &0/I         &protoBD    &S21        &2018.1.00126, S21\\
J154339         &Lupus I    &150    &15:43:39.9 &$-$33:35:54    &$0.005\pm0.001$  &574  &0.0003    &0.005  &I          &VeLLO      &K16         &$-$\\
J154548         &Lupus I    &150    &15:45:48.2 &$-$34:05:10    &$0.014\pm0.003$  &262  &0.0001    &0.011  &I          &VeLLO      &K16         &2015.1.01510, V21\\
J160754         &Lupus III  &200    &16:07:54.7 &$-$39:15:44    &$0.013\pm0.003$  &432  &0.0002    &0.020  &I          &VeLLO      &K16         &2013.1.00220, A16\\
\hline\noalign{\smallskip}
\end{tabular}
}
\end{center}

\end{table}
\end{landscape}
\pagestyle{plain}

\setcounter{table}{2}
\begin{landscape}
\begin{table}[htbp]

\caption{(cont.) 68 proto-BD candidates that fulfill criteria of Sec.~\ref{sec:mainprops-requirements}}
\label{tab:continued1}
\begin{center}
{\small
\begin{tabular}{llcccccc cccll}
\noalign{\smallskip}
\hline\noalign{\smallskip}
&&Dist.
&&
&$L_\mathrm{int}$
&$T_\mathrm{bol}$
&$M_\mathrm{env}$\supb
&$L_\mathrm{bol}$
&
&
\\
Source\supa
&Region
&(pc)
&R.A.
&Dec.
&(\lo)
&(K)
&(\mo)
&(\lo)
&Class
&Type
&Refs.\supc
&ALMA data\supc
\\
\noalign{\smallskip}
\hline\noalign{\smallskip}
J162145         &Ophiuchus  &138    &16:21:45.1 &$-$23:42:31    &$0.095\pm0.004$  &138  &0.0015    &0.105  &I          &VeLLO      &K16         &2013.1.00157, C17\\
J162648         &Ophiuchus  &138    &16:26:48.4 &$-$24:28:38    &$0.047\pm0.002$  &587  &0.013     &0.212  &Flat        &VeLLO      &A94,K16   &2015.1.00741, E21\\
I16253$-$2429   &Ophiuchus  &138    &16:28:21.6 &$-$24:36:23    &$0.129\pm0.005$  &37   &0.076     &0.237  &0           &protoBD    &D08,K16,H16 &2015.1.00741, H19b, E21\\
J163136         &Ophiuchus  &144    &16:31:36.8 &$-$24:04:20    &$0.057\pm0.006$  &32   &0.053     &0.656  &0/I         &protoBD    &RB21       &2015.1.00741, E21\\
ISO-Oph200      &Ophiuchus  &144    &16:31:43.8 &$-$24:55:24    &$0.076\pm0.007$  &329  &0.059    &0.600  &I          &protoBD    &RM21         &2015.1.00741, RM21, E21\\
J163152         &Ophiuchus  &144    &16:31:52.3 &$-$24:55:36    &$0.060\pm0.007$  &122  &0.100    &0.352  &0/I        &protoBD    &RB21,E09     &2015.1.00741, E21\\
\hline\noalign{\smallskip}
J171941         & Pipe Nebula & 130 &17:19:41.2 &$-$26:55:31    &$0.018\pm0.015$ & 175  &0.0037          &0.033 & I          & VeLLO     &F09, this work        &$-$\\
\hline\noalign{\smallskip}
J180439	        &Aqu-Rift   &436    &18:04:39.9 &$-$04:01:22    &$0.028\pm0.005$  &69\supd   &0.0007    &0.021\supd  &0/I        &VeLLO	    &K16          &$-$\\
J180449         &Aqu-Rift   &436    &18:04:49.3 &$-$04:36:39    &$0.128\pm0.010$  &474\supd  &0.007    &0.493\supd  &I          &VeLLO      &K16          &2019.1.00218, A22\\
J180941	        &Aqu-Rift   &436    &18:09:41.9 &$-$03:31:26    &$0.048\pm0.009$  &121\supd  &0.0007    &0.116\supd  &I          &VeLLO      &K16         &$-$\\
\hline\noalign{\smallskip}
MHO3257         & Aqu-Serp  & 436   &18:29:40.2 & +00:15:13     & 0.075$\pm$0.009 & 372 &0.027        & 0.212 &  I & Proto-BD & R18             &$-$\\
J182854         & Aqu-Serp  & 436   &18:28:54.9 & +00:18:32     & 0.116$\pm$0.010 & 49  &0.046        & 0.431 &  I & Proto-BD & R18             &2015.1.00310, F19\\
J182925         &Aqu-Serp   &436    &18:29:25.1 &$-$01:47:37    &$0.107\pm0.008$  &391  &0.012     &0.250  &Flat        &VeLLO      &K16        &$-$\\
J182933	        &Aqu-Serp   &436    &18:29:33.6 &$-$01:45:10    &$0.069\pm0.008$  &648  &0.150    &0.282  &Flat       &VeLLO	    &K16       &$-$\\
J182952         &Aqu-Serp   &436    &18:29:52.9 &$-$01:58:05    &$0.122\pm0.007$  &33   &0.034     &0.200  &0           &VeLLO      &K16        &2015.1.00223\\
J183047         &Aqu-Serp   &436    &18:30:47.6 &$-$02:43:56    &$0.085\pm0.014$  &82\supd   &0.0017    &0.059\supd  &0/I        &VeLLO      &K16         &$-$\\
J183245         &Aqu-Serp   &436    &18:32:45.6 &$-$02:46:57    &$0.069\pm0.009$  &186\supd  &0.022     &0.098\supd  &I           &VeLLO      &K16        &$-$\\
\hline\noalign{\smallskip}
L673-7-IRS      &L673   &300    &19:21:34.8 &$+$11:21:23    &$0.043\pm0.027$  &26   &0.100    &0.097  &0          &VeLLO      &K08,D10,K19      &2016.1.00039\\
\hline\noalign{\smallskip}
J210227         &Cepheus    &300    &21:02:27.3 &$+$67:54:18    &$0.090\pm0.005$  &100  &0.010     &0.161  &I           &VeLLO      &K16        &$-$\\
J210340         &Cepheus    &300    &21:03:40.4 &$+$68:26:31    &$0.064\pm0.005$  &181\supd  &0.010     &0.129\supd  &I           &VeLLO      &K16        &$-$\\
J215607         &Cepheus    &300    &21:56:07.3 &$+$76:42:29    &$0.023\pm0.002$  &458  &0.010     &0.053  &Flat        &VeLLO      &K16        &$-$\\
\hline
\end{tabular}
\begin{list}{}{} 
\item[$^\mathrm{a}$] etailed information about specific candidates (some of them controversial) can be found in \citet{PerezGarcia2024_SUCANES}. Here we only list the sub-sample of SUCANES that fulfill the criteria given in Sec.~\ref{sec:mainprops-requirements}.
\item[$^\mathrm{b}$] Envelope masses are taken from different papers of the literature as reported in \citet[]{PerezGarcia2024_SUCANES}. In the case of updated distances to the sources, these masses have also been updated \citep[see][]{PerezGarcia2024_SUCANES}. 
For those objects that do not have $\Menv$ estimated in the literature but do have a submillimetre flux, we calculated $\Menv$ in this work, assuming a dust temperature of 15 K, and a dust opacity per mass of dust and gas of 0.19~cm$^2$\,g$^{-1}$ at 250~\mum\ as in \citet{Kim2016_VeLLOs}, or 0.0175~cm$^2$\,g$^{-1}$ at 850~\mum. These objects are: J041740, J041828, J041836, J041938, J042019 (flux density at 250~\mum\ from \citealt{Morata2015_Jets_ProtoBD}), [GKH94]41 and L1YSO5 (flux density at 250~\mum\ measured in this work within an aperture of about $23\times23$~arcsec$^2$, using the {\it Herschel} images; the fluxes are $\sim0.5$ Jy, and $\sim0.7$ Jy, respectively), J171941 (flux density at 250~\mum\ is $0.49\pm0.08$~Jy, obtained from the Herschel/SPIRE Point Source Catalog: https://www.cosmos.esa.int/web/herschel/spire-point-source-catalogue), and J162648, which has no 250~\mum\ data. For this last object, $\Menv$ was estimated from SCUBA-2 data at 850~\mum, assuming a dust temperature of 10\,K according to \cite{Kim2016_VeLLOs}. 
\item[$^\mathrm{b}$] References and ALMA project number used in Fig.~\ref{fig:alma} and Fig.~\ref{fig:alma-nondetections}: 
A08: \cite{Andrews2008_I04158};
A16: \cite{Ansdell2016_Lupus-disks};
A94: \cite{Andre1994};
A99: \cite{Andre1999_Portostar_Taurus};
A22: \cite{Anderson2022_Serpens-disks};
B09: \cite{Barrado2009};
C16: \citet[]{Carney2016}; 
C17: \cite{Cox2017_Oph-disks};
D16: \cite{DangDuc2016_VeLLO_Taurus};
D08: \cite{Dunham2008_c2dVeLLOs};
D10: \cite{Dunham2010_VeLLO_L673_7};
dG16: \cite{deGregorio2016_PreProtoBD_ChaII};
E09:\cite{Enoch2009_PerSerOph};
E21: \cite{Encalada2021_Oph-disks};
F09: \cite{Forbrich2009ApJ_Pipe};
F19: \cite{Francis2019_ALMA-CARMA};
H16: \cite{Hsieh2016_Binary_ProtoBD_Jet_IRAS16253};
H19: \cite{Hsieh2019_Perseus-disks};
H19b: \cite{Hsieh2019_VeLLO_IRAS16253_ALMA};
K08: \cite{Kauffmann2008_MAMBO_SpitzerCores};
K16: \cite{Kim2016_VeLLOs}; 
K19: \cite{Kim2019_CO_Outflow_VeLLO}; 
M15: \cite{Morata2015_Jets_ProtoBD}; 
M17: \cite{Mowat2017_SCUBA2_LupusI};
P12: \cite{Palau2012_PreBD_Cores_Taurus};
P14: \cite{Palau2014};
P22: \cite{Palau2022_J041757};
PB14: \cite{PhanBao2014_Submm_ProtoBD_J041757};
R18: \cite{Riaz2018_chem-OphSer};
Rag21: \cite{Ragusa2021_I04158};
RB21: \cite{Riaz2021_AccretionOutflow_ProtoBD};
RM21: \cite{Riaz2021_Structure_ProtoBD};
S21: \cite{SantamariaMiranda2021_ALMA_Lupus_ProtoBD};
Tob20: \cite{Tobin2020_VANDAM-Orion};
Tok20: \cite{Tokuda2020_Taurus-disks};
vT22: \cite{vanTerwisga2022_Orion-disks};
V20: \cite{Villenave2020_edgeon-disks};
V21: \cite{Vazzano2021_outflows};
W04: \cite{White2004};
Y21: \cite{Yang2021_Perseus-PEACHES}.
\item[$^\mathrm{d}$] For these objects, $\Tbol$ and $\Lbol$ have been obtained using detections in two photometric tables only, instead of three as required in SUCANES \citep{PerezGarcia2024_SUCANES}.
\end{list}
}
\end{center}

\end{table}
\end{landscape}
\pagestyle{plain}


\setcounter{table}{3}
\begin{table*}[htbp]
\caption{26 pre-BD candidates from SUCANES database}
\label{tab:preBDs}
{\small
\begin{tabular}{llccccccc}
\hline
&&
  \multicolumn{1}{c}{Dist.} &
  \multicolumn{1}{c}{RA} &
  \multicolumn{1}{c}{DEC} &
  \multicolumn{1}{c}{$L_{\rm int}$\supa }&
  \multicolumn{1}{c}{$M_{\rm env}$} \\
  Source & Region & (pc) & (h m s) & (d m s)  &  (\lo) & (\mo) & Type & Refs.\supb \\ 
\hline
J041757-NE    &Taurus       & 140   & 04:18:00.30 & +27:41:36.3   & $-$       & 0.020   & Pre-BD         & P12   \\
\hline
B30-LB10      & Barnard30   & 400   & 05:31:09.29 & +12:11:08.8   & $-$       & 0.044   & Pre-BD         & H17   \\
B30-LB31      & Barnard30   & 400   & 05:31:15.32 & +12:03:38.2   & $-$       & 0.078   & Pre-BD         & H17   \\
B30-LB08      & Barnard30   & 400   & 05:31:22.97 & +12:11:34.7   & $<0.1$    & 0.101   & VeLLO/Pre-BD   & H17   \\
\hline
ChaII-APEX-N  & Cha II      & 198   & 12:51:55.8 & $-$77:08:43    & $-$       & 0.023   & Pre-BD         & dG16  \\
ChaII-APEX-A  & Cha II      & 198   & 12:53:30.5 & $-$77:11:07    & $-$       & 0.020   & Pre-BD         & dG16  \\
ChaII-APEX-J  & Cha II      & 198   & 12:53:44.8 & $-$77:04:43    & $-$       & 0.029   & Pre-BD         & dG16  \\
ChaII-APEX-G  & Cha II      & 198   & 12:54:12.3 & $-$77:06:47    & $-$       & 0.098   & Pre-BD         & dG16  \\  
ChaII-APEX-B  & Cha II      & 198   & 12:54:17.3 & $-$77:10:23    & $-$       & 0.020   & Pre-BD         & dG16  \\
ChaII-APEX-F  & Cha II      & 198   & 12:54:21.9 & $-$77:06:59    & $-$       & 0.026   & Pre-BD         & dG16  \\
ChaII-APEX-I  & Cha II      & 198   & 12:54:27.7 & $-$77:04:27    & $-$       & 0.023   & Pre-BD         & dG16  \\
ChaII-APEX-H  & Cha II      & 198   & 12:54:40.8 & $-$77:04:35    & $-$       & 0.034   & Pre-BD         & dG16  \\
ChaII-APEX-C  & Cha II      & 198   & 12:56:03.3 & $-$77:11:43    & $-$       & 0.027   & Pre-BD         & dG16  \\
\hline
ALMA-J153702  & Lupus I     & 94    & 15:37:02.65 & $-$33:19:24.9 & $<0.00075$& 0.00085 & Pre-BD         & S21   \\
ALMA-J154228  & Lupus I     & 153   & 15:42:28.67 & $-$33:42:30.2 & $<0.002$  & 0.039   & Pre-BD         & S21   \\
ALMA-J154456  & Lupus I     & 153   & 15:44:56.52 & $-$34:25:33.0 & $<0.003$  & 0.029   & Pre-BD         & S21   \\
ALMA-J154458  & Lupus I     & 153   & 15:44:58.06 & $-$34:25:28.5 & $<0.002$  & 0.017   & Pre-BD         & S21   \\
ALMA-J154506  & Lupus I     & 153   & 15:45:06.52 & $-$34:43:26.2 & $<0.002$  & 0.0485  & Pre-BD         & S21   \\
ALMA-J154634  & Lupus I     & 153   & 15:46:34.17 & $-$34:33:01.9 & $<0.002$  & 0.171   & Pre-BD         & S21   \\
ALMA-J160658  & Lupus III   & 155   & 16:06:58.60 & $-$39:04:07.9 & $<0.002$  & 0.00095 & Pre-BD         & S21   \\
ALMA-J160804  & Lupus III   & 155   & 16:08:04.17 & $-$39:04:52.8 & $<0.002$  & 0.002   & Pre-BD         & S21   \\
ALMA-J160920  & Lupus III   & 155   & 16:09:20.09 & $-$38:45:15.9 & $<0.002$  & 0.002   & Pre-BD         & S21   \\
ALMA-J160920  & Lupus III   & 155   & 16:09:20.17 & $-$38:44:56.4 & $<0.002$  & 0.002   & Pre-BD         & S21   \\
ALMA-J160932  & Lupus III   & 155   & 16:09:32.17 & $-$39:08:32.3 & $<0.003$  & 0.087   & Pre-BD         & S21   \\
ALMA-J161030  & Lupus III   & 155   & 16:10:30.27 & $-$38:31:54.5 & $<0.003$  & 0.001   & Pre-BD         & S21   \\
\hline
OphB-11       & Ophiuchus   & 138   & 16:27:13.96 & $-$24:28:29.3 & $<0.006$  & 0.009   & Pre-BD         & A12   \\
\hline\end{tabular}
\begin{list}{}{}
\item[$^\mathrm{a}$] `$-$' corresponds to no data or no measurement of the flux at 70~\mum\ provided in the reference. 
\item[$^\mathrm{b}$] Refs: 
A12: \cite{Andre2012_PreBD};
dG16: \cite{deGregorio2016_PreProtoBD_ChaII};
H17: \cite{Huelamo2017_B30};
P12: \cite{Palau2012_PreBD_Cores_Taurus};
S21: \cite{SantamariaMiranda2021_ALMA_Lupus_ProtoBD}.
\end{list}
}

\end{table*}

\setcounter{table}{4}
\begin{landscape}
\begin{table}[htbp]

\caption{Properties of nearby molecular clouds used in this work}
\label{tab:clouds1}
\begin{center}
{\small
\begin{tabular}{lccc rrrr cccc}
\noalign{\smallskip}
\hline\noalign{\smallskip}
&&Distance\supa
&Area\supb
&
&Mass\supc
&Mass$_\mathrm{Av7}$\supc
&SFR\supd
&YSO surf dens\supe
&log$F_\mathrm{FUV}$\supe
&Age\,\supf
\\
Source
&Project
&(pc)
&(pc$^2$)
&$\Nyso$\supb
&(\mo)
&(\mo)
&(10$^{-6}$\mo\,yr$^{-1}$)
&(pc$^{-2}$)
&(G$_0$)
&(Myr)
&Refs.\supf
\\
\noalign{\smallskip}
\hline\noalign{\smallskip}
Lupus I	    &c2d	&$150\pm20$	   &$8.9\pm2.4$    &13     &787	       &75	    &3      &1.47   &0.7   &1.6	&E09\\
Ophiucus	&c2d	&$125\pm25$	   &$30\pm12$      &290    &14165	   &1296    &73     &9.80   &3.3   &9.3	&E09\\
Perseus	    &c2d	&$293\pm22$	   &$98\pm29$      &385    &18438	   &1880	&96    &3.91   &3.0   &10.4&E09\\
Ser-Aquila	&GBS	&$436\pm00$	   &$196\pm15$     &1440   &67500	   &14700   &360    &7.35   &3.3   &21.	&P23\\
Taurus	    &c2d	&$137\pm10$	   &$252\pm50$     &267    &14964	   &1766	&67     &1.06   &1.3   &10.1&R10\\
OrionA-B	&$-$	&$420\pm42$	   &$432\pm86$     &3479   &139542	   &20982	&870    &8.05   &3.5   &22.	&M12\\
Lupus III   &c2d	&$200\pm20$	   &$15.4\pm3.1$   &68     &2157	   &163	    &17     &4.41   &0.7   &2.5	&E09\\
California  &GBS	&$450\pm23$	   &$646\pm13$     &173    &99930	   &3199	&43     &0.27   &$-$   &2.2	&B14\\
\noalign{\smallskip}
\hline\noalign{\smallskip}
\hline
\end{tabular}
\begin{list}{}{}
\item[$^\mathrm{a}$] The distance to California is taken as $450\pm23$~pc as this is the average distance found by \cite{Zucker2020_Distances_Clouds}. We note that \cite{AlvarezGutierrez2021_California} report a distance of $511\pm17$~pc, but this is mainly towards the L1482 cloud, which lies at the eastern side of the entire cloud, while \cite{Zucker2019_Distances_Clouds} report a distance gradient from the east (further away) to the west (closer), and the proto-BD candidates studied lie outside L1482.
\item[$^\mathrm{b}$] Areas were adopted from \cite{Heiderman2010_Nyso} or estimated from \cite{Carpenter2000} and assuming 20\% of uncertainty. 
$\Nyso$ includes young stellar objects of all evolutionary stages, from Class 0 to Class III. For most of the clouds, this number corresponds to $N_\mathrm{YSO,tot}$ reported by \cite{Heiderman2010_Nyso}. For Taurus, OrionA-B, and California, $\Nyso$ is taken from \citet{Rebull2010_Taurus-Spitzer}, \citet{Megeath2012_Orion-Spitzer} and \citet{Kim2016_VeLLOs}, respectively. For the particular case of Taurus, $\Nyso$ is obtained by summing the number of previously identified young stellar objects by \citet[][see their Table~3]{Rebull2010_Taurus-Spitzer}, 215, plus the number of new candidate young stellar objects including only the confirmed, probable and possible, 52, reported in Table~8 of the same paper.
\item[$^\mathrm{c}$] Masses are taken from \cite{Lada2010_SFRs}, with the exception of Serpens-Aquila. In this case, the mass above the $A_\mathrm{V}$=7~mag threshold (which corresponds to $N$(H$_2$)=$6.9\times10^{21}$~cm$^{-2}$ according to the relation of \citealt{Bohlin1978} and \citealt{Konyves2015_Cores_Aquila}) was estimated using the {\it Herschel} column density map provided by \citet[]{Andre2010_CMF} and \citet{Konyves2015_Cores_Aquila}, and following \cite{Konyves2015_Cores_Aquila} and \citet{Shimoikura2020_Aquila}, assuming a distance of 436~pc. Our mass is consistent with the mass from \citet[]{Schneider2022_MassiveMolecularClouds}. The total mass for this cloud was taken from \citet[][of 24000~\mo]{Konyves2015_Cores_Aquila}, corrected to our adopted distance.
\item[$^\mathrm{d}$] SFRs are calculated following equation (1) of \cite{Lada2010_SFRs}, $\mathrm{SFR}=0.25\,\Nyso\times10^{-6}$~\mo\,yr$^{-1}$, and using $\Nyso$ given in this table.
%
\item[$^\mathrm{e}$] The YSO surface density is calculated using $\Nyso$ and the area given in this table. The logarithm of the FUV flux is taken from \cite{Gupta2024_IC1396} and \cite{Xia2022_fluxuv}.
\item[$^\mathrm{f}$] The cloud age is estimated following \citet{VazquezSemadeni2018_cloudages}, who model the evolution of molecular clouds undergoing global and hierarchical collapse. The model inputs are the total mass of the cloud and the mass fraction in dense gas (taken as the mass above $A_\mathrm{V}$=7~mag as in \citealt{Lada2010_SFRs}, over the total mass). Since Serpens-Aquila is not included in the work of \citet{VazquezSemadeni2018_cloudages}, we estimated the age using our measured values for the mass and mass above a threshold.
\item[$^\mathrm{g}$] 
E09: \citet[]{Evans2009_Lifetimes};
R10: \cite{Rebull2010_Taurus-Spitzer};
M12: \cite{Megeath2012_Orion-Spitzer};
B14: \cite{BroekhovenFiene2014_California}; 
P23: \cite{Pokhrel2023_Aquila}.
\end{list}
}
\end{center}

\end{table}
\end{landscape}
\pagestyle{plain}

\setcounter{table}{5}
\begin{landscape}
\begin{table}[htbp]
\caption{Numbers of protostars and proto-BD candidates in nearby molecular clouds}
\label{tab:clouds2}
\begin{center}
{\small
\begin{tabular}{lccc cccc cc}
\noalign{\smallskip}
\hline\noalign{\smallskip}
&&&&&&Proto-
&&Proto-
\\
Source
&$\Nproto$\supa
&$N_\mathrm{proto-Lrestrict}$\supa
&$\Nbd$\supb
&$N_\mathrm{proBD-c004}$\supb
&$N_\mathrm{expected}$\supc
&star-to-BD\supc
&$N_\mathrm{proBD-c0004}$\supb
&star-to-BD0004\supc
&Refs.\supd
\\
\noalign{\smallskip}
\hline\noalign{\smallskip}
Lupus I	    &2	    &0	   &4  &0  &0     &$-$  &4    &0.25   &E09\\
Ophiucus	&57	    &13	   &3  &3  &2.6   &4.3  &3    &4.3    &E09\\
Perseus	    &111    &32	   &3  &2  &6.4   &16   &$-$  &$-$    &E09\\
Ser-Aquila	&171	&47	   &5  &5  &9.4   &9.4  &$-$  &$-$    &P23\\
Taurus	    &27	    &15	   &9  &2  &3.0   &7.5  &8    &1.9    &R10\\
OrionA-B	&400	&167   &8  &6  &33    &27.8 &$-$  &$-$    &M12\\
Lupus III   &2	    &1	   &1  &0  &0.2   &$-$  &1    &1.0    &E09\\
California  &37	    &19	   &4  &2  &3.8   &9.5  &$-$  &$-$    &B14\\
\noalign{\smallskip}
\hline\noalign{\smallskip}
\hline
\end{tabular}
\begin{list}{}{}
\item[$^\mathrm{a}$] $\Nproto$ corresponds to the number of YSOs for which $\Tbol\leq650$~K, thus including both Class 0 and Class I. For the catalogs where $\Tbol$ is not given, we included those objects classified as Class I. 
$N_\mathrm{proto-Lrestrict}$ is obtained by restricting $\Nproto$ to those objects whose $\Lint$ is in the range 0.13--1~\lo. When no $\Lint$ is given, we estimated $\Lint$ from the flux at 70~\mum\ and following eq.~\ref{eq:LintFlux70mic}.
\item[$^\mathrm{b}$] $\Nbd$ corresponds to the number of proto-BD listed in Table~\ref{tab:protoBDs} that belong to the catalog of \cite{Kim2016_VeLLOs}. $N_\mathrm{proBD-c004}$ corresponds to $\Nbd$ restricted only to those objects with $\Lint \geq 0.04$~\lo, which is the minimum $\Lint$ detectable at the distance of the most distant cloud. Thus, $N_\mathrm{proBD-c004}$ is complete down to $0.04$~\lo. $N_\mathrm{proBD-c0004}$ corresponds to $\Nbd$ complete down to $0.004$~\lo, which is only possible to obtain for the most nearby clouds Ophiucus, Taurus, Lupus I and Lupus III.
\item[$^\mathrm{c}$] $N_\mathrm{expected}$ is the number of expected proto-BDs, given $\Nproto$, and is calculated as $N_\mathrm{expected}=N_\mathrm{proto-Lrestrict}/5$. In this expression, we have used $N_\mathrm{proto-Lrestrict}$ to make it comparable to the analysis of \cite{Muzic2019_NGC2244} and \cite{AlmendrosAbad2023_NGC2264}, who calculate the star-to-BD ratio by restricting the stars to the range of 0.075--1~\mo. Similarly, the factor of 5 applied is taken from \cite{Muzic2019_NGC2244} and \cite{AlmendrosAbad2023_NGC2264}, which corresponds to the upper-end of star-to-BD ranges found in these works for more evolved objects. Proto(star-to-BD) is estimated as $N_\mathrm{proto-Lrestrict}/N_\mathrm{proBD-c004}$, and is comparable to the star-to-BD ratio measured by \cite{Muzic2019_NGC2244} and  \cite{AlmendrosAbad2023_NGC2264}. Proto(star-to-BD0004) is estimated as $N_\mathrm{proto-Lrestrict}/N_\mathrm{proBD-c0004}$, and is presumably sensitive to the planetary-mass regime. 
\item[$^\mathrm{d}$] References: $\Nproto$ has been obtained from 
E09: \citet[]{Evans2009_Lifetimes};
R10: \cite{Rebull2010_Taurus-Spitzer};
M12: \cite{Megeath2012_Orion-Spitzer};
B14: \cite{BroekhovenFiene2014_California}; 
P23: \cite{Pokhrel2023_Aquila}.
\end{list}
}
\end{center}

\end{table}
\end{landscape}
\pagestyle{plain}


\appendix


\section{ALMA archive images of the proto-BD candidates of Table~3 that have not been detected with ALMA}\label{app:alma}

In Sec.~\ref{sec:mainprops-sample}, a sample of 68 proto-BD candidates was presented, and the ALMA archive was searched to explore the properties of the millimetre continuum emission in each case. Out of the 68 candidates, 33 were found to have ALMA data available in the archive, and only four out of the 34 are non-detections. We present in this Appendix these four ALMA non-detections.

\begin{figure}
\begin{tabular}[b]{cc}
    \epsfig{file=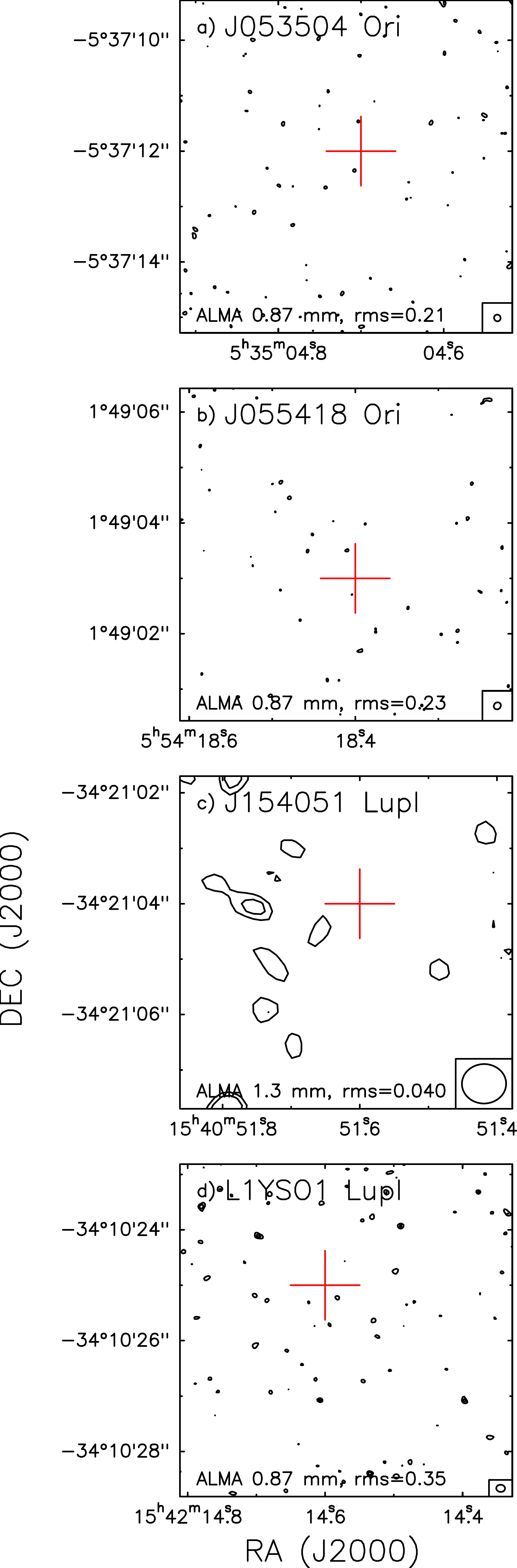, width=7cm,angle=0}\\
\end{tabular}
\caption{Archive ALMA emission of the four proto-BD candidates of Table~\ref{tab:protoBDs} that are non-detections. In all the panels, the field of view is $6''\times6''$, the beam is shown in the bottom-right corner, and the contours correspond to the 1.3 mm or 0.87~mm continuum emission (3, and 4 $\times$ the rms noise). The rms noise of each image is indicated at the bottom of each panel and is given in m\jpb. In all the panels, the red plus sign corresponds to the coordinates of the proto-BD candidate as provided in the SUCANES database \citep{PerezGarcia2024_SUCANES}. The ALMA project codes along with the corresponding references are listed in Table~\ref{tab:protoBDs}.
}
\label{fig:alma-nondetections}
\end{figure}

\section{Data used for Fig.~\ref{fig:correls}}\label{app:tables}

In Sec.~\ref{sec:mainprops-comparison2protostellar} the well-known correlations that hold for protostellar objects were explored in the substellar regime. In this Appendix we provide the tables with the data used to make this comparison.

\begin{table*}[htbp]
\caption{Data used for the $\Minfrate$ vs $\Lbol$ relation of Fig.~\ref{fig:correls}a}
\label{tab:correlMinfrateLbol}
{\small
\begin{tabular}{lrrrrl}
\hline
&
  \multicolumn{1}{c}{$\Lbol$\supa} &
  \multicolumn{1}{c}{error} &
  \multicolumn{1}{c}{$\Minfrate$} &
  \multicolumn{1}{c}{error} \\
  \multicolumn{1}{l}{Source} &
  \multicolumn{1}{c}{(\lo)} &
  \multicolumn{1}{c}{(\lo)} &
  \multicolumn{1}{c}{(\mo\,yr$^{-1}$)} &
  \multicolumn{1}{c}{(\mo\,yr$^{-1}$)} &
  \multicolumn{1}{l}{Refs.} \\
\hline
J033032 & 0.106 & 0.014 & 3.1E-6 & 5.9E-6       &\citet[][Tables 1 and 3]{Kim2021_infallVeLLO}\\
L1521F-IRS & 0.038 & 0.004 & 1.7E-6 & 2.9E-6    &\citet[][Tables 1 and 3]{Kim2021_infallVeLLO}\\
J0430149 & 0.013 & 0.002 & 1.1E-6 & 2.5E-6      &\citet[][Tables 1 and 3]{Kim2021_infallVeLLO}\\
L328-IRS & 0.172 & 0.046 & 7.2E-7 & 2.9E-7      &\citet[][Tables 1 and 3]{Kim2021_infallVeLLO}\\
J1830144 & 0.25 & 0.095 & 8.1E-6 & 2.7E-6       &\citet[][Tables 1 and 3]{Kim2021_infallVeLLO}\\
J1830156 & 0.735 & 0.278 & 5.5E-5 & 2.3E-5      &\citet[][Tables 1 and 3]{Kim2021_infallVeLLO}\\
J1830162 & 0.519 & 0.194 & 2.0E-5 & 7.2E-5      &\citet[][Tables 1 and 3]{Kim2021_infallVeLLO}\\
J1832424 & 0.677 & 0.253 & 4.3E-6 & 1.7E-6      &\citet[][Tables 1 and 3]{Kim2021_infallVeLLO}\\
L1148-IRS & 0.127 & 0.014 & 2.3E-6 & 1.0E-6     &\citet[][Tables 1 and 3]{Kim2021_infallVeLLO}\\
J2102212 & 0.381 & 0.037 & 1.7E-5 & 2.9E-5      &\citet[][Tables 1 and 3]{Kim2021_infallVeLLO}\\
J2144570 & 0.066 & 0.015 & 1.2E-5 & 3.7E-5      &\citet[][Tables 1 and 3]{Kim2021_infallVeLLO}\\
J2229594 & 0.33 & 0.032 & 1.1E-5 & 2.8E-5       &\citet[][Tables 1 and 3]{Kim2021_infallVeLLO}\\
GF9-2 & 1.7 & 0.8 & 2.5E-5 & 1.0E-5             &\cite{Furuya2009_GF9-2}\\
G34.26+0.2 & 46000.0 & 20000.0 & 0.009 & 0.004      &\citet[][Tables 1 and 5]{Wyrowski2016_infall}\\
G327.29$-$0.6 & 82000.0 & 41000.0 & 0.004 & 0.002     &\citet[][Tables 1 and 5]{Wyrowski2016_infall}\\
G351.58$-$0.4 & 240000.0 & 120000.0 & 0.016 & 0.008   &\citet[][Tables 1 and 5]{Wyrowski2016_infall}\\
G23.21$-$0.3 & 13000.0 & 6000.0 & 0.008 & 0.004       &\citet[][Tables 1 and 5]{Wyrowski2016_infall}\\
G35.20$-$0.7 & 25000.0 & 12000.0 & 3.0E-4 & 1.5E-4    &\citet[][Tables 1 and 5]{Wyrowski2016_infall}\\
G34.41+0.2 & 4800.0 & 2400.0 & 7.0E-4 & 3.0E-4      &\citet[][Tables 1 and 5]{Wyrowski2016_infall}\\
L1521F & 0.037 & 0.001 & 2.7E-5 & 1.3E-5    &\cite{Keown2016ApJ_L1521F-infall}\\
I09018 & 52000.0 & 20000.0 & 0.0475 & 0.02 &\citet[][Table 1]{Yue2021_infall}\\
I13134 & 40000.0 & 20000.0 & 0.0135 & 0.005    &\citet[][Table 1]{Yue2021_infall}\\
I14382 & 160000.0 & 80000.0 & 0.0149 & 0.007   &\citet[][Table 1]{Yue2021_infall}\\
I14498 & 25000.0 & 12000.0 & 0.0156 & 0.007    &\citet[][Table 1]{Yue2021_infall}\\
I15520 & 130000.0 & 60000.0 & 0.0314 & 0.015   &\citet[][Table 1]{Yue2021_infall}\\
I15596 & 320000.0 & 150000.0 & 0.0448 & 0.022  &\citet[][Table 1]{Yue2021_infall}\\
I16060 & 630000.0 & 310000.0 & 0.0767 & 0.035  &\citet[][Table 1]{Yue2021_infall}\\
I16071 & 63000.0 & 31000.0 & 0.0583 & 0.029    &\citet[][Table 1]{Yue2021_infall}\\
I16076 & 200000.0 & 100000.0 & 0.0537 & 0.027  &\citet[][Table 1]{Yue2021_infall}\\
I16272 & 20000.0 & 10000.0 & 0.0149 & 0.007    &\citet[][Table 1]{Yue2021_infall}\\
I16348 & 250000.0 & 120000.0 & 0.0581 & 0.029  &\citet[][Table 1]{Yue2021_infall}\\
I16351 & 79000.0 & 40000.0 & 0.0212 & 0.01 &\citet[][Table 1]{Yue2021_infall}\\
I17160 & 1000000.0 & 500000.0 & 0.0379 & 0.019 &\citet[][Table 1]{Yue2021_infall}\\
I17204 & 16000.0 & 8000.0 & 0.0049 & 0.002 &\citet[][Table 1]{Yue2021_infall}\\
I17220 & 500000.0 & 200000.0 & 0.1194 & 0.05   &\citet[][Table 1]{Yue2021_infall}\\
I18056 & 500000.0 & 200000.0 & 0.1022 & 0.05   &\citet[][Table 1]{Yue2021_infall}\\
I18182 & 20000.0 & 10000.0 & 0.0108 & 0.005    &\citet[][Table 1]{Yue2021_infall}\\
I18507 & 63000.0 & 30000.0 & 0.0533 & 0.02 &\citet[][Table 1]{Yue2021_infall}\\
AGAL008.206+00.191 & 127.6 & 60.0 & 0.0021 & 0.001  &\citet[][Table 3]{Pillai2023_infall}, \citet[][Table 5]{Urquhart2018_ATLASGAL}\\
AGAL008.706$-$00.414 & 1452.0 & 700.0 & 0.0147 & 0.007    &\citet[][Table 3]{Pillai2023_infall}, \citet[][Table 5]{Urquhart2018_ATLASGAL}\\
AGAL022.376+00.447 & 263.0 & 130.0 & 0.0068 & 0.0034    &\citet[][Table 3]{Pillai2023_infall}, \citet[][Table 5]{Urquhart2018_ATLASGAL}\\
AGAL024.373$-$00.159 & 160.0 & 80.0 & 0.0101 & 0.005  &\citet[][Table 3]{Pillai2023_infall}, \citet[][Table 5]{Urquhart2018_ATLASGAL}\\
BG009.212 & 3981.0 & 2000.0 & 5.8E-4 & 2.0E-4    &\citet[][Tables 1 and 5]{Xu2023_infall}\\
BG012.889 & 15849.0 & 7000.0 & 0.00447 & 0.002   &\citet[][Tables 1 and 5]{Xu2023_infall}\\
BG014.606 & 6310.0 & 3100.0 & 1.5E-4 & 7.0E-5    &\citet[][Tables 1 and 5]{Xu2023_infall}\\
BG025.400 & 316228.0 & 150000.0 & 0.00141 & 7.0E-4   &\citet[][Tables 1 and 5]{Xu2023_infall}\\
BG028.565 & 200.0 & 100.0 & 0.0321 & 0.015   &\citet[][Tables 1 and 5]{Xu2023_infall}\\
BG030.719 & 50119.0 & 25000.0 & 0.00791 & 0.004  &\citet[][Tables 1 and 5]{Xu2023_infall}\\
BG033.740 & 3162.0 & 1500.0 & 0.00458 & 0.0022   &\citet[][Tables 1 and 5]{Xu2023_infall}\\
BG081.721 & 3981.0 & 2000.0 & 0.0215 & 0.011 &\citet[][Tables 1 and 5]{Xu2023_infall}\\
BG133.748 & 6310.0 & 3100.0 & 0.00426 & 0.0021   &\citet[][Tables 1 and 5]{Xu2023_infall}\\
BG133.949 & 1000.0 & 500.0 & 0.00548 & 0.0022    &\citet[][Tables 1 and 5]{Xu2023_infall}\\
\hline
J032832/Per-Bolo25\supb & 0.061 & 0.008 & 6.5E-6 &1.4E-6    &\citet[][Table 3]{Kim2021_infallVeLLO}, this work (Table~\ref{tab:protoBDs})\\
J041840\supb & 0.004 & 0.001 & 3.5E-7 &2.5E-7              &\citet[][Table 3]{Kim2021_infallVeLLO}, this work (Table~\ref{tab:protoBDs})\\
J210227\supb & 0.090 & 0.017 & 1.0E-5 &1.5E-5              &\citet[][Table 3]{Kim2021_infallVeLLO}, this work (Table~\ref{tab:protoBDs})\\
IRAM\,04191\supb & 0.042 & 0.001 & 3.0E-6 &1.5E-6          &\cite{Belloche2002_IRAM04191}, this work (Table~\ref{tab:protoBDs})\\
\hline
\end{tabular}
\begin{list}{}{}
\item[$^\mathrm{a}$] $\Lbol$ for the proto-BD candidates is taken as $\Lint$ because the least luminous objects are most affected by the ISRF and the accretion/ejection processes must be driven by the mass of the central object, more closely related to $\Lint$ than to $\Lbol$.
\item[$^\mathrm{b}$] Proto-BD candidate included in Table~\ref{tab:protoBDs}. This object is not included to perform the fit shown in Fig.~\ref{fig:correls}a.
\end{list}
}

\end{table*}

\begin{table*}[htbp]
\caption{Data used for the $\Lcm$ vs $\Lbol$ relation of Fig.~\ref{fig:correls}b}
\label{tab:correlLcmLbol}
{\small
\begin{tabular}{lrrrl}
\hline
  &
  \multicolumn{1}{c}{D} &
  \multicolumn{1}{c}{$\Lbol$\supa} &
  \multicolumn{1}{c}{$F_\mathrm{3.6cm}$\supb} \\
  \multicolumn{1}{l}{Source} &
  \multicolumn{1}{c}{(kpc)} &
  \multicolumn{1}{c}{(\lo)} &
  \multicolumn{1}{c}{(mJy)} &
  \multicolumn{1}{l}{Refs.} \\
\hline
  L1014IRS & 0.258 & 0.1 & 0.11     &\citet[][Table 2]{Anglada2018}, \citet{PerezGarcia2024_SUCANES}\\
  L1148-IRS & 0.33 & 0.112 & 0.08   &\citet[][Table 2]{Anglada2018}, \citet{PerezGarcia2024_SUCANES}\\
  L1521F-IRS & 0.137 & 0.037 & 0.07 &\citet[][Table 2]{Anglada2018}, \citet{PerezGarcia2024_SUCANES}\\
  HH30 & 0.14 & 0.42 & 0.042        &\citet[][Table 2]{Anglada2018}\\
  L1262 & 0.2 & 1.0 & 0.33          &\citet[][Table 2]{Anglada2018}\\
  VLA1623 & 0.16 & 1.5 & 0.6        &\citet[][Table 2]{Anglada2018}\\
  L723 & 0.3 & 2.4 & 0.4            &\citet[][Table 2]{Anglada2018}\\
  L1489 & 0.14 & 4.4 & 0.5          &\citet[][Table 2]{Anglada2018}\\
  B335 & 0.25 & 4.0 & 0.2           &\citet[][Table 2]{Anglada2018}\\
  NGC2264G & 0.8 & 5.0 & 0.3        &\citet[][Table 2]{Anglada2018}\\
  AS353A & 0.3 & 8.4 & 0.1          &\citet[][Table 2]{Anglada2018}\\
  L1448C & 0.35 & 9.0 & 0.1         &\citet[][Table 2]{Anglada2018}\\
  L1448N(A) & 0.35 & 10.0 & 0.9     &\citet[][Table 2]{Anglada2018}\\
  RNO43 & 0.4 & 12.0 & 0.5          &\citet[][Table 2]{Anglada2018}\\
  L483 & 0.2 & 14.0 & 0.31          &\citet[][Table 2]{Anglada2018}\\
  L1251B & 0.2 & 14.0 & 1.2         &\citet[][Table 2]{Anglada2018}\\
  TTAU & 0.14 & 17.0 & 5.8          &\citet[][Table 2]{Anglada2018}\\
  HH111 & 0.46 & 24.0 & 0.9         &\citet[][Table 2]{Anglada2018}\\
  IRAS16293 & 0.16 & 27.0 & 2.9     &\citet[][Table 2]{Anglada2018}\\
  L1251A & 0.3 & 27.0 & 0.47        &\citet[][Table 2]{Anglada2018}\\
  HARO4-255FIR & 0.48 & 28.0 & 0.2  &\citet[][Table 2]{Anglada2018}\\
  L1551-IRS5 & 0.14 & 20.0 & 0.8    &\citet[][Table 2]{Anglada2018}\\
  HLTAU & 0.16 & 44.0 & 0.3         &\citet[][Table 2]{Anglada2018}\\
  L1228 & 0.2 & 7.0 & 0.15          &\citet[][Table 2]{Anglada2018}\\
  PVCEP & 0.5 & 80.0 & 0.2          &\citet[][Table 2]{Anglada2018}\\
  L1641N & 0.42 & 170.0 & 0.6       &\citet[][Table 2]{Anglada2018}\\
  FIRSSE101 & 0.45 & 123.0 & 1.1    &\citet[][Table 2]{Anglada2018}\\
  HH7-11VLA3 & 0.35 & 150.0 & 0.8   &\citet[][Table 2]{Anglada2018}\\
  RE50 & 0.46 & 295.0 & 0.84        &\citet[][Table 2]{Anglada2018}\\
  SERPENS & 0.415 & 98.0 & 2.0      &\citet[][Table 2]{Anglada2018}\\
  IRAS22198 & 1.3 & 1240.0 & 0.57   &\citet[][Table 2]{Anglada2018}\\
  NGC2071-IRS3 & 0.39 & 520.0 & 2.9 &\citet[][Table 2]{Anglada2018}\\
  HH34 & 0.42 & 15.0 & 0.16         &\citet[][Table 2]{Anglada2018}\\
  AF5142CM1 & 2.14 & 10000.0 & 1.3  &\citet[][Table 2]{Anglada2018}\\
  AF5142CM2 & 2.14 & 1000.0 & 0.35  &\citet[][Table 2]{Anglada2018}\\
  CepAHW2 & 0.725 & 10000.0 & 6.9   &\citet[][Table 2]{Anglada2018}\\
  IRAS20126 & 1.7 & 13000.0 & 0.2   &\citet[][Table 2]{Anglada2018}\\
  HH80--81 & 1.7 & 20000.0 & 5.0    &\citet[][Table 2]{Anglada2018}\\
  V645Cyg & 3.5 & 40000.0 & 0.6     &\citet[][Table 2]{Anglada2018}\\
  IRAS16547 & 2.9 & 60000.0 & 8.7   &\citet[][Table 2]{Anglada2018}\\
  IRAS18566B & 6.7 & 80000.0 & 0.077    &\citet[][Table 2]{Anglada2018}\\
  IRAS04579 & 2.5 & 3910.0 & 0.15   &\citet[][Table 2]{Anglada2018}\\
  GGD14-VLA7 & 0.9 & 1000.0 & 0.18  &\citet[][Table 2]{Anglada2018}\\
  IRAS23139 & 4.8 & 20000.0 & 0.53  &\citet[][Table 2]{Anglada2018}\\
  N7538-IRS9 & 2.8 & 40000.0 & 3.8  &\citet[][Table 2]{Anglada2018}\\
  N7538-IRS9A1 & 2.8 & 40000.0 & 1.4    &\citet[][Table 2]{Anglada2018}\\
  IRAS16562 & 1.6 & 70000.0 & 21.1  &\citet[][Table 2]{Anglada2018}\\
  I18264-1152F & 3.5 & 10000.0 & 0.28   &\citet[][Table 2]{Anglada2018}\\
  G31.41 & 7.9 & 200000.0 & 0.32    &\citet[][Table 2]{Anglada2018}\\
  AF2591-VLA3 & 3.3 & 230000.0 & 1.52   &\citet[][Table 2]{Anglada2018}\\
  I18182-1433b & 4.5 & 20000.0 & 0.6    &\citet[][Table 2]{Anglada2018}\\
  I18089-1732(1)a & 3.6 & 32000.0 & 1.1 &\citet[][Table 2]{Anglada2018}\\
\hline
\end{tabular}
}

\end{table*}

\setcounter{table}{7}
\begin{table*}[htbp]
\caption{(cont.) Data used for the $\Lcm$ vs $\Lbol$ relation of Fig.~\ref{fig:correls}b}
\label{tab:correlLcmLbol}
{\small
\begin{tabular}{lrrrl}
\hline
  &
  \multicolumn{1}{c}{D} &
  \multicolumn{1}{c}{$\Lbol$\supa} &
  \multicolumn{1}{c}{$F_\mathrm{3.6cm}$\supb} \\
  \multicolumn{1}{l}{Source} &
  \multicolumn{1}{c}{(kpc)} &
  \multicolumn{1}{c}{(\lo)} &
  \multicolumn{1}{c}{(mJy)} &
  \multicolumn{1}{l}{Refs.} \\
\hline
  G28S-JVLA1 & 4.8 & 100.0 & 0.02       &\citet[][Table 2]{Anglada2018}\\
  G28N-JVLA2N & 4.8 & 1000.0 & 0.06     &\citet[][Table 2]{Anglada2018}\\
  G28N-JVLA2S & 4.8 & 1000.0 & 0.03     &\citet[][Table 2]{Anglada2018}\\
  L1287 & 0.85 & 1000.0 & 0.5           &\citet[][Table 2]{Anglada2018}\\
  DGTauB & 0.15 & 0.9 & 0.31            &\citet[][Table 2]{Anglada2018}\\
  ZCMa & 1.15 & 3000.0 & 1.74           &\citet[][Table 2]{Anglada2018}\\
  W3IRS5(d) & 1.83 & 200000.0 & 1.5     &\citet[][Table 2]{Anglada2018}\\
  YLW16A & 0.16 & 13.0 & 0.78           &\citet[][Table 2]{Anglada2018}\\
  YLW15\_VLA1 & 0.12 & 1.0 & 1.5        &\citet[][Table 2]{Anglada2018}\\
  L778 & 0.25 & 0.93 & 0.69             &\citet[][Table 2]{Anglada2018}\\
  W75N(B)VLA1 & 1.3 & 19000.0 & 4.0     &\citet[][Table 2]{Anglada2018}\\
  NGC1333\_VLA2 & 0.235 & 1.5 & 2.5     &\citet[][Table 2]{Anglada2018}\\
  NGC1333\_IRAS4A1 & 0.235 & 8.0 & 0.32 &\citet[][Table 2]{Anglada2018}\\
  NGC1333\_IRAS4B & 0.235 & 1.1 & 0.33  &\citet[][Table 2]{Anglada2018}\\
  L1551NE-A & 0.14 & 4.0 & 0.39         &\citet[][Table 2]{Anglada2018}\\
  Haro6-10\_VLA1 & 0.14 & 0.5 & 1.1     &\citet[][Table 2]{Anglada2018}\\
  L1527\_VLA1 & 0.14 & 1.9 & 1.1        &\citet[][Table 2]{Anglada2018}\\
  OMC2/3\_VLA4 & 0.414 & 40.0 & 0.83    &\citet[][Table 2]{Anglada2018}\\
  HH1-2\_VLA1 & 0.414 & 23.0 & 1.2    &\citet[][Table 2]{Anglada2018}\\
  HH1-2\_VLA3 & 0.414 & 84.0 & 0.4    &\citet[][Table 2]{Anglada2018}\\
  OMC2\_VLA11 & 0.414 & 360.0 & 2.16    &\citet[][Table 2]{Anglada2018}\\
  HH46/47 & 0.45 & 12.0 & 1.8           &\citet[][Table 2]{Anglada2018}\\
  IRAS20050 & 0.7 & 260.0 & 1.4         &\citet[][Table 2]{Anglada2018}\\
\hline
J041757\supc &0.137  &$<0.002$  &0.14      &\citet[][Table 2]{Anglada2018}, \citet{PerezGarcia2024_SUCANES}\\
J041836\supc &0.137  &$<0.003$  &0.13      &\citet[][Table 2]{Anglada2018}, \citet{PerezGarcia2024_SUCANES}\\
J041938\supc &0.137  &0.006  &0.10      &\citet[][Table 2]{Anglada2018}, \citet{PerezGarcia2024_SUCANES}\\
IRAM04191\supc &0.160 &0.042 &0.14      &\citet[][Table 2]{Anglada2018}, \citet{PerezGarcia2024_SUCANES}\\
IC348-SMM2E\supc &0.320 &0.109 &0.027   &\citet[][Table 2]{Anglada2018}, \citet{PerezGarcia2024_SUCANES}\\
\hline
\end{tabular}
\begin{list}{}{}
\item[$^\mathrm{a}$] $\Lbol$ for the proto-BD candidates is taken as $\Lint$ because the least luminous objects are most affected by the ISRF and the accretion/ejection processes must be driven by the mass of the central object, more closely related to $\Lint$ than to $\Lbol$ \citep{AMIConsortium2011_AMIobs-c2d-smallcores}.
\item[$^\mathrm{b}$] Flux density at 3.6 cm.
\item[$^\mathrm{c}$] Proto-BD candidate included in Table~\ref{tab:protoBDs}. This object is not included to perform the fit shown in Fig.~\ref{fig:correls}b.
\end{list}
}

\end{table*}

\begin{table*}[htbp]
\caption{Interferometric data used for the $\Fout$ vs $\Lbol$ and $\Fout$ vs $\Menv$ relations of Fig.~\ref{fig:correls}c,d}
\label{tab:correlFout}
{\small
\begin{tabular}{lcccl}
\hline
  &
  \multicolumn{1}{c}{$\Lbol$\supa} &
  \multicolumn{1}{c}{$\Fout$\supb} &
    \multicolumn{1}{c}{$\Menv$\supb} \\
  \multicolumn{1}{l}{Source} &
  \multicolumn{1}{c}{(\lo)} &
  \multicolumn{1}{c}{(\mo\,\kms\,yr$^{-1}$)} &
  \multicolumn{1}{c}{(\mo)} &
  \multicolumn{1}{l}{Refs.} \\
\hline
IRAS\,20343+4129\,IRS1 & 2000.0 & 1.3E-4 & 0.8   &\citet[][Table 5]{Beltran2008_outflows}\\
L1206 & 1200.0 & 7.7E-5 & 14.2                  &\citet[][Table 5]{Beltran2008_outflows}\\
Cepheus\,E & 70.0 & 2.3E-4 & 13.6               &\citet[][Table 5]{Beltran2008_outflows}\\
IRAS\,21391+5802 & 370.0 & 0.0014 & 5.1         &\citet[][Table 5]{Beltran2008_outflows}\\
IRAS\,20050+2720\,A & 280.0 & 5.3E-4 & 6.5      &\citet[][Table 5]{Beltran2008_outflows}\\
OMC-2/3\,MM7 & 76.0 & 0.0011 & 0.54             &\citet[][Table 5]{Beltran2008_outflows}\\
IRAS\,20293+3952\,A & 1050.0 & 0.022 & 4.0      &\citet[][Table 5]{Beltran2008_outflows}\\
IRAS\,22171+5549 & 1800.0 & 0.012 & 25.0        &\citet[][Table 5]{Beltran2008_outflows}\\
IRAS\,21307+5049 & 4000.0 & 0.0029 & 53.0       &\citet[][Table 5]{Beltran2008_outflows}\\
S235\,NE-SW & 1000.0 & 0.014 & 16.0             &\citet[][Table 5]{Beltran2008_outflows}\\
IRAS\,23385+6053 & 1500.0 & 0.009 & 61.0        &\citet[][Table 5]{Beltran2008_outflows}\\
HH\,288 & 500.0 & 0.014 & 18.0                  &\citet[][Table 5]{Beltran2008_outflows}\\
IRAS\,00117\,MM1 & 500.0 & 7.5E-4 & 1.5         &\citet[][Tables 3 and 6]{Palau2010_I00117}\\
HH\,211 & 3.6 & 2.8E-5 & 0.04                   &\citet[][Table 1]{Palau2006_HH211}\\
CB17-IRS & 0.5 & 1.1E-7 & 0.023                 &\citet[][Tables 3 and 5]{Chen2012ApJ_BokGlobule_CB17_FHC}\\
CygX-N3\,MM1 & 106.0 & 0.00131 & 12.5           &\citet[][Table 2]{DuarteCabral2013_outflows}\\
CygX-N3\,MM2 & 121.0 & 7.2E-4 & 13.8            &\citet[][Table 2]{DuarteCabral2013_outflows}\\
CygX-N48\,MM1 & 102.0 & 0.00135 & 17.0          &\citet[][Table 2]{DuarteCabral2013_outflows}\\
CygX-N48\,MM2 & 85.0 & 4.5E-4 & 8.1             &\citet[][Table 2]{DuarteCabral2013_outflows}\\
CygX-N53\,MM1 & 199.0 & 0.00412 & 34.2          &\citet[][Table 2]{DuarteCabral2013_outflows}\\
CygX-N63\,MM1 & 339.0 & 0.00291 & 44.3          &\citet[][Table 2]{DuarteCabral2013_outflows}\\
Per-Bolo58 & 0.012 & 5.6E-8 & 0.11              &\citet[][Tables 2, 3 and 4]{Dunham2011_Outflow_FHC}\\
L1451mm & 0.045 & 8.3E-9 & 0.024                &\citet[][Tables 11 and 12]{Pineda2011_FHC_VeLLO_L1451}\\
B1b-S & 0.058 & 2.4E-6 & 0.36                   &\citet[][Tables 6 and 7]{Hirano2014_Barnard1}, this work (Table~\ref{tab:rejected})\\
L1521F-IRS & 0.037 & 3.3E-7 & 0.065             &\citet[][Sec. 3.1, Table 2]{Takahashi2013_Outflow_VeLLO_L1521F-IRS}, this work (Table~\ref{tab:rejected})\\
L1148-IRS & 0.112 & 1.0E-7 & 0.02\supc          &\citet[][Sec.~3.3, Table 1]{Kauffmann2011_VeLLO_L1148-IRS}, this work (Table~\ref{tab:rejected})\\
L1014-IRS & 0.100 & 5.3E-8 & 0.010              &\citet[][Secs.~3, 4.1]{Bourke2005_LowMass_Protostars}, this work (Table~\ref{tab:rejected})\\
082012 & 6.3 & 7.0E-4 & 9.4                     &\citet[][Tables 4 and 5]{Tobin2016_outflows}, \citet[][Table 2]{Tobin2015_YoungestHerschelProtosars}\\
093005 & 1.7 & 1.1E-5 & 5.4                     &\citet[][Tables 4 and 5]{Tobin2016_outflows}\\
090003 & 2.71 & 2.0E-6 & 7.0                    &\citet[][Tables 4 and 5]{Tobin2016_outflows}\\
135003 & 12.0 & 6.2E-6 & 3.0                    &\citet[][Tables 4 and 5]{Tobin2016_outflows}\\
119019 & 1.56 & 2.4E-6 & 0.6                    &\citet[][Tables 4 and 5]{Tobin2016_outflows}\\
302002 & 0.85 & 3.0E-7 & 2.9                    &\citet[][Tables 4 and 5]{Tobin2016_outflows}\\
019003A & 3.16 & 4.8E-6 & 2.4                   &\citet[][Tables 4 and 5]{Tobin2016_outflows}\\
HOPS68 & 5.7 & 8.8E-5 & 2.5                     &\citet[][Tables 4 and 5]{Tobin2016_outflows}, \citet[][Table 2]{Tobin2015_YoungestHerschelProtosars}\\
HOPS223 & 28.0 & 5.8E-6 & 3.0                   &\citet[][Tables 4 and 5]{Tobin2016_outflows}, \citet[][Table 2]{Tobin2015_YoungestHerschelProtosars}\\
L328-IRS & 0.121 & 3.3E-6\supd & 0.01           &\citet{Lee2018_L328_ProtoBD_ALMA}, this work (Table~\ref{tab:rejected})\\
GF9-2\supe & 1.7 & 7.4E-6 & 0.067               &\citet[][Table 5]{Busch2020_ChaMMS1-outflows}, \citet[][Table 2]{Furuya2019_GF9-2}\\
Cha-MMS1 & 0.041 & 4.1E-7 & 0.003               &\citet[][Sec. 3.1 and Table 5]{Busch2020_ChaMMS1-outflows}, this work (Table~\ref{tab:rejected})\\
SerpM-S68Nb & 1.8 & 7.0E-6 & 0.30\supf          &\citet[][Table 5]{Podio2021_Class0}, \citet[][Table 4]{Maury2019_CALYPSO}\\
IRAS4B1 & 2.3 & 3.5E-5 & 1.76\supf              &\citet[][Table 5]{Podio2021_Class0}, \citet[][Table 4]{Maury2019_CALYPSO}\\
L1157 & 4.0 & 4.5E-5 & 0.60\supf                &\citet[][Table 5]{Podio2021_Class0}, \citet[][Table 4]{Maury2019_CALYPSO}\\
L1448C & 11.0 & 2.35E-4 & 0.48\supf             &\citet[][Table 5]{Podio2021_Class0}, \citet[][Table 4]{Maury2019_CALYPSO}\\
IRAS15398-3359 & 1.34 & 7.4E-6 & 0.0023         &\citet[][Tables 6 and 8]{Vazzano2021_outflows}, this work (footnote a of Table~\ref{tab:LintMdyn}) \\
IRAS16059-3857 & 0.17 & 2.1E-6 & 0.02           &\citet[][Tables 6, 8 and 11]{Vazzano2021_outflows}\\
J160115-4152 & 0.099 & 3.1E-7 & 0.016           &\citet[][Tables 6 and 8]{Vazzano2021_outflows}, this work (Table~\ref{tab:rejected})\\
G191.9S & 0.4 & 2.992E-5 & 0.51        &\citet[][Tables 2 and 3]{Dutta2024_EpisodicAccretion}\\
G203.2W2 & 0.5 & 7.07E-6 & 0.32        &\citet[][Tables 2 and 3]{Dutta2024_EpisodicAccretion}\\
G205.4M1B & 4.8 & 1.0E-6 & 5.8          &\citet[][Tables 2 and 3]{Dutta2024_EpisodicAccretion}\\
G205.4S1A & 22.0 & 0.00165 & 0.88       &\citet[][Tables 2 and 3]{Dutta2024_EpisodicAccretion}\\
G209.5S2 & 3.4 & 1.12E-6 & 0.61        &\citet[][Tables 2 and 3]{Dutta2024_EpisodicAccretion}\\
G205.4S3 & 6.4 & 3.25E-5 & 0.7         &\citet[][Tables 2 and 3]{Dutta2024_EpisodicAccretion}\\
G206.1 & 3.0 & 1.62E-6 & 2.44          &\citet[][Tables 2 and 3]{Dutta2024_EpisodicAccretion}\\
\hline
\end{tabular}
}

\end{table*}

\setcounter{table}{8}
\begin{table*}[htbp]
\caption{(cont.) Interferometric data used for the $\Fout$ vs $\Lbol$ and $\Fout$ vs $\Menv$ relations of Fig.~\ref{fig:correls}c,d}
\label{tab:correlFout}
{\small
\begin{tabular}{lcccl}
\hline
  &
  \multicolumn{1}{c}{$\Lbol$\supa} &
  \multicolumn{1}{c}{$\Fout$\supb} &
    \multicolumn{1}{c}{$\Menv$\supb} \\
  \multicolumn{1}{l}{Source} &
  \multicolumn{1}{c}{(\lo)} &
  \multicolumn{1}{c}{(\mo\,\kms\,yr$^{-1}$)} &
  \multicolumn{1}{c}{(\mo)} &
  \multicolumn{1}{l}{Refs.} \\
\hline
G206.9W2 & 6.3 & 3.83E-4 & 4.56        &\citet[][Tables 2 and 3]{Dutta2024_EpisodicAccretion}\\
G208.6N1 & 37.0 & 2.54E-6 & 8.26       &\citet[][Tables 2 and 3]{Dutta2024_EpisodicAccretion}\\
G208.8E & 2.2 & 3.242E-5 & 0.28        &\citet[][Tables 2 and 3]{Dutta2024_EpisodicAccretion}\\
G208.8Walma & 0.8 & 3.2E-8 & 0.38      &\citet[][Tables 2 and 3]{Dutta2024_EpisodicAccretion}\\
G210.3S & 0.6 & 1.6E-7 & 0.54          &\citet[][Tables 2 and 3]{Dutta2024_EpisodicAccretion}\\
G210.4W & 60.0 & 1.644E-4 & 1.4        &\citet[][Tables 2 and 3]{Dutta2024_EpisodicAccretion}\\
G200.3N & 1.5 & 1.85E-6 & 0.51         &\citet[][Tables 2 and 3]{Dutta2024_EpisodicAccretion}\\
G192.1 & 9.5 & 1.045E-5 & 1.21         &\citet[][Tables 2 and 3]{Dutta2024_EpisodicAccretion}\\
G209.5N1 & 9.0 & 2.05E-6 & 1.98        &\citet[][Tables 2 and 3]{Dutta2024_EpisodicAccretion}\\
G208.6S & 49.0 & 4.1E-7 & 1.55         &\citet[][Tables 2 and 3]{Dutta2024_EpisodicAccretion}\\
G211.0N & 4.5 & 1.08E-7 & 0.81         &\citet[][Tables 2 and 3]{Dutta2024_EpisodicAccretion}\\
G212.8N & 3.0 & 2.28E-7 & 0.52         &\citet[][Tables 2 and 3]{Dutta2024_EpisodicAccretion}\\
\hline
I16253$-$2429\supg &0.129 &6.5E-7 &0.018 &\citet[][Sec. 4.2.1 and Table 3]{Hsieh2016_Binary_ProtoBD_Jet_IRAS16253}, this work (Table~\ref{tab:protoBDs})\\
ISO-Oph200\supg &0.076 &7E-7 &0.006      &\citet[][Sec. 5.2 and Table 1]{Riaz2021_Structure_ProtoBD}, this work (Table~\ref{tab:protoBDs})\\
IC348-SMM2E\supg &0.109 &2.7E-7 &0.051     &\citet[][Table 4]{Palau2014}, this work (Table~\ref{tab:protoBDs})\\
J041757B\supg &$<0.002$ &3.7E-8 &0.00015       &\citet{Palau2022_J041757}, this work (Table~\ref{tab:protoBDs})\\
SMA1627-2441\suph &$<0.004$ &7.9E-10 &0.0024    &\citet{PhanBao2022_protoBD}\\
J041757NE-MM1\suph &$<0.006$ &1.2E-9 &0.012    &\citet{Palau2022_J041757}\\
\hline
\end{tabular}
\begin{list}{}{}
\item[$^\mathrm{a}$] $\Lbol$ for the proto-BD candidates is taken as $\Lint$ because the least luminous objects are most affected by the ISRF and the accretion/ejection processes must be driven by the mass of the central object, more closely related to $\Lint$ than to $\Lbol$ \citep{AMIConsortium2011_AMIobs-c2d-smallcores}.
\item[$^\mathrm{b}$] $\Fout$ corresponds to the outflow momentum rate or outflow force (uncorrected for inclination/opacity in most of the cases), while $\Menv$ corresponds to the envelope mass of the driving source, all of them observed with interferometers. $\Menv$ is obtained from the same work that estimate $\Fout$. Thus the interferometric configuration was the same in both measurements, being both equally affected by missing flux.
\item[$^\mathrm{c}$] Upper limit for $\Menv$ from \citet[][Sec. 3.3]{Kauffmann2011_VeLLO_L1148-IRS}.
\item[$^\mathrm{d}$] $\Fout$ for L328-IRS has been estimated from equation (2) of \citet{Lee2018_L328_ProtoBD_ALMA}, $\Fout=0.025\,\dot{M}_\mathrm{acc}\,V_\mathrm{W}$, and adopting a mass accretion rate of $8.9\times10^{-7}$~\mo\,yr$^{-1}$ and $V_\mathrm{W}\sim150$~\kms\ \citep{Lee2018_L328_ProtoBD_ALMA}.
\item[$^\mathrm{e}$] For GF9-2, \citet{Busch2020_ChaMMS1-outflows} adopted the most recent distance of 474 pc (instead of 200 pc as previously used), and $\Fout$ is given for this new distance. For $\Menv$, we have taken the value from Table 2 of \cite{Furuya2019_GF9-2} for the frequency of the SMA data used to estimate the outflow parameters (345~GHz), and re-scaled it to the new distance. $\Lbol$ is taken from \cite{Furuya2006_Class0_GF9_2} and has also been re-scaled to the new distance.
\item[$^\mathrm{f}$] $\Menv$ has been estimated using the flux density reported by \citet[][Table 4]{Maury2019_CALYPSO}, assuming 20~K and the dust opacity from \citet[][thin ice mantles at 10$^6$~\cmt]{Ossenkopf1994_dustopacity} and the distance given in Table 1 of \cite{Podio2021_Class0}. For SerpM-S68Nb, the peak intensity at 94 GHz reported in Table 3 of \citet{Maury2019_CALYPSO} was used because this source is not included in their Table 4.
\item[$^\mathrm{g}$] Proto-BD candidate included in Table~\ref{tab:protoBDs}. This object is not included to perform the fits shown in Fig.~\ref{fig:correls}c and d. For the case of IC348-SMM2E, all the values have been updated to the new distance of 320~pc, and an inclination with respect to the plane-of-sky of 40$^{\circ}$ has been assumed.
\item[$^\mathrm{h}$] Tentative proto-BD candidate for which signs of outflow have been reported but there are no infrared counterparts. This object is not included to perform the fits shown in Fig.~\ref{fig:correls}c and d. 
\end{list}
}

\end{table*}

\section{Images of spatial distribution of VeLLOs/proto-BD candidates within their molecular clouds}\label{app:spat-distr}

In this Appendix, the {\it Herschel}/SPIRE images at 250~\mum\ for different molecular clouds are presented, with black open circles indicating the positions of the SUCANES candidates \citep{PerezGarcia2024_SUCANES}, white open circles indicating the positions of the \cite{Kim2016_VeLLOs} candidates, and red filled circles indicating the positions of the proto-BD candidates selected in this work and listed in Table~\ref{tab:protoBDs}.

\setcounter{figure}{0}

\begin{figure*}
\begin{center}
\begin{tabular}[b]{cc}
    \epsfig{file=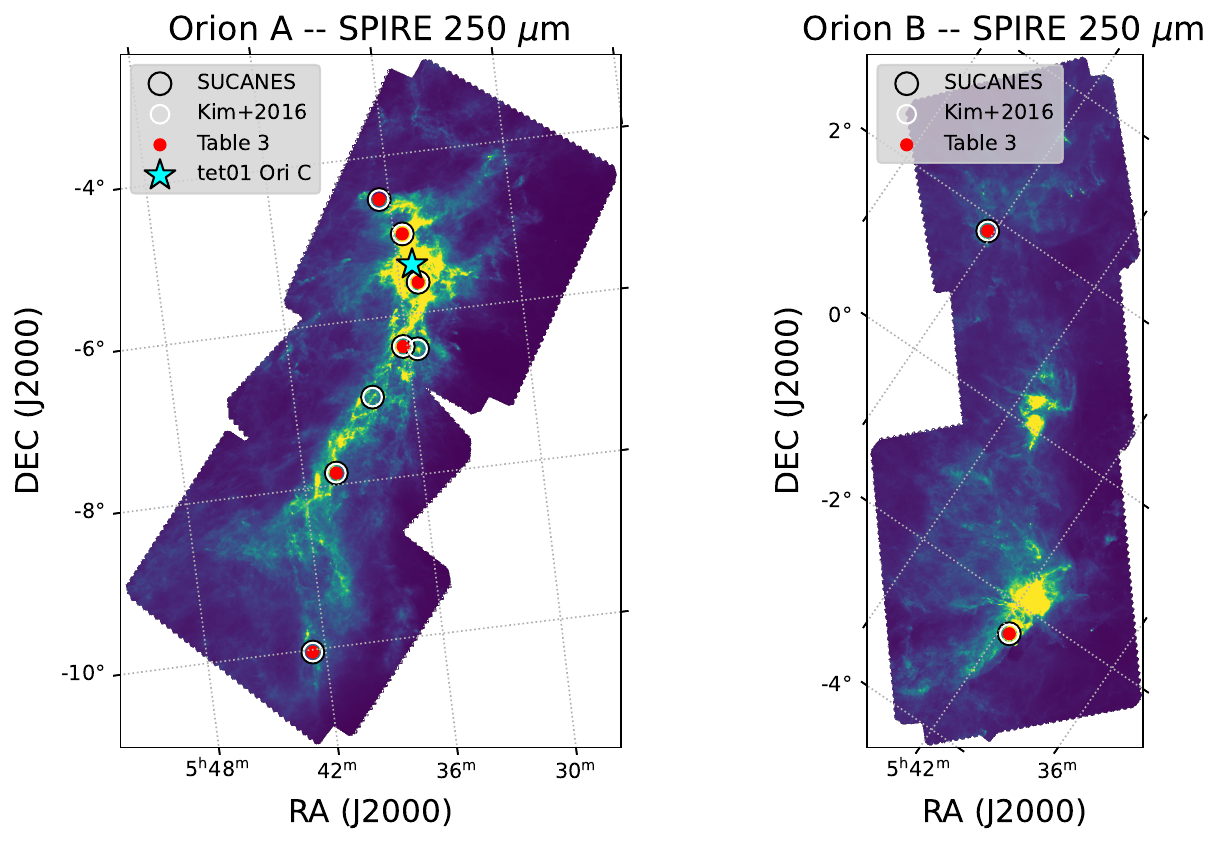, width=17cm,angle=0}\\
    \epsfig{file=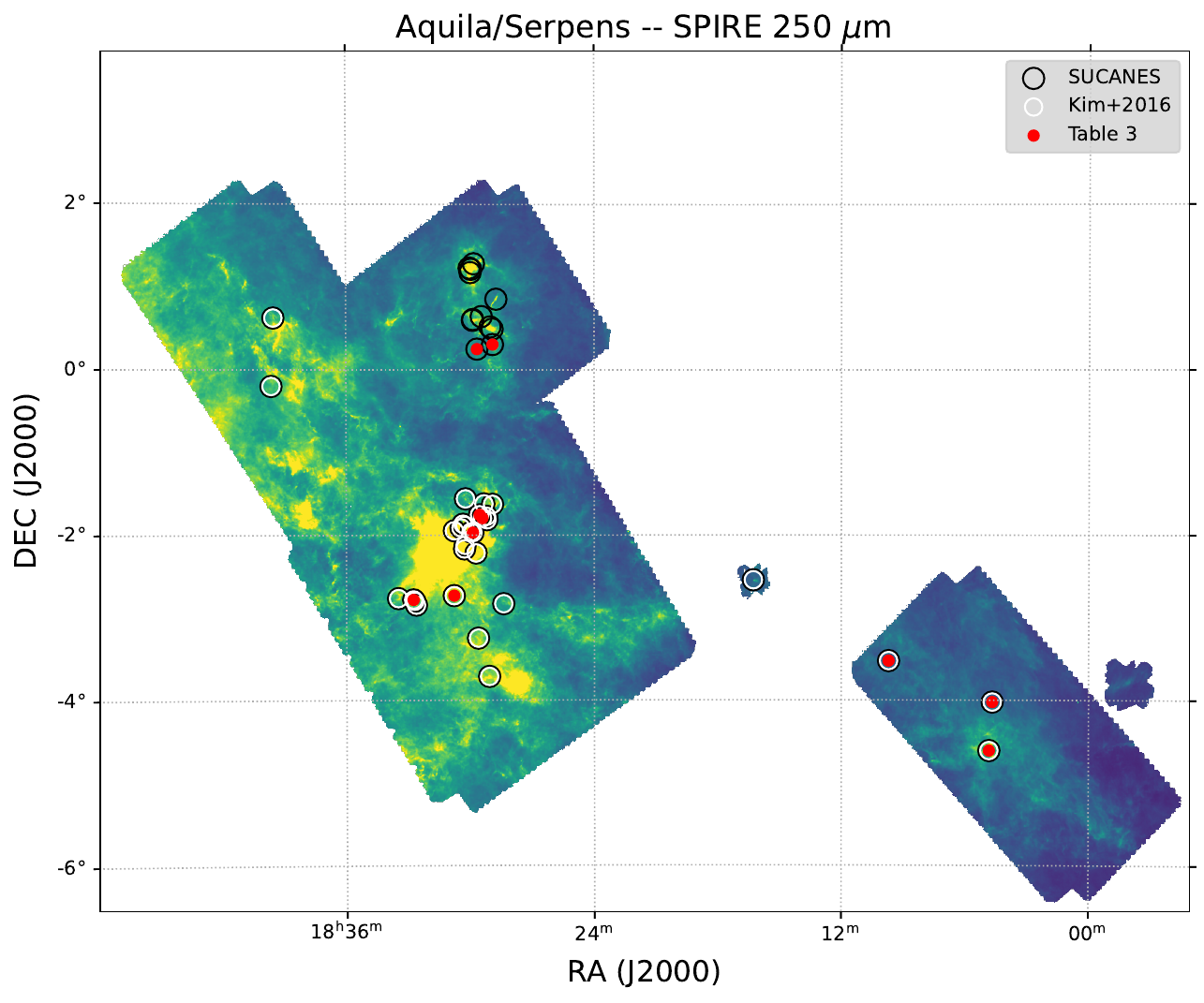, width=14cm,angle=0}\\
\end{tabular}
\caption{RGB Herschel/SPIRE image of Orion (top) and Aquila/Serpens (bottom) molecular clouds, with the open black circles indicating the SUCANES objects, the white open circles indicating the subsample of SUCANES that belong to \cite{Kim2016_VeLLOs} groups A+B, and the red circles corresponding to the subsample of SUCANES that include the proto-BD candidates of Table~\ref{tab:protoBDs} selected in this work.
}
\label{fig:spadistribOrionSerpens}
\end{center}
\end{figure*} 

\begin{figure*}
\begin{center}
\begin{tabular}[b]{cc}
    \epsfig{file=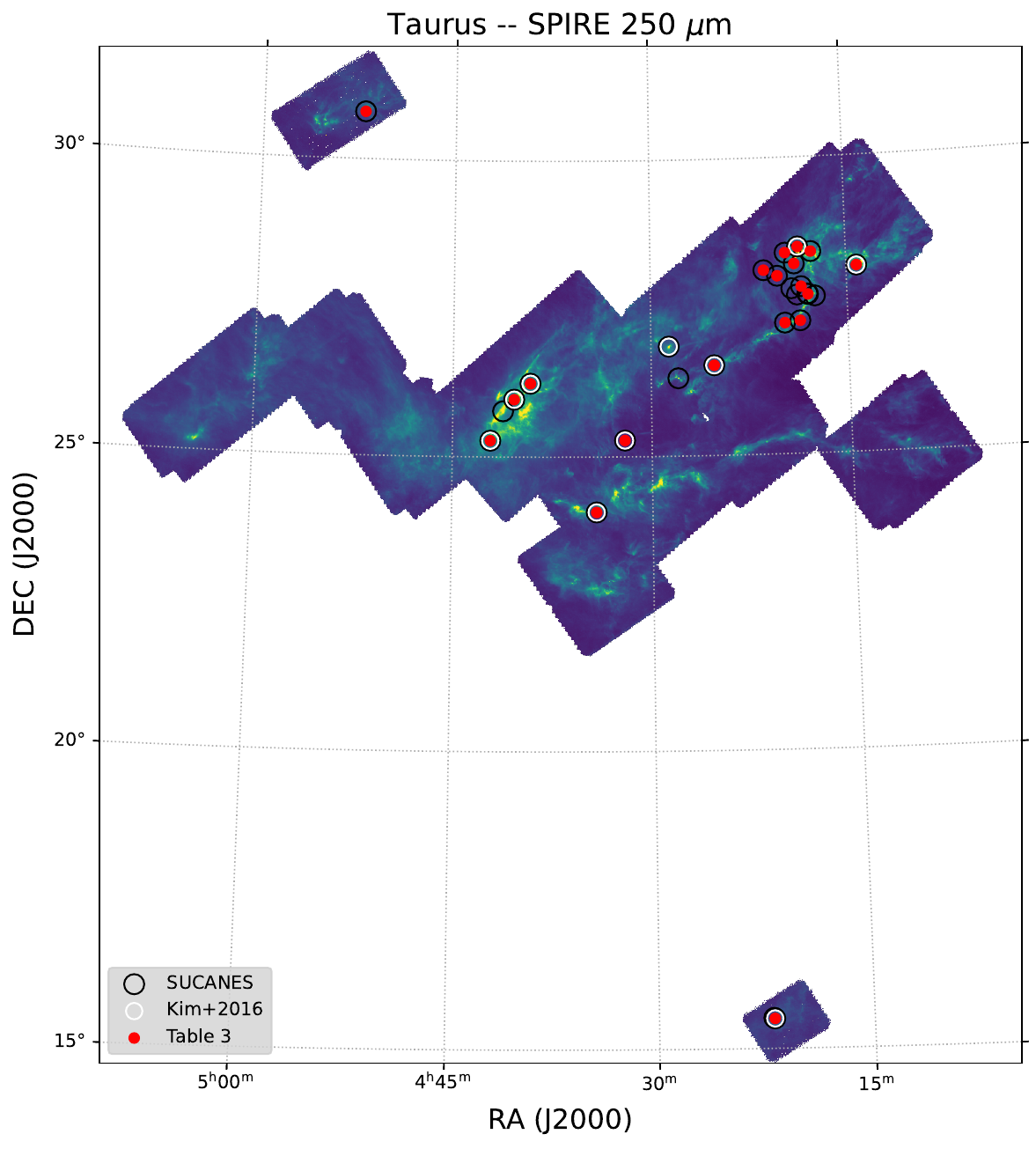, width=12cm,angle=0}\\
    \epsfig{file=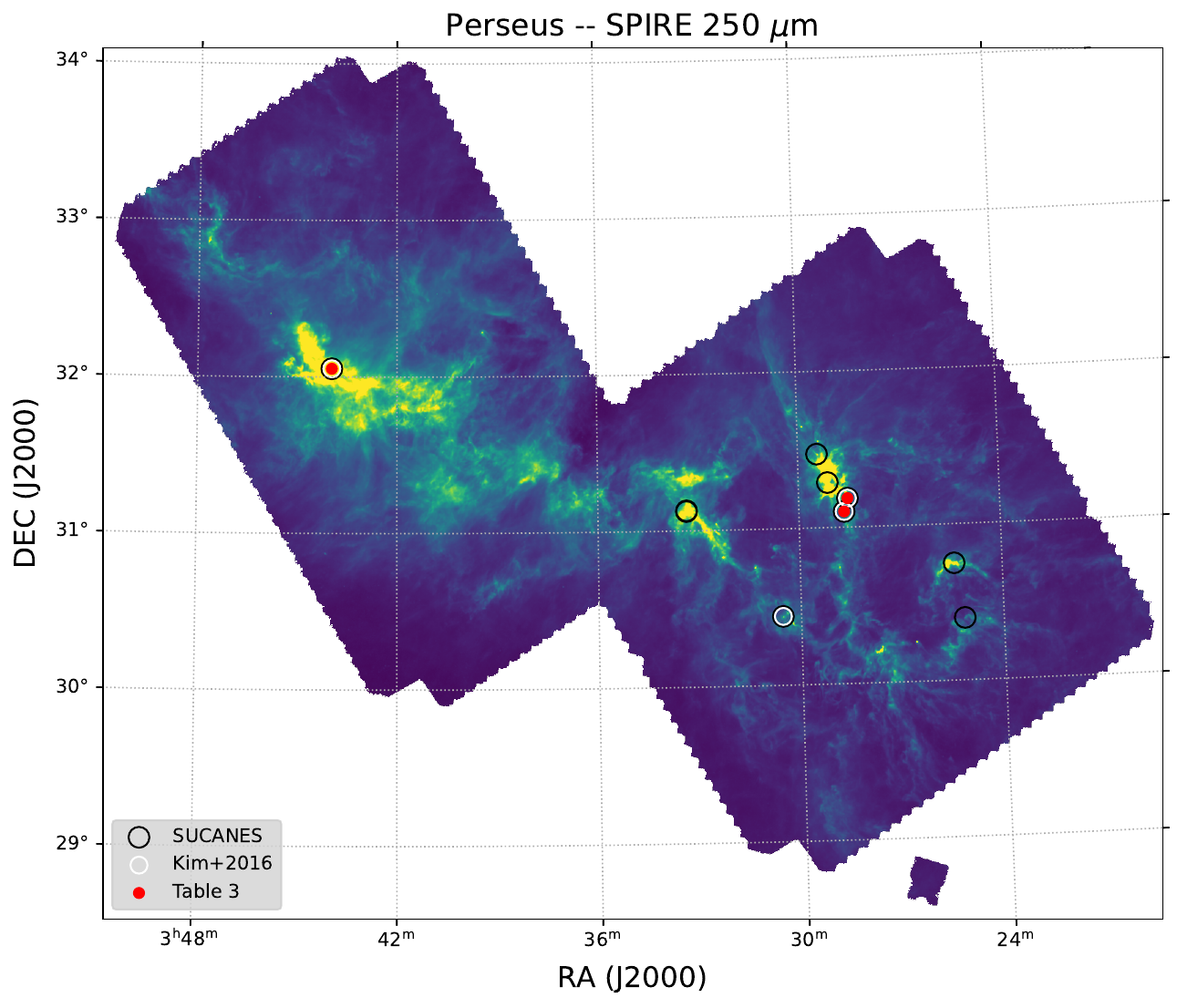, width=12cm,angle=0}\\
\end{tabular}
\caption{RGB Herschel/SPIRE image of Taurus (top) and Perseus (bottom) molecular clouds, with the open black circles indicating the SUCANES objects, the white open circles indicating the subsample of SUCANES that belong to \cite{Kim2016_VeLLOs} groups A+B, and the red circles corresponding to the subsample of SUCANES that include the proto-BD candidates of Table~\ref{tab:protoBDs} selected in this work.
}
\label{fig:spadistribTaurusPerseus}
\end{center}
\end{figure*} 

\begin{figure*}
\begin{center}
\begin{tabular}[b]{cc}
    \epsfig{file=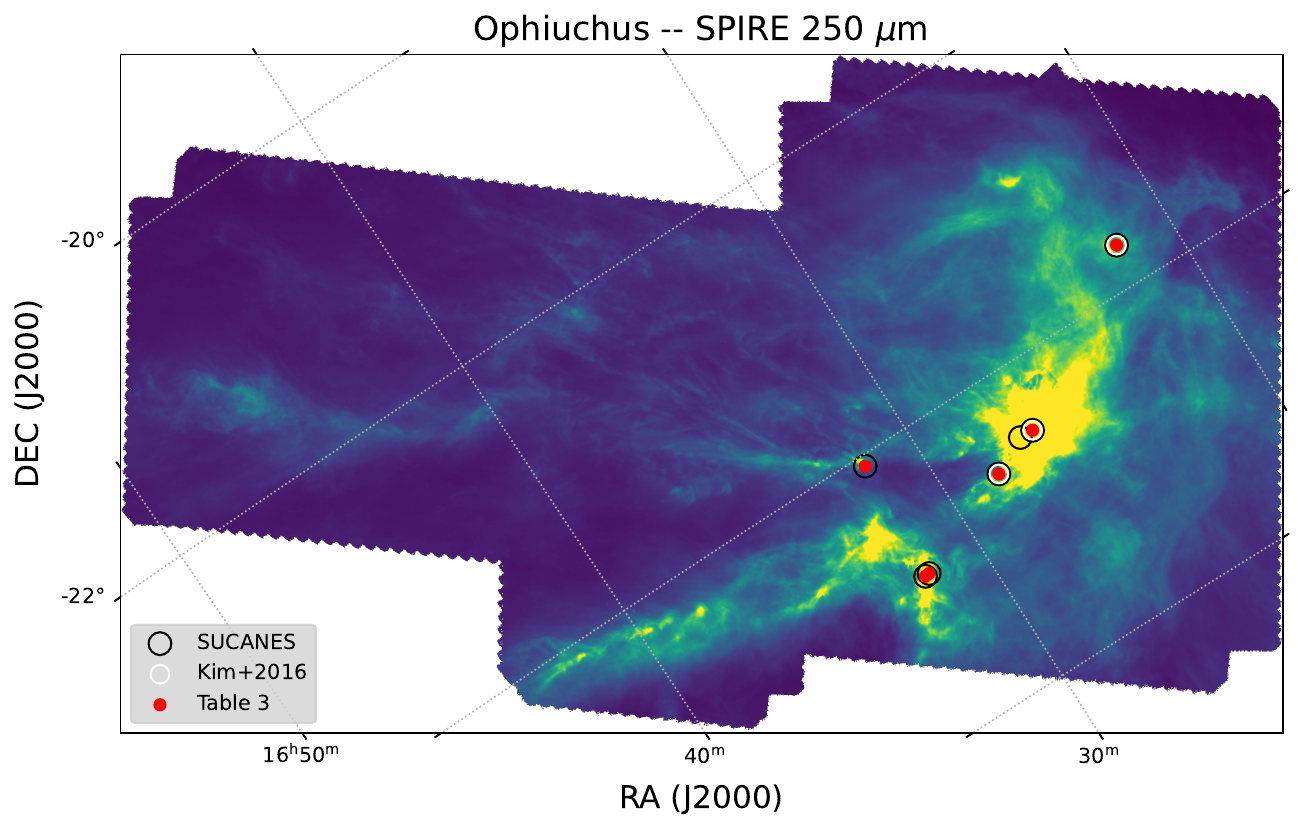, width=14cm,angle=0}\\
    \epsfig{file=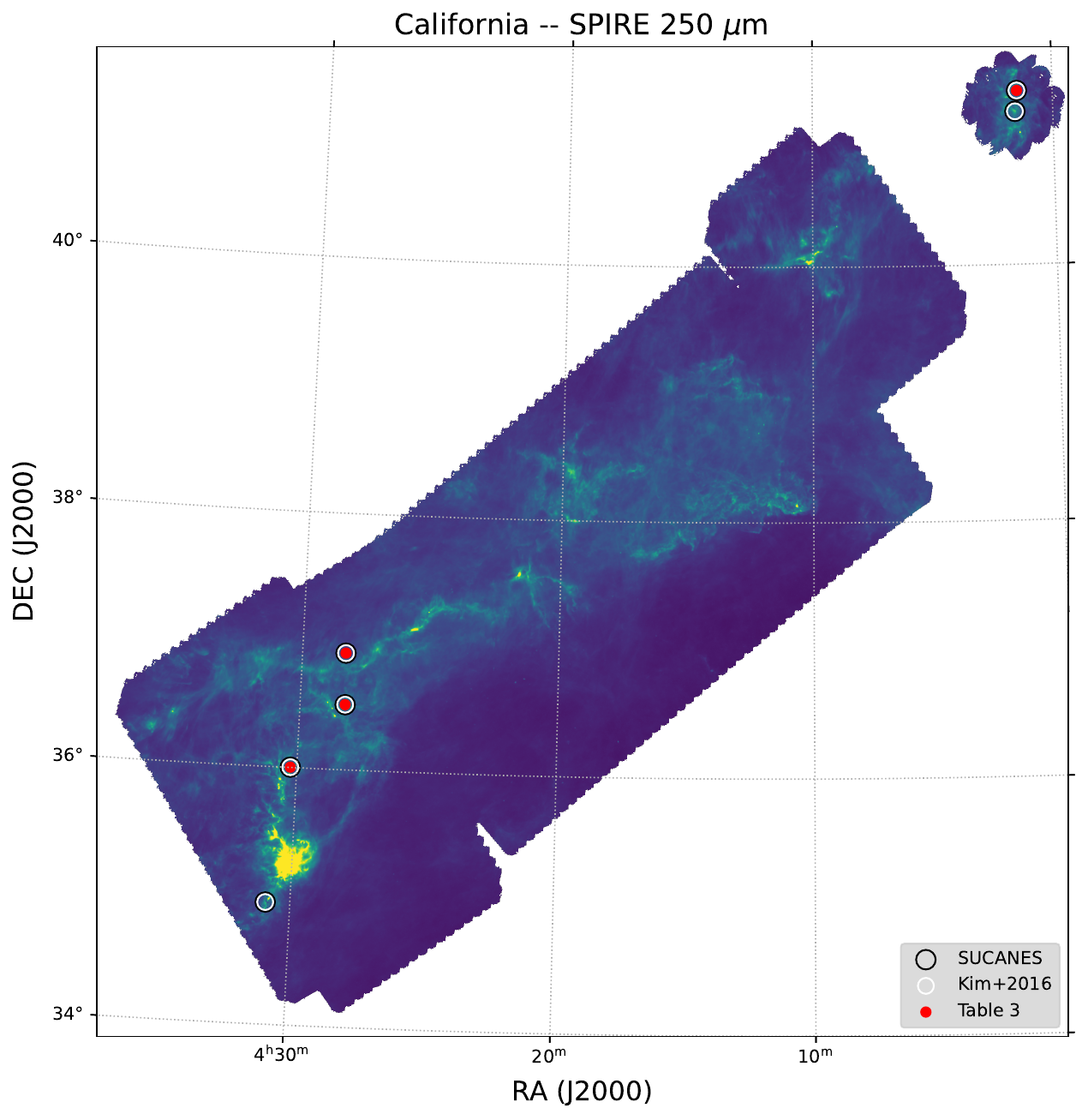, width=13cm,angle=0}\\
\end{tabular}
\caption{
RGB Herschel/SPIRE image of Ophiuchus (top) and California (bottom) molecular clouds, with the open black circles indicating the SUCANES objects, the white open circles indicating the subsample of SUCANES that belong to \cite{Kim2016_VeLLOs} groups A+B, and the red circles corresponding to the subsample of SUCANES that include the proto-BD candidates of Table~\ref{tab:protoBDs} selected in this work.
}
\label{fig:spadistribOphCal}
\end{center}
\end{figure*}

\begin{figure*}
\begin{center}
\begin{tabular}[b]{cc}
    \epsfig{file=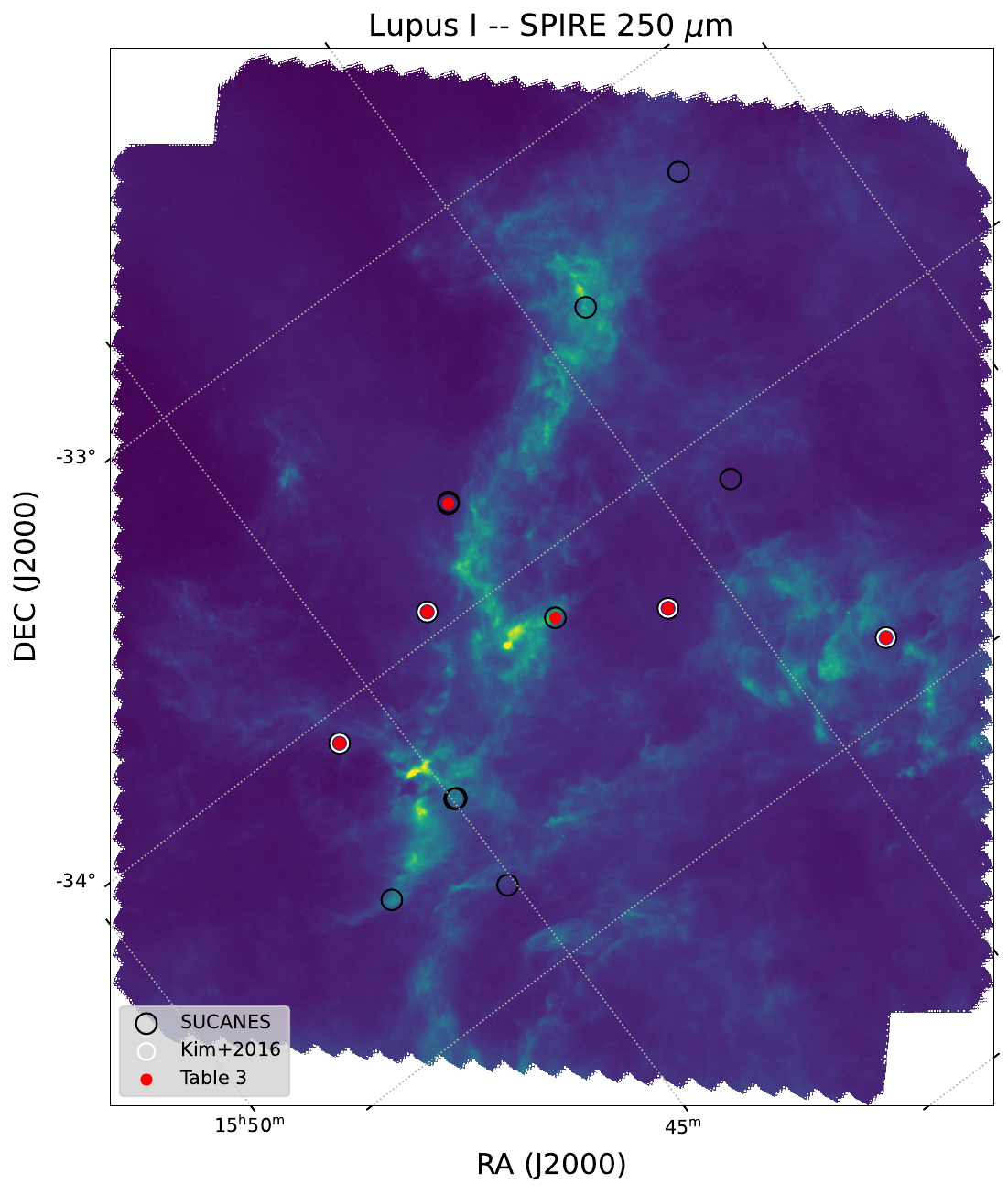, width=11cm,angle=0}\\
    \epsfig{file=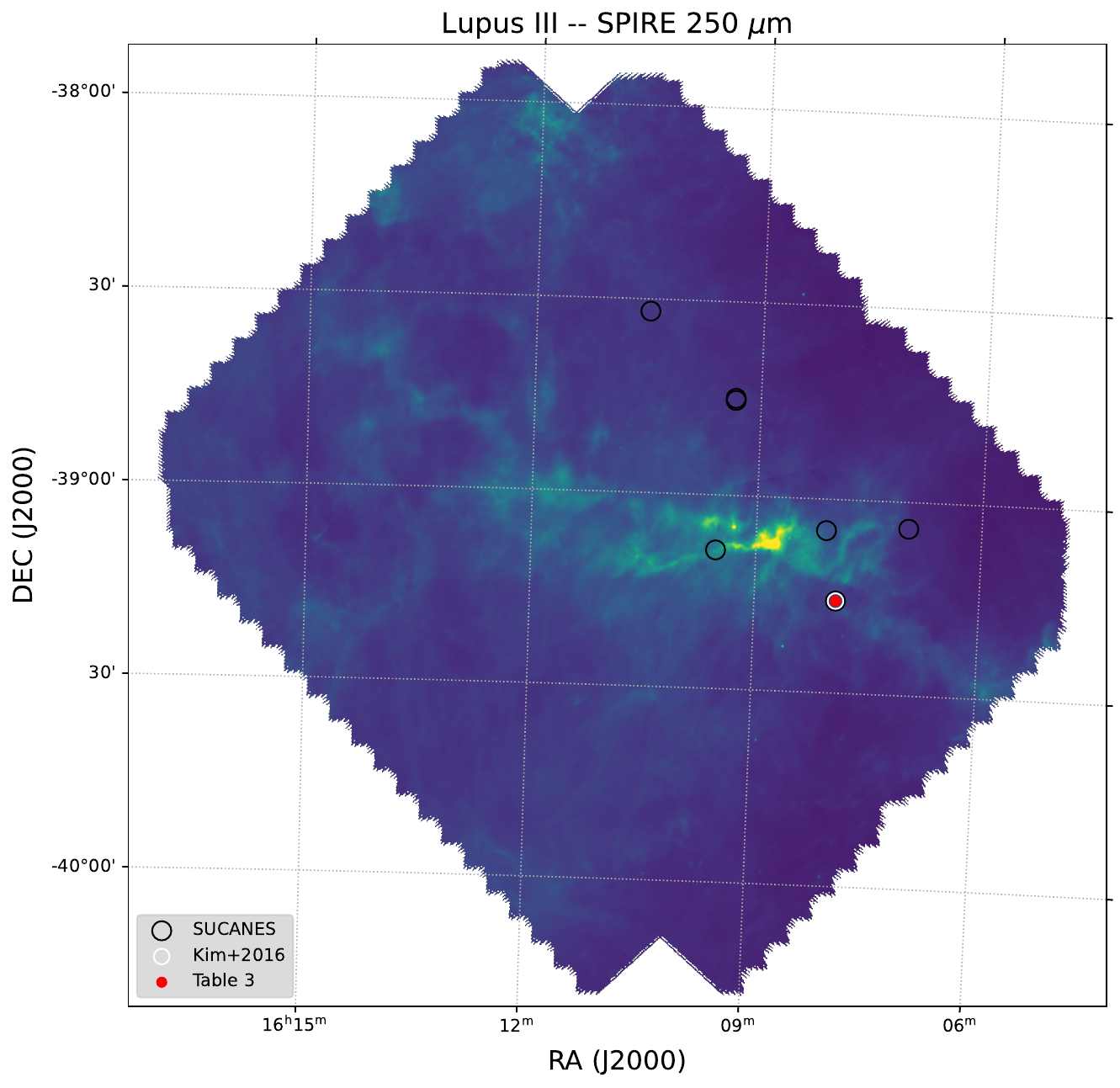, width=11cm,angle=0}\\
\end{tabular}
\caption{
RGB Herschel/SPIRE image of Lupus I (top) and Lupus III (bottom) molecular clouds, with the open black circles indicating the SUCANES objects, the white open circles indicating the subsample of SUCANES that belong to \cite{Kim2016_VeLLOs} groups A+B, and the red circles corresponding to the subsample of SUCANES that include the proto-BD candidates of Table~\ref{tab:protoBDs} selected in this work.
}
\label{fig:spadistribLupus}
\end{center}
\end{figure*}

\begin{figure*}
\begin{center}
\begin{tabular}[b]{c}
    \epsfig{file=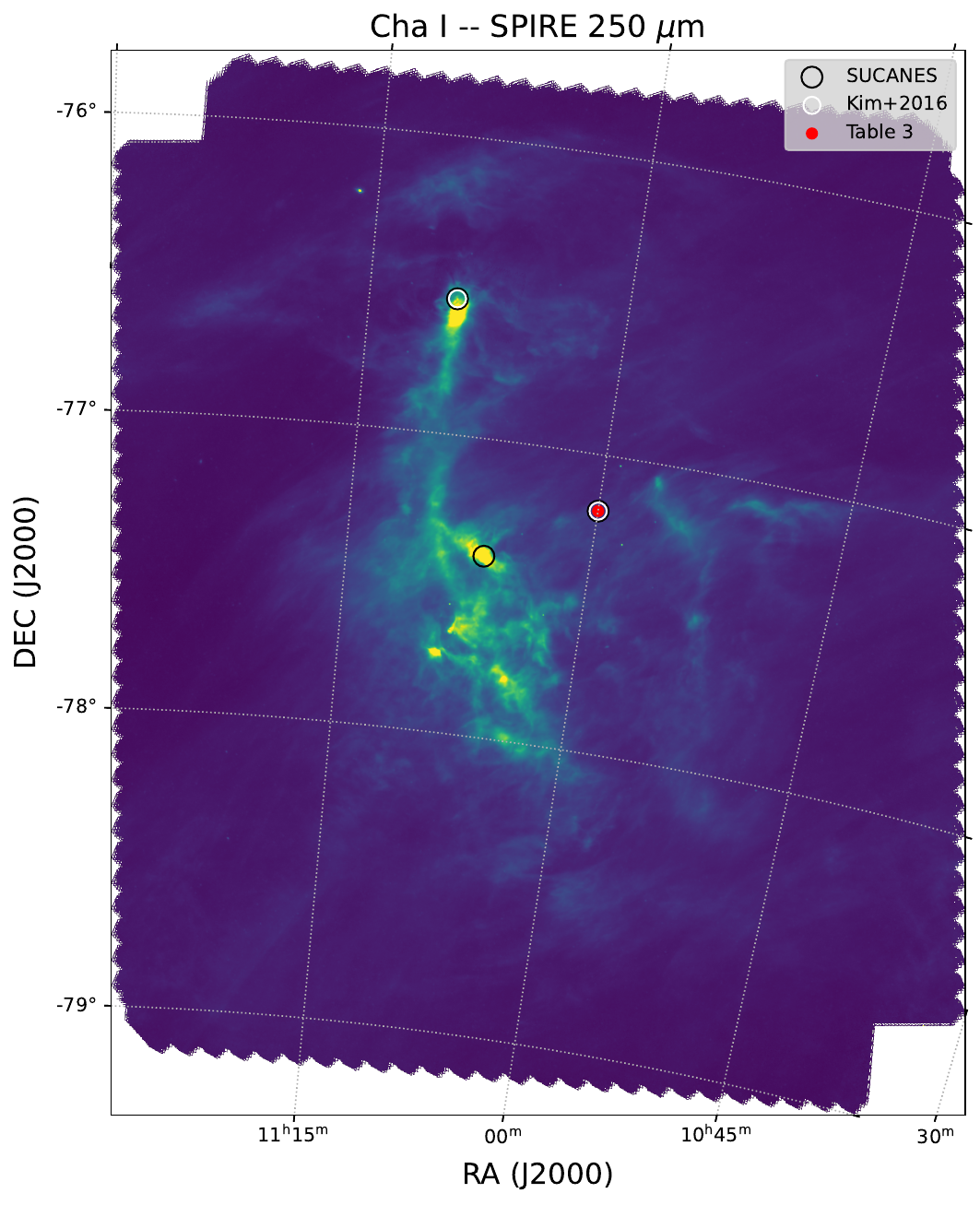, width=11cm,angle=0}\\
    \epsfig{file=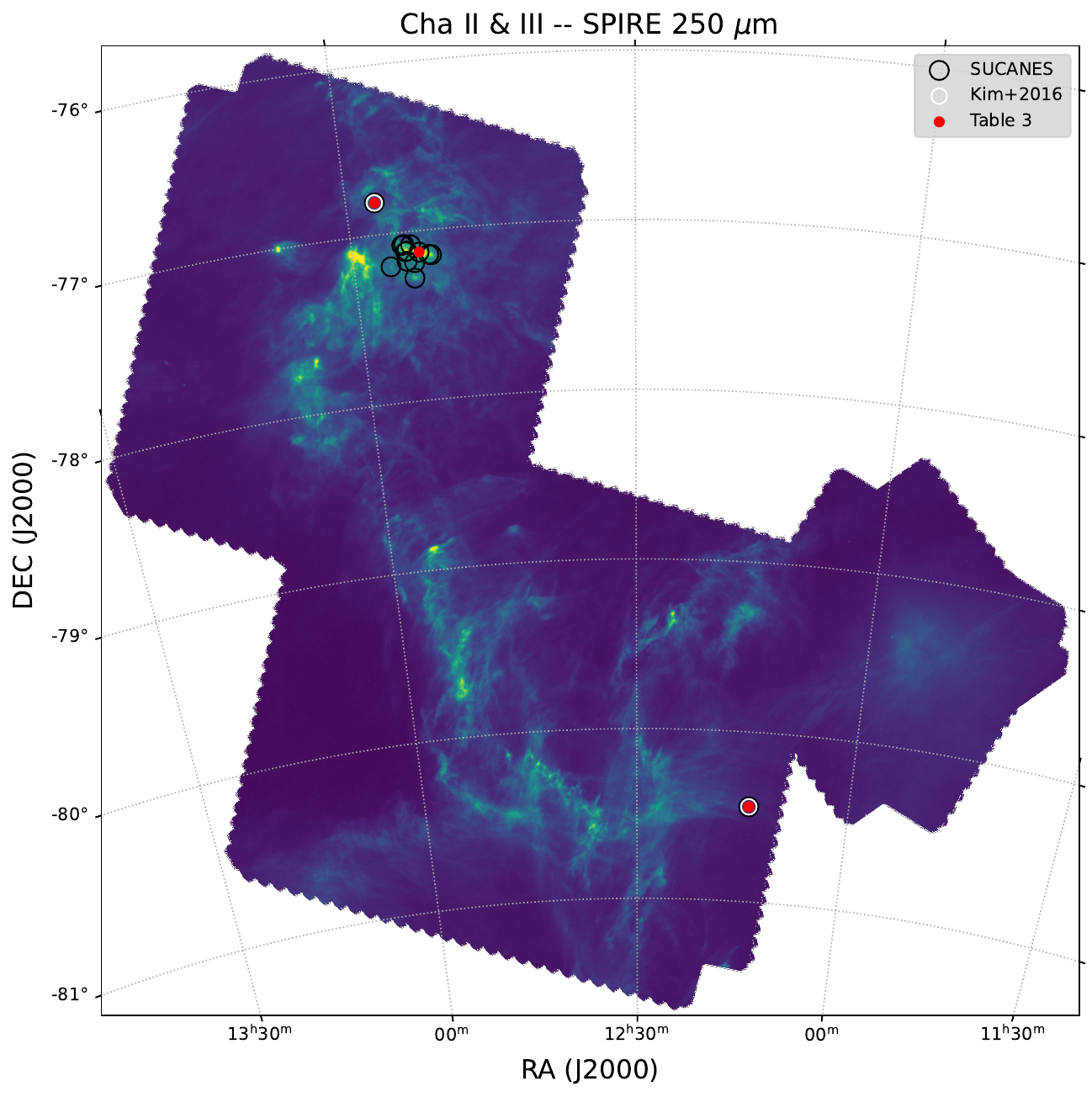, width=11cm,angle=0}\\
\end{tabular}
\caption{
RGB Herschel/SPIRE image of Chamaleon I (top) and Chamaleon II/III (bottom) molecular clouds, with the open black circles indicating the SUCANES objects, the white open circles indicating the subsample of SUCANES that belong to \cite{Kim2016_VeLLOs} groups A+B, and the red circles corresponding to the subsample of SUCANES that include the proto-BD candidates of Table~\ref{tab:protoBDs} selected in this work.
}
\label{fig:spadistribCha}
\end{center}
\end{figure*}

\begin{figure*}
\begin{center}
\begin{tabular}[b]{c}
    \epsfig{file=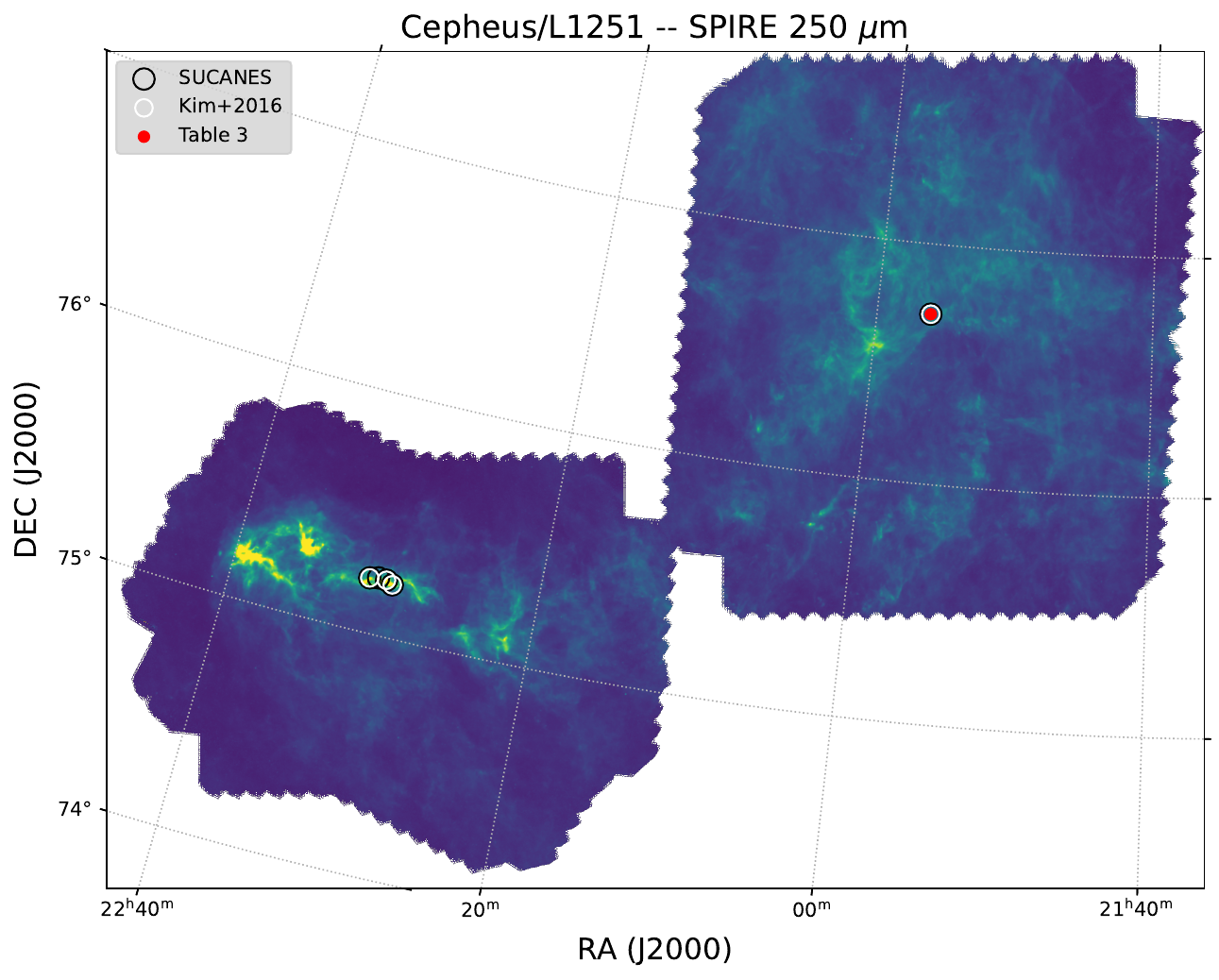, width=17cm,angle=0}\\
    \epsfig{file=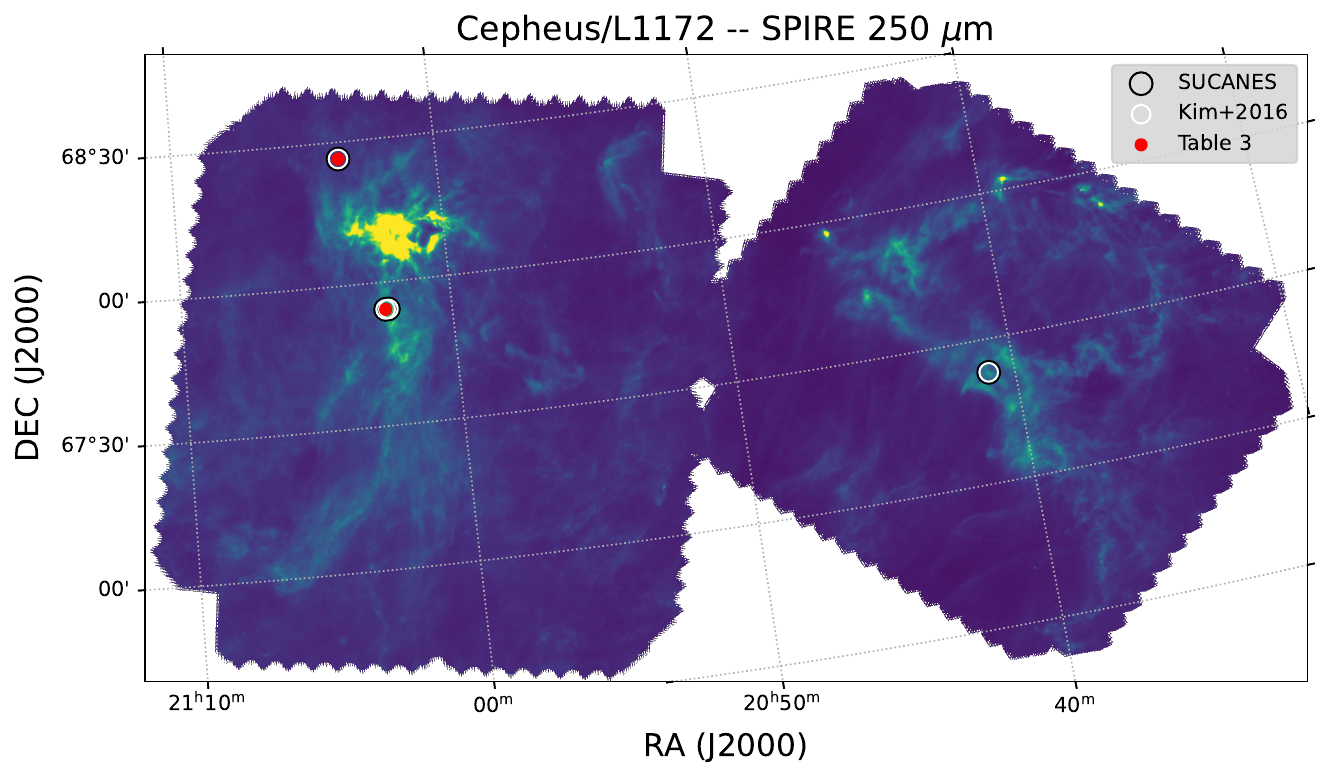, width=17cm,angle=0}\\
\end{tabular}
\caption{
RGB Herschel/SPIRE image of L1251 (top) and L1172 (bottom) Cepheus molecular clouds, with the open black circles indicating the SUCANES objects, the white open circles indicating the subsample of SUCANES that belong to \cite{Kim2016_VeLLOs} groups A+B, and the red circles corresponding to the subsample of SUCANES that include the proto-BD candidates of Table~\ref{tab:protoBDs} selected in this work.
}
\label{fig:spadistribCepheus}
\end{center}
\end{figure*} 

\begin{figure*}
\begin{center}
\begin{tabular}[b]{c}
    \epsfig{file=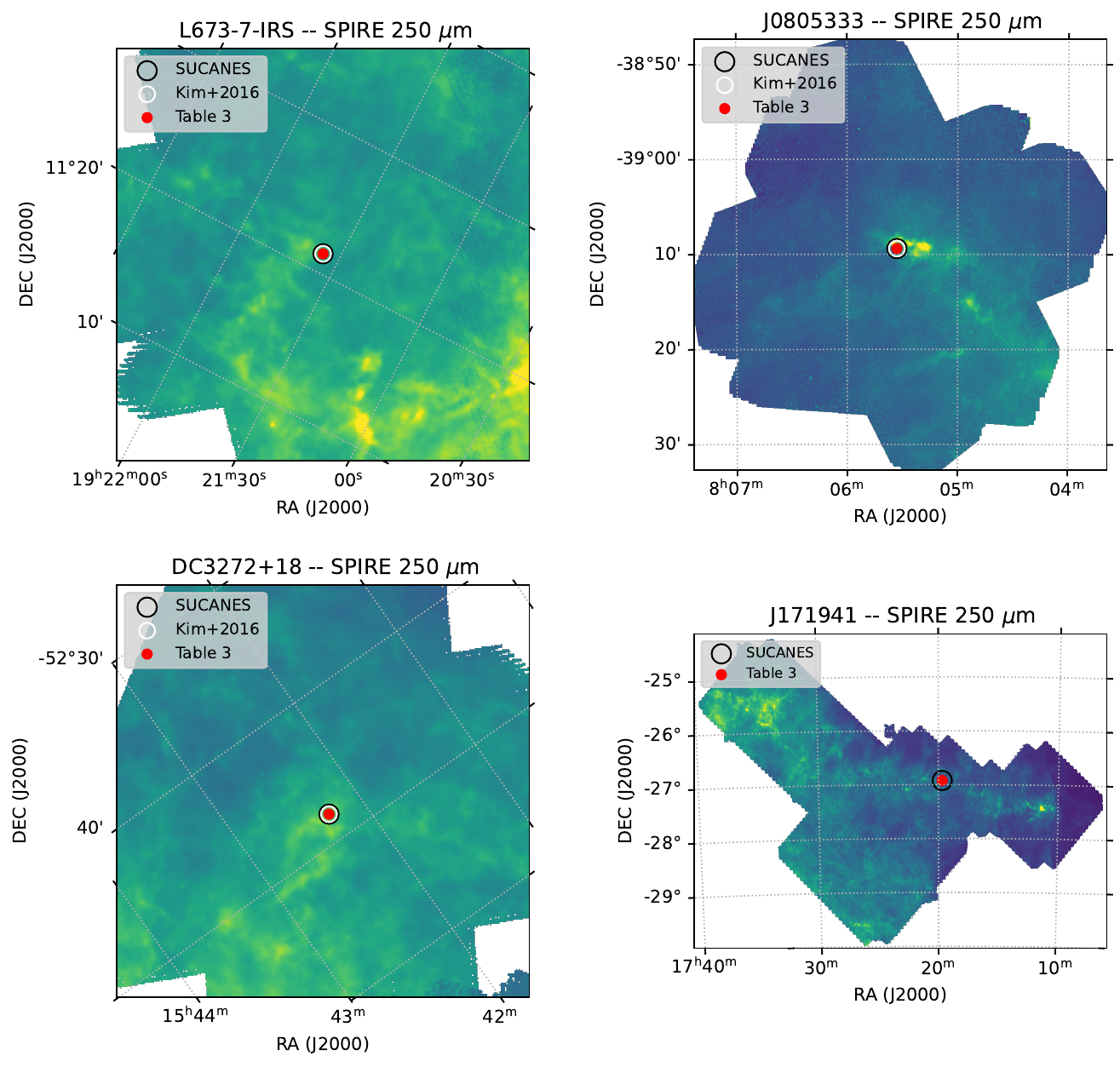, width=17cm,angle=0}\\
\end{tabular}
\caption{
RGB Herschel/SPIRE image of L673-7 (top-left), J080533 (top-right), DC3272 (bottom-left) and J171941 (bottom-right) isolated molecular clouds, with the open black circles indicating the SUCANES objects, the white open circles indicating the subsample of SUCANES that belong to \cite{Kim2016_VeLLOs} groups A+B, and the red circles corresponding to the subsample of SUCANES that include the proto-BD candidates of Table~\ref{tab:protoBDs} selected in this work.
}
\label{fig:spadistribisolated}
\end{center}
\end{figure*}

\bibliographystyle{elsarticle-harv} 
\bibliography{0_RefsProtoBD}






\end{document}